\documentclass[epj]{svjour}
\usepackage{graphics}
\def\Oi{\Omega_i}
\def\Qi{{\bf Q}_i}
\def\Qk{{\bf Q}_k}
\def\Qz{{\bf Q}}
\def\Q1{{\bf Q}_1}

\def\Qis{{\bf Q}_{i;{\rm s}}}
\def\qj{{\bf q}_j}
\def\phiv{\mbox{\boldmath$\phi$}}
\def\phivs{\mbox{\boldmath${\scriptstyle{\phi}}$}}
\def\piv{\mbox{\boldmath${\scriptstyle{\pi}}$}}

\def\etav{\mbox{\boldmath$\eta$}}
\def\etavs{\mbox{\boldmath${\scriptstyle{\eta}}$}}
\def\Ph{{\hat P}}
\def\LL{{\sf L}}
\def\lala{{\sf \Lambda}}

\def\ui{u_{Ti}}
\def\uqi{u^2_{Ti}}
\def\utm{{\bar u}_T}
\def\utms{\overline{u_T^2}}
\def\gmt{{\bar \gamma}_T}
\def\btt{{\bar \beta}_T}
\def\d{{\rm d}}
\def\D{{\rm D}}
\def\e{{\rm e}}
\def\i{{\rm i}}
\def\vl{\Big\vert}
\def\oV{\overline V}

\def\ee{e$^+$e$^-$}
\def\ss{${\rm s}\bar{\rm s}\;$}
\def\ssb{\langle {\rm s}\bar{\rm s}\rangle}
\def\uub{\langle {\rm u}\bar{\rm u}\rangle}
\def\ddb{\langle {\rm d}\bar{\rm d}\rangle}
\def\ppb{${\rm p}\bar{\rm p}\;$}

\begin{document}

\title{
\vspace*{-3cm} 
\begin{small}
\begin{flushright}
{DFF 376/09/2001} \\
\end{flushright}
\end{small}
\vspace*{2.0cm}
Statistical hadronisation model and transverse momentum 
spectra of hadrons in high energy collisions}
\titlerunning{Statistical hadronisation model...}


\author{F. Becattini\inst{1} \and G. Passaleva\inst{2}}

\institute{University of Florence and INFN Sezione di Firenze 
\and INFN Sezione di Firenze\\
Via G. Sansone 1, I-50019, Sesto F.no, Firenze (Italy)\\
e-mail: becattini@fi.infn.it, passaleva@fi.infn.it}
%
%
\abstract{A detailed analysis of transverse momentum spectra of several 
identified hadrons in high energy collisions within the canonical framework of 
the statistical model of hadronisation is performed. The study of particle 
momentum spectra requires an extension of the statistical model formalism 
used to handle particle
multiplicities, which is described in detail starting from a microcanonical 
treatment of single hadronising clusters. Also, a new treatment of extra
strangeness suppression is presented which is based on the enforcement 
of fixed numbers of \ss pairs in the primary hadrons.   
The considered centre-of-mass energies range from $\simeq$ 10 to 30 GeV
in hadronic collisions ($\pi$p, pp and Kp) and from $\simeq$ 15 to 35 
GeV in \ee collisions. The effect of the decay chain following hadron 
generation is accurately and exhaustively taken into account by a newly
proposed numerical method. The exact ${\bf p}_T$ conservation at low energy 
and the increasing hard parton emission at
high energy bound the validity of the presently taken approach within 
a limited centre-of-mass energy range. However, within this region,
a clear consistency is found between the temperature parameter extracted
from the present analysis and that obtained from fits to average hadron 
multiplicities in the same collision systems. This finding indicates 
that in the hadronisation process the production of different particle species
and their momentum spectra are two closely related phenomena governed
by one parameter.
}
\maketitle
%

\section{Introduction}

The idea of a statistical approach to hadron production in high energy 
collisions dates back to '50s \cite{fermi} and '60s \cite{hage} and 
it has been recently revived by the observation that hadron multiplicities 
in \ee and pp collisions agree very well with a thermodynamical-like ansatz 
\cite{beca1,beca2,beca3}. 
This finding has also been confirmed in hadronic collisions and it has 
been interpreted in terms of a pure statistical filling of multi-hadronic 
phase space of assumed pre-hadronic clusters (or fireballs) formed in 
high-energy collisions, at a critical value of energy density 
\cite{beca3,heinz,stock}. 
In this framework, temperature and other thermodynamical quantities 
have a purely statistical meaning and do not involve the existence 
of a hadronic thermalisation process through multiple collisions on an 
event-by-event basis. Stated otherwise, statistical equilibrium is an 
intrinsic feature of the hadronisation process and hadrons are directly 
created in such a state \cite{beca3,heinz,stock}, as it was envisaged 
by Hagedorn \cite{hage2}.

So far, this proposed statistical cluster hadronisation model has been 
mainly tested against measured abundances of different hadron species 
for a twofold reason. Firstly, unlike momentum spectra, they are quantities 
which are not affected by hard (perturbative) QCD dynamical effects but are 
only determined by the hadronisation process; indeed, in the framework of
a multi-cluster model, they are Lorentz invariant quantities which are 
independent of cluster's overall momentum. Secondly, they are fairly easy 
to calculate and provide a very sensitive test of the model yielding an 
accurate determination of the temperature. However, in order to establish 
the validity of the model, it is necessary to test further observables 
and to assess their consistency with the results obtained for multiplicities. 
One of the best suited observables in this regard is the transverse momentum of 
identified hadrons (where transverse is meant to be with respect to beam 
line in high energy hadronic collision, and thrust or event axis in high 
energy \ee collisions) because, amongst all projections of particle 
momentum, this is supposed to be the most sensitive to hadronisation or, 
conversely, the least sensitive to perturbative QCD dynamics.
 
Actually, it has been known for a long time that transverse momentum 
spectra follow a Boltzmann distribution in hadronic collisions and this very 
observation was pointed out by Hagedorn as a major indication in favour 
of statistical hadron production \cite{hage3}. It must be emphasized 
that the prediction of a thermal-like shape in principle only applies to 
particles directly emitted from the hadronising source, whereas measured 
spectra also include particles produced by decays of heavier hadrons. Yet,
due to involved calculations into play, most analyses do not take into 
account the distortion of primordial hadronisation spectrum induced by 
hadronic decays and try to fit the data straight through it. This 
problem has been discussed in literature \cite{hage4} and an analytical 
calculation has been developed to take into account the effect of two and 
three body decays \cite{sollfran,sollfran2}, which has then been used both 
for pp \cite{sollfran2} and heavy ion collision 
\cite{sollfran,sollfran2,wiedheinz,peitzmann} including most abundant resonances. 
In this paper we introduce a new method allowing to rigorously and exhaustively
determine the contribution of all particle decays. Thence, by taking 
advantage of this technique, we have performed an analysis of many measured 
transverse momentum spectra of identified hadrons in a wide range of 
centre-of-mass energies for several kind of collisions.

The paper is organised as follows: in Sect.~2 the statistical hadronisation
model is described in detail starting from a basic microcanonical treatment 
of clusters. In Sect.~3 the analytical formulae for transverse momentum spectra
of hadrons within the statistical hadronisation model are derived whilst in 
Sect.~4 they are worked out for the primary and secondary component separately
and the numerical method used to calculate them is described. In Sect.~5 and
6 the data analysis is presented and discussed. Conclusions are drawn in Sect.~7.     

\section{Statistical hadronisation model}

The statistical hadronisation model assumes that in high energy 
collisions, as a consequence of strong interaction dynamics, a set of 
colour singlet clusters (or fireballs) endowed with mass, volume, internal 
quantum numbers and momentum, whose distribution is governed by the dynamical
stage of the process, is formed. These clusters are supposed to give rise 
to hadrons according to a pure statistical law in the multi-hadronic phase 
space defined by their four-momentum, volume and quantum numbers. This approach 
differs from another popular cluster hadronisation model \cite{webber} mainly because 
clusters are provided with a volume, so that hadron production is governed by 
proper phase space rather than relativistic momentum space. In this framework, 
the use of statistical mechanics and thermodynamical quantities, such as 
temperature, which needs spacial dimension besides momentum space in order to 
be meaningful, is allowed. We emphasize once more that the introduction of 
such quantities does not entail any thermalisation process of hadrons after 
their formation and that statistical equilibrium simply means that all final
available quantum mechanical states are equally likely.
 
In refs.~\cite{beca3,beca4} the statistical hadronisation model was described 
within a canonical framework, with clusters characterized by temperature and 
volume instead of mass and volume. Therein, it was shown that a particular 
choice of the probabilities of distributing quantum numbers among 
the clusters and the assumption of a common temperature lead to a very simple 
expression for particle multiplicities and global particle correlations which
are in fact the same as those relevant to one cluster having as volume the
sum of the volumes of all clusters in their own rest frame. Furthermore, no 
dependence on single cluster properties nor on their number is left. The 
question arises 
whether such clusters are actually large enough to allow a canonical description. 
In fact, in principle, their mass and volume might be so small to require a 
more appropriate microcanonical framework, enforcing the exact conservation of 
energy and momentum in the calculation of the available multi-hadronic phase space. 
In the following, we will prove that a similar reduction of the expression of particle 
multiplicities holds in the microcanonical case, provided that suitable clusters 
configuration probabilities occur. 

\subsection{From microcanonical to canonical ensemble} 

In the canonical framework of the statistical hadronisation model, the main tool 
to derive physical observables is the partition function $Z$, which is the sum
over all physical states with fixed quantum numbers weighted by $\exp[-E/T]$,
where $E$ is the energy of the state and $T$ the temperature \footnote{in a covariant
formulation this weight is to be replaced by $\exp[-\beta \cdot P]$ where $P$
is the four-momentum and $\beta$ the four-temperature vector.}. 
Similarly, in the microcanonical case, it is possible to introduce the sum over 
all physical states with fixed values of energy-momentum and quantum numbers, 
i.e. the density of states $\Omega$ which, for the $i^{\rm th}$ cluster, reads: 

\begin{eqnarray}\label{omega1}
\Oi && \equiv \Oi(P_i,\Qi,V_i) \nonumber \\
&& = \sum_{\rm states} \delta^4(P_i - P_{i;{\rm s}})\,\,\delta_{\Qi,\Qis} 
\qquad i=1,...,N
\end{eqnarray}
where $P_i$ is the four-momentum of the cluster, $V_i$ is its volume in the
frame where four-momentum is $P_i$ and $\Qi = (Q_{i1},\ldots,Q_{in})$ is a vector 
of its $n$ relevant quantum numbers; $P_{i;{\rm s}}$ and $\Qis$ are the corresponding 
quantities of a general multi-hadronic state which, in the ideal hadron gas 
approximation, is described by a set $\{n_{jk}\}$ 
\footnote{Throughout the paper, by $\{A_i\}$ we mean a shorthand for the vector 
$(A_1,\ldots,A_n)$, either finite or infinite-dimensional.} of occupation numbers 
for each hadronic species $j$ and for each phase space cell $k$, running from 
0 to 1 for fermions and from 0 to $\infty$ for bosons. Hence:

\begin{eqnarray}\label{state}
&& P_{i;{\rm s}} = \sum_{jk} p_{jk} n_{jk}  \nonumber \\
&& \Qis = \sum_{jk} \qj n_{jk}  
\end{eqnarray}
where $\qj$ is the quantum number vector for the $j^{\rm th}$ hadron and $p_{jk}$
its four-momentum in the $k^{\rm th}$ phase space cell. The quantum numbers are 
supposed to be either integer-valued additive conserved quantities in strong 
interactions (namely electric charge, baryon number, strangeness, charm and 
beauty) or positive integer-valued absolute numbers of valence quarks plus 
antiquarks. It is worth remarking that $\Oi$ defined in Eq.~(\ref{omega1}) is 
a number of states per unit four-momentum cell, thus it is a Lorentz invariant 
quantity and has to depend on Lorentz scalars only. Otherwise stated:

\begin{equation}\label{lorentz1}
 \Oi(P_i,\Qi,V_i) = \Oi(P'_i,\Qi,V'_i) = \Oi(P^*_i,\Qi,V^*_i)
\end{equation}  
where $P'_i$ is the Lorentz-transformed of $P_i$ and $V'_i$ is the volume in
the corresponding frame, while $P^*_i$ and $V^*_i$ are the four-momentum 
$(M_i,{\bf 0})$ and the volume respectively in the cluster's rest frame. 

$\Oi$ in Eq.~(\ref{omega1}) can be transformed by means of the integral 
representations of Dirac's and Kronecker's delta:

\begin{eqnarray}\label{deltin}
 && \delta^4(P_i - P_{i;{\rm s}}) = \frac{1}{(2\pi)^4} \int \d^4 x_i \,\,
 \exp \, [\i (P_i - P_{i;{\rm s}})\cdot x_i] \nonumber \\
 && \delta_{\Qi,\Qis} = \frac{1}{(2\pi)^n} \int_{-\pi}^{\pi}\ldots\int_{-\pi}^{\pi} 
 \!\! \d^n \phi_i \,\,
 \exp \, [\i (\Qi -\Qis) \cdot \phiv_i] \nonumber \\
\end{eqnarray}  
By plugging Eqs.~(\ref{deltin}) into Eq.~(\ref{omega1}), the density of states
reads:

\begin{eqnarray}\label{omega2}
\Oi && = \frac{1}{(2\pi)^{4+n}}\int \d^4 x_i \int_{-\piv}^{\piv} 
   \!\!\! \d^n \phi_i \; \exp \big[ \i (P_i\cdot x_i + \Qi \cdot \phiv_i) \big] 
   \nonumber \\ 
&& \times \sum_{\{n_{jk}\}} \prod_{jk} 
 \exp \big[ -\i \, n_{jk} (p_{jk} \cdot x_i + \qj \cdot \phiv_i) \big] 
\end{eqnarray}
The calculation now proceeds by taking advantage of the commutability between
sum and product in Eq.~(\ref{omega2}). However, the sum over occupation number
does not converge to a finite value for bosons as $n_{jk}$ runs from 0 to $\infty$.
The convergence is recovered if the time component of $x_i$ is provided with a 
negative imaginary part $-\i \varepsilon$. If we introduce such a term in  
Eq.~(\ref{omega2}), then the bosonic sums can be performed and $\Oi$ reads:

\begin{eqnarray}\label{omega3}
 \Oi && = \lim_{\varepsilon \to 0} \frac{1}{(2\pi)^{4+n}} 
\int_{-\infty-\i \varepsilon}^{+\infty-\i \varepsilon} \!\!\!\!\!\!\!\!\!\! 
 \d x^0_i  \int \d^3 {\rm x}_i \int_{-\piv}^{\piv} \!\!\! \d^n \phi_i \nonumber \\
&& \times \exp \big[ \i \,(P_i\cdot x_i + \Qi \cdot \phiv_i)+F(x_i,\phiv_i) \big] 
\end{eqnarray}   
where:

\begin{equation}\label{f}
F(x_i,\phiv_i) = \sum_{jk} 
\log \{ 1\pm\exp \big[ -\i\,(p_{jk}\cdot x_i + \qj\cdot\phiv_i) \big]\}^{\pm 1}  
\end{equation}   
In the above and following equations the upper sign applies to fermions and 
the lower to bosons. By taking the continuum limit of the sum over phase 
space cell:

\begin{equation}
  \sum_{k} \rightarrow (2J_j + 1) \frac{V_i}{(2\pi)^3} \int \d^3{\rm p}   
\end{equation} 
the function $F$ finally reads:

\begin{eqnarray}\label{f2}
 && F(x_i,\phiv_i) = \nonumber \\ 
 && = \frac{V_i}{(2\pi)^3} \sum_j (2J_j + 1) \int \d^3{\rm p}
 \,\, \log \, [1\pm \e^{\,-\i\,(p_{j}\cdot x_i + \qj\cdot \phivs_i)}]^{\pm 1} 
   \nonumber \\
 && = \frac{V_i}{(2\pi)^3} \sum_j (2J_j + 1) \sum_{n=1}^\infty 
 (\mp 1)^{n+1} \!\!\! \int \d^3{\rm p} \,\, \e^{\,-\i\,n(p_{j}\cdot x_i + 
 \qj\cdot \phivs_i)} \nonumber \\
 &&
\end{eqnarray}

The density of states $\Oi$ can now be used to obtain physical quantities of
interest, for instance the {\em primary} (i.e. directly emitted by the hadronising
source and not by subsequent hadronic decays) average multiplicity of the 
$j^{\rm th}$ hadron species. Since every multi-hadronic state in the cluster 
has the same probability, it can be easily proved that this can be derived 
multiplying by a fictitious fugacity $\lambda_j$ the exponential factor 
$\exp[-\i \, (p_{jk} \cdot x_i + \qj \cdot \phiv_i)]$ for all $k$ 
in Eq.~(\ref{omega2}), taking the derivative of $\log \Oi$ with respect
to $\lambda_j$ and finally setting $\lambda_j=1$:

\begin{equation}\label{mult1}
\langle n_j^i \rangle = \frac{\partial}{\partial \lambda_j}
\log \Oi (\lambda_j)\Bigg|_{\lambda_j=1}
\end{equation}

Most often, the physical quantities to be compared with experimental measurements 
are not those relevant to individual clusters, rather global ones, that is 
summed over all clusters in the event. This implies that sums over 
clusters with different four-momenta, quantum numbers and proper volumes 
must be performed and we are then led to consider all possible 
cluster configurations in terms of $P$, $V^*$ and $\bf Q$ and their 
probabilities. In the most general picture, one envisages the formation 
of a variable number $N$ of clusters with probability $P_N$. For a fixed 
$N$, there shall be a conditional probability $f_*(\{P_i,\Qi,V^*_i\})$ 
relevant to the configuration $\{P_i,\Qi,V^*_i\}$ of four-momenta $P_i$, 
quantum numbers $\Qi$ and proper volumes $V^*_i$ ($i=1,\ldots N$); 
of course, any configuration must fulfill conservation laws, i.e. 
$\sum_i P_i = P$ and $\sum_i \Qi = \Qz$ where $P$ and $\Qz$ are the initial 
four-momentum and quantum number vector respectively. The corresponding 
distribution function in the variables $\{P_i,\Qi,V_i\}$ is:

\begin{equation}\label{fdis}
 f(\{P_i,\Qi,V_i\}) = f_*(\{P_i,\Qi,V_i \frac{P^0_i}{M_i}\}) 
 \prod_{i=1}^N \frac{P^0_i}{M_i}
\end{equation}
where $M_i=\sqrt{P^2_i}$ is the mass of the $i^{\rm th}$ cluster and $P^0_i$ 
its energy. Thus, the overall average primary multiplicity of hadron $j$ should
be written as the sum of average primary multiplicities of single clusters 
(see Eq.~(\ref{mult1})), in a fixed configuration, weighted by the 
configuration probability $f$ and summed up over all configurations:

\begin{eqnarray}\label{mult2}
\langle n_j \rangle && = \sum_N P_N \Big[ \prod_{i=1}^N \int \d^4 P_i \, \d V_i
 \sum_{\Qi} \Big] f(\{P_i,\Qi,V_i\})  \nonumber \\
&& \times \sum_{i=1}^N \frac{\partial}{\partial \lambda_j}
  \log \Oi (\lambda_j,P_i,\Qi,V_i)\Bigg|_{\lambda_j=1}
\end{eqnarray}          
where the symbol $[ \prod_{i=1}^N \int \d^4 P_i \d V_i \sum_{\Qi} ]$ stands 
for a $N^{\rm th}$-uple integration and sum.
The function $f$ is in general unknown and depends on the dynamics of the 
cluster formation process. It can be further expanded according to the 
well known conditional probability decomposition $P(AB) = P(A|B)P(B)$:

\begin{equation}\label{fcon}
 f(\{P_i,\Qi,V_i\}) = g(\{P_i,\Qi\}|\{V_i\}) \, H(\{V_i\})
\end{equation}
where $H(\{V_i\})$ is the probability distribution for the volumes $\{V_i\}$ 
and $g(\{P_i,\Qi\}|\{V_i\})$ the conditional probability distribution of
four-momenta and quantum numbers $\{P_i,\Qi\}$ once $\{V_i\}$ are fixed. A 
considerable simplification would occur if $g$ were equal to the $w$ function
below:

\begin{eqnarray}\label{w}
 && w(\{P_i,\Qi\}|\{V_i\})_{P} = \nonumber \\
 && = \frac{\delta^4(P - \Sigma_i P_i) \,\, \delta_{\Qz,\Sigma_i\Qi}
 \prod_i \theta(P^0_i) \Oi(P_i,\Qi,V_i)}{\Big[ \prod_k \sum_{\Qk}
 \int \d^4 P_k \, \theta(P^0_k) \Omega_k \Big] \delta^4(P - \Sigma_k P_k)
 \,\, \delta_{\Qz,\Sigma_k\Qk}} \nonumber \\
 && 
\end{eqnarray}    
where $\theta$ is the Heavyside step function. Because of the identity:

\begin{eqnarray}\label{log}
 && \sum_{i=1}^N \frac{\partial}{\partial \lambda_j}\log \Oi (\lambda_j)
 \Bigg|_{\lambda_j=1} = \frac{\partial}{\partial \lambda_j}
 \log \prod_{i=1}^N \Oi(\lambda_j)\vl_{\lambda_j=1} \nonumber \\
 && = \frac{1}{\prod_{i=1}^N \Oi} \frac{\partial}{\partial \lambda_j}
 \prod_{i=1}^N \Oi(\lambda_j)\vl_{\lambda_j=1} 
\end{eqnarray}
the substitution of $g$ in Eq.~(\ref{fcon}) with $w$ in Eq.~(\ref{w}), 
turns Eq.~(\ref{mult2}) into:

\begin{equation}\label{mult3}
 \langle n_j \rangle = \sum_N P_N \Big[ \prod_{i=1}^N \int \d V_i \Big]
 H(\{V_i\}) \frac{\partial}{\partial \lambda_j}\log \Omega(\lambda_j)
 \Bigg|_{\lambda_j=1}
\end{equation}
where $\Omega (\lambda_j)$ is defined as:

\begin{eqnarray}\label{Omega}
\!\!\!\!\!\!\!\!\!\!\! && \Omega (\lambda_j) \equiv \nonumber \\ 
\!\!\!\!\!\!\!\!\!\!\! && \equiv \Big[ \prod_{i=1}^N \sum_{\Qi} 
 \int \d^4 P_i \, \theta(P^0_i) \, \Oi (\lambda_j) \Big] 
 \delta^4(P \! - \!\!\Sigma_i P_i) \,\, \delta_{\Qz,\Sigma_i\Qi} 
\end{eqnarray}
It is a remarkable fact that $\Omega(1)\equiv \Omega$ is exactly the density 
of states of a single large cluster, here defined as {\em equivalent global cluster} 
(EGC) with four-momentum $P = \sum_i P_i$, quantum numbers $\Qz = \sum_i \Qi$ 
and volume (in the reference frame where four-momentum is $P$) $V = \sum_i V_i$
and can thus be written in the same fashion as $\Oi$ in Eq.~(\ref{omega3}):

\begin{eqnarray}\label{Omega2}
\Omega && = \lim_{\varepsilon \to 0} \frac{1}{(2\pi)^{4+n}} 
\int_{-\infty-\i \varepsilon}^{+\infty-\i \varepsilon} \!\!\!\!\!\!\!\!\!\! 
 \d x^0  \int \d^3 {\rm x} \int_{-\piv}^{\piv} \!\!\! \d^n \phi \nonumber \\
&& \times \exp \big[ \i \,(P \cdot x + \Qz \cdot \phiv)+F (x,\phiv) \big]  
\end{eqnarray}   
where:
\begin{eqnarray}\label{f3}
  && F(x,\phiv) = \nonumber \\ 
  && = \frac{\sum_i V_i}{(2\pi)^3} \sum_j (2J_j + 1) \int \d^3{\rm p}
  \,\, \log \, [1\pm \e^{\,-\i\,(p_{j}\cdot x + \qj\cdot \phivs)}]^{\pm 1} 
  \nonumber \\
  &&
\end{eqnarray}   
The fact that $\Omega$ is the density of states of the EGC can be proved in an 
elegant way by showing that the $w$'s in Eq.~(\ref{w}) are just the probabilities 
of getting a set of four-momenta $\{P_i\}$ and of quantum numbers $\{\Qi\}$ if a 
cluster of volume $V$, four-momentum $P$ and quantum numbers $\Qz$ is {\em randomly} 
split into $N$ sub-clusters with fixed volumes $\{V_i\}$ such that 
$V = \sum_i V_i$, for they maximise total entropy. We stress that the additivity 
of volumes in Eq.~(\ref{f3}) applies to $V_i$'s and not to proper ones $V^*_i$'s 
because the splitting ideally takes place with no spacial overlap between sub-clusters 
and Lorentz contraction must be taken into account. Let now $w(\{P_i,\Qi\}|\{V_i\})_{P}$ 
be such unknown probabilities and let us calculate the probability $p$ of a full 
microscopic state assuming, for sake of simplicity, that $\{P_i\}$ are discrete 
variables. According to the basic law of statistical mechanics, $p$ turns out to be: 

\begin{equation}
   p = w(\{P_i,\Qi\}|\{V_i\})_P \prod_{i=1}^N 
   \frac{\delta^4(P_i - P_{i;{\rm s}}) \,\, \delta_{\Qi,\Qis}}{\Oi(P_i,\Qi,V_i)}
\end{equation} 
where $w(\{P_i,\Qi\}|\{V_i\})_{P}$ vanishes if $P \neq \sum_i P_i$ or $P^0_i<0$
or $\Qz \neq \sum_i \Qi$ and: 

\begin{equation}
   \sum_{\{P_i,\Qi\}} \!\! w(\{P_i,\Qi\}|\{V_i\})_{P} = 1
\end{equation}   
The entropy $S = - \sum p \log p$ should be calculated by summing over all 
possible configurations $\{P_i,\Qi\}$ fulfilling the constraints on total 
four-momentum and quantum numbers, and, once a configuration is fixed, over all 
possible microscopic states of the clusters. Therefore: 

\begin{eqnarray}
\!\!\!\!\!\!\!\! S && = - \!\!\!\! \sum_{\{P_i,\Qi\}} \!\! w \prod_{i=1}^N 
 \sum_{{\rm states}_i} \frac{\delta^4(P_i - P_{i;{\rm s}}) \delta_{\Qi,\Qis}}{\Oi} 
  \nonumber \\
\!\!\!\!\!\!\!\! && \times \log \Bigg[ w \prod_{i=1}^N \frac{\delta^4(P_i - P_{i;{\rm s}}) 
 \delta_{\Qi,\Qis}}{\Oi} \Bigg] = \nonumber \\
\!\!\!\!\!\!\!\! && =  - \!\!\!\! \sum_{\{P_i,\Qi\}} \!\! w \Bigg[ \log w + 
 \log \prod_{i=1}^N \frac{\delta^4(P_i - P_{i;{\rm s}}) \delta_{\Qi,\Qis}}{\Oi} 
 \Bigg]
\end{eqnarray}
where the arguments of $w$ and $\Oi$ are implied. In the above equation 
advantage has been taken of the fact that logarithm's argument is actually
independent of the microscopic states of the clusters. In order to determine 
the $w$'s, $S$ must be maximised with respect to all of them with the constraint 
$\sum w = 1$. This can be done by means of the Lagrange multiplier method which
leads to the equation: 

\begin{eqnarray}
&& \frac{\partial S}{\partial w(\{P_i,\Qi\}|\{V_i\})_{P}} + \mu =    \nonumber \\
&& = - 1 - \log w - \log \prod_{i=1}^N  \delta^4(P_i - P_{i;{\rm s}}) 
 + \log \prod_{i=1}^N \Oi + \mu = 0 \nonumber \\
&&
\end{eqnarray} 
implying that:

\begin{equation}
 w \propto \prod_{i=1}^N \Oi 
\end{equation}
The above equation, after a due normalization and taking into account that
the $w$'s must vanish if $P \neq \sum_i P_i$ or $P_i^0 < 0$ or $\Qz \neq 
\sum_i \Qi$, coincides with Eq.~(\ref{w}) and this proves our statement. A 
different proof based on a direct calculation starting from Eq.~(\ref{Omega}) 
can be found in Appendix A.

By using Eq.~(\ref{Omega}) with $\lambda_j=1$, Eq.~(\ref{w}) can be written as:

\begin{eqnarray}\label{w2}
&& w(\{P_i,\Qi\}|\{V_i\})_{P} = \nonumber \\
&& = \frac{\delta^4(P - \Sigma_i P_i) \,\, 
 \delta_{\Qz,\Sigma_i\Qi} \prod_i \theta(P^0_i) \Oi(P_i,\Qi,\Sigma_i V_i)}
 {\Omega(P,\Qz,V)}
\end{eqnarray} 
and Eq.~(\ref{mult2}) as:

\begin{eqnarray}\label{mult4}
 \langle n_j \rangle && = \sum_N P_N \int \d V 
 \Big[ \prod_{i=1}^N \int \d V_i \Big] H(\{V_i\}) \delta (V -\Sigma_i V_i)
 \nonumber \\
 && \times \frac{\partial}{\partial \lambda_j} \log \Omega (\lambda_j,P,\Qz,V)
 \Bigg|_{\lambda_j=1} \nonumber \\ 
 &&= \sum_N P_N \int \d V \; h_N(V) \, 
 \frac{\partial}{\partial \lambda_j}\log \Omega (\lambda_j,P,\Qz,V)
 \Bigg|_{\lambda_j=1}
\end{eqnarray}
where $h_N(V) \equiv [\prod_i \int \d V_i] H(\{V_i\}) \delta (V -\Sigma_i V_i)$
is, by definition, the probability distribution for the EGC with volume $V$ split
into $N$ sub-clusters. Since there is no further effective dependence left on $N$ 
in the integrand of Eq.~(\ref{mult4}), one can define $h(V) = \sum_N P_N h_N(V)$ 
so that the average primary multiplicity of the $j^{\rm th}$ hadron species finally reads:

\begin{equation}\label{mult5}
  \langle n_j \rangle = \int \d V \; h(V) \, 
  \frac{\partial}{\partial \lambda_j}\log \Omega (\lambda_j,P,\Qz,V)
  \Bigg|_{\lambda_j=1}
\end{equation}
This a noteworthy result because multiplicity in Eq.~(\ref{mult5}) no longer 
depends on the configuration of all clusters in the event nor on 
their number. Instead, it depends on much fewer parameters, only those 
of the EGC (its total four-momentum, volume and quantum numbers).

It is apparent from the previous derivation that the equivalence between a 
many-cluster system and one global cluster ultimately rests on the occurrence 
of configuration probabilities (\ref{w}). Furthermore, it can be realised by 
inspecting Eqs.~(\ref{mult2}-\ref{Omega}) that the equivalence holds in
general for any observable $A$ (not necessarily a Lorentz scalar), which can 
be written, for a given event configuration $\{P_i,\Qi,V_i\}$, as follows:

\begin{equation}\label{linear}
  A(\{P_i,\Qi,V_i\}) =  \frac{{\cal L} \left(\prod_{i=1}^N \Oi(P_i,\Qi,V_i)\right)}
  {\prod_{i=1}^N \Oi(P_i,\Qi,V_i)}
\end{equation}  
where ${\cal L}$ is a linear operation, e.g. derivation or integration. If this
is the case, the observable averaged over all configurations with the probabilities
(\ref{w}) reads:

\begin{equation}\label{genobs}
  \langle A \rangle = \int \d V \; h(V) \, \frac{{\cal L}
  \left(\Omega(P,\Qz,V)\right)}{\Omega(P,\Qz,V)}
\end{equation} 
Amongst such observables, one of the most important and general is the 
multi-hadronic probability distribution. For a given event configuration 
$\{P_i,\Qi,V_i\}$ it reads:

\begin{eqnarray}\label{multiprob}
 && P(N_1,\ldots,N_K) = \nonumber \\
 && = \Bigg[ \frac{1}{2\pi \i} \prod_{j=1}^K \oint \frac{\d \lambda_j}{\lambda_j^{N_j+1}}
 \Bigg] \prod_{i=1}^N \frac{\Oi(\{\lambda_j\},P_i,\Qi,V_i)}{\Oi(P_i,\Qi,V_i)}
\end{eqnarray}
In the above expression $K$ is meant to be the total number of hadron species; 
$\lambda_j$ are complex variables and the integration is taken on a closed 
path around the origin; the function $\Oi(\{\lambda_j\})$ is a 
generalisation of the previously used $\Oi(\lambda_j)$ with the insertion of 
fictitious fugacities of all hadron species at the same time. In fact, 
Eq.~(\ref{multiprob}) has the form required in Eq.~(\ref{linear}) for the
equivalence with EGC to apply. Eq.~(\ref{multiprob}) can be obtained by 
inverting $P$'s generating function $G(\lambda_1,\ldots,\lambda_K)$:

\begin{eqnarray}
\!\!\!\!\!\!\!\!\!\!\! && G(\lambda_1,\ldots,\lambda_K) \equiv \sum_{N_1,\ldots,N_K} \!\!\! 
 P(N_1,\ldots,N_K) \, \lambda_1^{N_1} \ldots \lambda_K^{N_K} \nonumber \\
\!\!\!\!\!\!\!\!\!\!\! && = \sum_{N_1,\ldots,N_K} \!\!\! \Big[ \sum_{\rm states} 
  P({\rm state})\Big|_{{\rm fixed}
  (N_1,\ldots,N_K)} \Big] \lambda_1^{N_1} \ldots \lambda_K^{N_K} \nonumber \\
\!\!\!\!\!\!\!\!\!\!\! && = \sum_{\rm states} \!\!\! P({\rm state}) \, \lambda_1^{N_1} 
 \ldots \lambda_K^{N_K} 
 \!\! = \prod_{i=1}^N \frac{\Oi(\{\lambda_j\},P_i,\Qi,V_i)}{\Oi(P_i,\Qi,V_i)}   
\end{eqnarray}
The last equality follows from:

\begin{equation}
 P({\rm state}) \prod_{j=1}^K \lambda_j^{N_j} = 
 \prod_{j=1}^K \lambda_j^{N_j} \prod_{i=1}^N \frac{\delta^4(P_i - P_{i;{\rm s}}) 
 \, \delta_{\Qi,\Qis}}{\Oi(P_i,\Qi,V_i)} 
\end{equation}
which is to be worked out thereafter as in Eqs.~(\ref{omega1}-\ref{omega3}). 

Besides the reduction in the number of parameters, the equivalence with EGC 
has another attracting feature: for the canonical formalism to be a sufficiently 
accurate approximation, only the EGC has to be large, while there is no need to 
enforce a similar request for each individual cluster. Therefore, one would be 
allowed to treat canonically (as long as cluster-integrated quantities are 
concerned) even hadronising systems in wich single clusters are {\em a priori} known 
to be too small for a canonical treatment to apply individually. In fact, the 
transition from microcanonical to canonical ensemble is based on an asymptotic 
expansion of $\Omega$ for large values of EGC volume and mass, through the
saddle-point method, in which only the leading order is retained. To show
this, first a rotation $z = \i x$ in the complex hyperplane for the four-dimensional 
integration in Eq.~(\ref{Omega2}) is performed:
 
\begin{eqnarray}\label{Omegac}
  && \Omega = \nonumber \\ 
  && = \lim_{\varepsilon \to 0} \frac{1}{(2\pi\i)^{4}} 
  \int_{-\i\infty + \varepsilon}^{+\i\infty + \varepsilon} \!\!\!\!\!\!\!\!\!\!\!\! 
  \d^4 z \int_{-\piv}^{\piv} \! \frac {\d^n \phi}{(2\pi)^n} \;
  \e^{P \cdot z + \i \Qz \cdot \phivs + F(-\i z,\phivs)} = \nonumber \\
  && = \lim_{\varepsilon \to 0} \frac{1}{(2\pi\i)^{4}} 
  \int_{-\i\infty + \varepsilon}^{+\i\infty + \varepsilon} \!\!\!\!\!\!\!\!\!\!\!\! 
  \d^4 z \,\, \exp \, [P \cdot z + \log Z(z,\Qz)]
\end{eqnarray} 
where:

\begin{equation}\label{zcan}
 Z(z,\Qz) = \frac{1}{(2\pi)^n} \int_{-\piv}^{\piv} \!\!\! \d^n \phi \; 
 \exp \, [\i \, \Qz \cdot \phiv + F(-\i z,\phiv)]
\end{equation}
One can recognize in Eq.~(\ref{zcan}), by looking at Eq.~(\ref{f3}) for $F$, 
the expression of the canonical partition function \cite{beca2,beca3,beca4} of an 
ideal hadron gas calculated for a {\em complex four-temperature} $z$. Now, 
the saddle-point asymptotic expansion of $\Omega$ in Eq.~(\ref{Omegac}) can be 
performed which, at the leading order, reads:

\begin{equation}\label{Omegacan}
 \Omega \simeq \exp \,[P \cdot \beta + \log Z (\beta, \Qz) ] 
 \sqrt{\frac{1}{(2\pi)^4 \det {\sf H (\beta,\Qz)}}} 
\end{equation}
where $\beta$ is a four-vector such that:

\begin{eqnarray}\label{encan}
 && \!\!\!\!\!\!\!\!\!\! \frac{\partial}{\partial z^\mu}\Big[P \cdot z + 
 \log Z (z,\Qz)\Big]\Big|_{z=\beta} = \nonumber \\
 && \!\!\!\!\!\!\!\!\!\! = P_{\mu} + \frac{\partial}{\partial z^\mu} 
   \log Z(z,\Qz)\Big|_{z=\beta} = 0 \qquad \mu=0,1,2,3
\end{eqnarray}
and $\sf H$ is the hessian matrix $\partial \log Z /\partial z^\mu \partial z^\nu$
calculated for $z=\beta$. Since $F (-\i z, \phiv)$ is real valued (see Eq.~(\ref{f3})) 
if $z$ is a real four-vector and $\log Z (z, \Qz)$ is real too, according to 
Eq.~(\ref{zcan}), then Eq.~(\ref{encan}) states that $\beta$ must be a real 
four-vector. Moreover, it has to be a timelike vector for the momentum integration
in Eq.~(\ref{f3}) to converge. If $\beta$ is identified with the four-temperature
vector $(1/T) \hat u$, $\hat u$ being a unit timelike vector, the second term in 
Eq.~(\ref{encan}) is nothing but the expression, with negative sign, of the mean 
energy-momentum of a canonical system with four-temperature $\beta$. In other words, 
the temperature can be defined for a microcanonical system (at the lowest order of 
an asymptotic saddle-point expansion) equating the given total energy to its 
expression for a canonical system having same volume $V$ and the same set of 
quantum numbers $\Qz$. The canonical expression of entropy 
$S= P \cdot \beta + \log Z$ can be recognized in Eq.~(\ref{Omegacan}) so that 
the well known Boltzmann formula is recovered:

\begin{equation}
 \Omega \propto \exp S
\end{equation} 
The square root factor in Eq.~(\ref{Omegacan}) can be neglected in the 
derivation of most physical observables, such as multiplicities, because it
depends on a fractional power of the partition function $Z$ whilst the first 
factor is an exponential of it. Altogether, the expressions of multiplicities, 
multiplicity distributions etc. in the canonical ensemble can be recovered. 
The canonical partition function~(\ref{zcan}) can also be written as follows:

\begin{eqnarray}\label{zpart}
 && Z(\beta,\Qz,V) = \sum_{\rm states} \e^{-\beta_{\utm} \cdot P_{\rm state}} 
  \, \delta_{\Qz,{\bf Q}_{\rm state}} \nonumber \\
 && = \sum_{\rm states} \int \d^4 P \; \e^{-\beta \cdot P} \,
 \delta^4(P-P_{\rm state}) \,\, \delta_{\Qz,{\bf Q}_{\rm state}} \nonumber \\
 && = \int \d^4 P \; \e^{-\beta \cdot P} \, \Omega(P,\Qz,V)
\end{eqnarray}  
As has been mentioned, the volume $V$ argument of $\Omega$ in Eq.~(\ref{zpart}) is 
measured in the frame where four-momentum is $P$. Indeed, there is a subtle 
difference between what is meant as proper volume in canonical and microcanonical
ensemble. In the microcanonical ensemble the definition is clear because all states
have a definite total momentum and the reference frame where it vanishes can be 
chosen. On the other hand, in the canonical ensemble, we can choose a reference 
frame where $\beta = (1/T, {\bf 0})$ but this does not ensure that total momentum
exactly vanish for all the states (it does so {\em on the average} with
a small broadening around $0$); hence, the actual proper volume (measured with 
${\bf P} = 0$) does not coincide with the parameter volume used in the canonical 
partition function~(\ref{zpart}) with $\beta = (1/T, {\bf 0})$. Nevertheless, if 
we think of a canonical ensemble as an approximation of a microcanonical ensemble 
in its rest frame with associated volume $V^*$ (according to Eq.~(\ref{Omegacan})), 
the four-vector $\beta$ solution of Eq.~(\ref{encan}) has vanishing spacial part. 
This means that the use of proper frame and proper volume $V^*$ in $\Omega$ and $Z$ in 
Eqs.~(\ref{Omegacan}) goes together with proper four-temperature $\beta = (1/T, 
{\bf 0})$.
   
Two major points are worth being stressed. Firstly, as we have emphasised, 
the reduction to an EGC possibly allows a canonical treatment for Lorentz invariant
quantities even though physical cluster must be treated microcanonically and 
this implies that temperature may be a well defined quantity only in a global
sense (at the level of EGC) while locally, at the level of single cluster, one 
should stick to microcanonically well defined concepts such as energy density. 
Secondly, a quantitative estimate of how large an EGC should be in terms
of volume and mass for the canonical approximation to be satisfactory is highly
desirable but it is not available by now. Nevertheless, as the hadron gas has a 
huge number of degrees of freedom, it can be reasonably expected that the 
validity of a canonical treatment should set in at relatively low values of 
volume and mass, though settling this issue definitely requires very involved 
microcanonical calculations. At present, the legitimacy of the canonical approximation 
essentially relies on the agreement with the data.

\begin{figure}
 \resizebox{0.4\textwidth}{!}{%
 \includegraphics{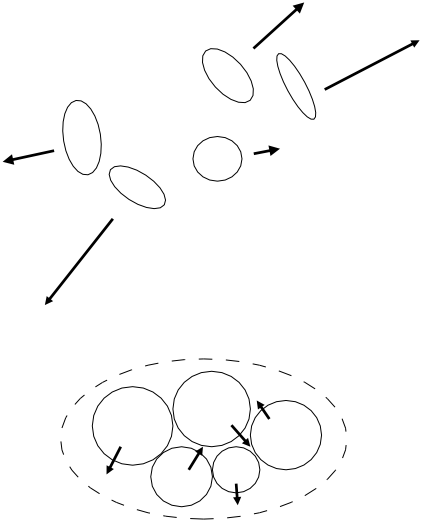}}
\caption{Top: the typical configuration of the momenta of clusters 
in an actual high energy collision. Bottom: the configuration of the
momenta of the same clusters originated from the splitting of one global 
cluster at rest. The two configurations are indeed equivalent for Lorentz
invariant observables like hadron multiciplities.}
\label{clust}
\end{figure}

\subsection{Back-boosting clusters}

As we have seen, the choice of probabilities~(\ref{w}) for the configurations 
of the produced set of clusters is essential for the canonical approximation 
to apply. However, those configuration 
probabilities are unrealistic because hadrons emitted in a high energy
\ee or hadronic collision should look like coming from one source at rest in the 
centre-of-mass frame whereas they typically emerge in a two (or more) jet-like 
structure (see Fig.~\ref{clust}). Nevertheless, as long as one is interested in 
Lorentz invariant quantities such as particle multiplicities, it is possible to 
rearrange cluster momenta at leisure because there is no dependence on them as 
demonstrated by Eqs.~(\ref{lorentz1},\ref{mult1}): the effective arguments of 
density of states and, consequently, average multiplicities, are mass and proper 
volume of the cluster. Then, the question arises whether it is possible to find a 
suitable rearrangements of cluster momenta so as to get expression for particle 
primary multiplicities equal to those obtained before in 
Eqs.~(\ref{mult3},\ref{mult4},\ref{mult5}) for the EGC starting from a configuration 
probability distribution whatsoever instead of the particular one (\ref{w}).\\
This question may be restated more quantitatively as follows: whether, given a 
general distribution of cluster four-momenta such as $f_*$ in Eq.~(\ref{fdis}), 
it is possible to find a suitable rearrangement of cluster momenta in each event 
such that:

\begin{eqnarray}\label{backb}
  && \Bigg[\prod_{i=1}^N \int \d^4 P_i \Bigg] f_*(\{P_i,\Qi,V^*_i\}) \,\,
    O(\{M_i,\Qi,V^*_i\}) = \nonumber \\
  = &&\Bigg[\prod_{i=1}^N \int \d^4 P_i \Bigg] 
    u_*(\{P_i,\Qi,V^*_i\}) \,\, O(\{M_i,\Qi,V^*_i\})
\end{eqnarray}
where $u_*$ is a known distribution, $M_i=\sqrt{P^2_i}$ is the mass of the 
$i^{\rm th}$ cluster and $O$ an arbitrary Lorentz invariant observable
dependent on cluster configurations. If the answer was affirmative, the task 
of reducing calculations of average Lorentz invariant quantities to those of the 
EGC could be accomplished by choosing $u_*$ suitably related to conditional
probabilities $w$ in Eq.~(\ref{w}). Thereby, an effective cluster back-boosting 
from configurations as in Fig.~\ref{clust} top to Fig.~\ref{clust} bottom
could be achieved. It must be pointed out, however, that $f_*$ and $u_*$ cannot 
be completely independent of each other. In fact, since $O$ in Eq.~(\ref{backb}) 
is an arbitrary Lorentz invariant observable depending on $\{P^2_i\}$, the 
marginal distributions $f_{*M}(\{M_i,\Qi,V^*_i\})$ and 
$u_{*M}(\{M_i,\Qi,V^*_i\})$ obtained by integrating out the directions of all 
$P_i$'s in Minkowski space, must be equal, i.e.:

\begin{equation}\label{mass}
  f_{*M}(\{M_i,\Qi,V^*_i\}) = u_{*M}(\{M_i,\Qi,V^*_i\})
\end{equation} 
and this means that the actual $f_*$ cannot be completely arbitrary for the 
reduction to EGC to apply. This point will be discussed more in detail later on.
 
In principle, the distributions $f_*$ and $u_*$ can be linked by a linear 
transformation through an unknown matrix $X$ depending only on four-momenta 
directions $\Ph, \Ph'$ in view of Eq.~(\ref{mass}):

\begin{eqnarray}\label{eqint1}
 && f_*(\{P_i,\Qi,V^*_i\}) = \nonumber \\
 && = \Bigg[\prod_{i=1}^N \int \d^4 \Ph'_i \Bigg] X(\Ph_i,\Ph'_i) \,\, 
    u_*(\{P'_i,\Qi,V^*_i\})
\end{eqnarray} 
Provided that Eq.~(\ref{mass}) is fulfilled, there are actually infinitely many 
matrices $X$ satisfying Eq.~(\ref{eqint1}) and this enables us setting further 
requirements. For this purpose, it will be assumed that the dependence of $X$ on 
$\Ph_i$ and $\Ph'_i$ is realised only through Lorentz transformations $\LL$ of the 
type (see ref.~\cite{group}):

\begin{equation}\label{boost}
  {\sf L} = {\sf \hat R}_3 (\varphi) \, {\sf \hat R}_2 (\theta) \, 
  {\sf \hat L}_3 (\xi) 
\end{equation}
transforming $\Ph'_i$ into $\Ph_i$; ${\sf \hat L}_3(\xi)$ is a Lorentz boost along 
the $z$ axis with positive hyperbolic sector $\xi$ and ${\sf \hat R}_k$ are 
rotations around $k^{\rm th}$ axis with angles $\varphi \in [0,2\pi)$ and 
$\theta \in [0,\pi]$ \footnote{The axis index $k$ here is not to be confused with 
the index $i$ in following equations referring to the cluster which the boost is 
applied to.}. Lorentz transformations of the type~(\ref{boost}) allow to transform 
the time axis unit vector $\hat t= (1,0,0,0)$ into a given arbitrary timelike vector
\cite{group}. Conversely, there is only one Lorentz transformation of this type 
transforming $\Ph'_i$ into $\Ph_i$ (see Appendix B). Hence, Eq.~(\ref{eqint1}) 
is rewritten as: 

\begin{eqnarray}\label{eqint3}
 && f_*(\{P_i,\Qi,V^*_i\}) = \nonumber \\
 && = \Bigg[ \prod_{i=1}^N \int \d \LL_i \Bigg] X(\{\LL_i\}) \,\,
  u_*(\{\LL^{-1}_i(P_i),\Qi,V^*_i\}) 
\end{eqnarray} 
where $\d \LL$ is the group measure for the Lorentz transformations of the 
type~(\ref{boost}), namely \cite{group2}:

\begin{equation}\label{measure}
    \d \LL = \frac{\d [\LL({\hat t})]^1 \d [\LL({\hat t})]^2 
    \d [\LL({\hat t})]^3}{\LL({\hat t})^0}
\end{equation}
Therefore, the problem stated at the beginning of the this subsection in 
Eq.~(\ref{backb}) has been transformed into the quest of a solution $X$ of the 
multiple-integral equation~(\ref{eqint3}), which is formally a Fredholm integral 
equation of the first kind. If a solution exists, then Eq.~(\ref{backb}) is fulfilled 
for any Lorentz invariant observable $O$. It should be noted that if $X$ satisfies 
Eq.~(\ref{eqint3}), then:

\begin{equation}\label{norm2}
 \Bigg[ \prod_{i=1}^N \int \d \LL_i \Bigg] X(\{\LL_i\}) = 1
\end{equation}
which can be obtained from Eq.~(\ref{eqint3}) by enforcing the normalization 
constraints on both $f_*$ and $u_*$:

\begin{equation}
  \Bigg[\prod_{i=1}^N \sum_{\Qi} \int \d^4 P_i \, \d V^*_i \Bigg] 
  \left\{ \begin{array}{l} f_* \\ u_* \end{array} \right\}
   (\{P_i,\Qi,V^*_i\}) = 1
\end{equation}
Henceforth, we will assume that a solution $X$ of the Fredholm integral equation 
(\ref{eqint3}) exists and that it is a positive definite function. In fact, if
$X > 0$, Eq.~(\ref{eqint3}) means that $f_*$ is a probability distribution
of cluster four-momenta obtained by boosting clusters, primordially generated with
a four-momenta probability distribution $u_*$, with Lorentz transformations 
distributed with probability $X$. The requirement of positive definite $X$ 
makes integral equation (\ref{eqint3}) not always solvable and it is not difficult
to devise actual examples. For instance, if $u_*$ is the distribution associated
with the splitting of an EGC into $N$ sub-clusters and $f_*$ is a distribution
characterized by constantly vanishing momenta of the clusters, no probability 
distribution $X$ is able to transform $u_*$ into $f_*$ through~(\ref{eqint3}) 
because random boosts to the (small) non-vanishing momenta involved in $u_*$ 
yield non-vanishing momenta as well.      

In order to accomplish the reduction of calculations of average Lorentz invariant 
quantities to the EGC, we now have to specify the relationship between $u_*$ and 
$w$ in Eq.~(\ref{w}). As a first step, let us define the new variables:

\begin{equation}\label{varia}  
   P'_i = \LL_i^{-1}(P_i) \qquad  V'_i = V^*_i \frac{M_i}{\LL_i^{-1}(P_i)^0} 
   = V^*_i \frac{M'_i}{P'^0_i} 
\end{equation}
along with the distribution $u$ corresponding to $u_*$ in the $V'_i$ volumes
(in analogy with Eq.~(\ref{fdis})): 

\begin{equation}\label{udis}
  u(\{P'_i,\Qi,V'_i\}) = u_*(\{P'_i,\Qi,V'_i P'^0_i/M'_i\}) \prod_{i=1}^N 
  \frac{P'^0_i}{M'_i} 
\end{equation} 
Evidently, $V'_i$ are the cluster volumes measured in the frame where their 
four-momentum are $P'_i$ and the very reason of their introduction resides in 
their additivity, as it will become clear later. The distribution $u$ can be 
expanded by means of conditional probabilities, in the same fashion as 
Eq.~(\ref{fcon}):

\begin{equation}\label{ucon}
   u(\{P'_i,\Qi,V'_i\}) = v(\{P'_i,\Qi\}|\{V'_i\} ) \, Y(\{V'_i\})
\end{equation}   
In the second step, let us calculate the mean value of a general observable 
$A$ by using Eq.~(\ref{eqint3}):

\begin{eqnarray}\label{genaver}
\!\!\!\! \langle A \rangle && = \sum_N P_N \Bigg[\prod_{i=1}^N \sum_{\Qi} \int 
 \d^4 P_i \, \d V^*_i \Bigg] f_*(\{P_i,\Qi,V^*_i\}) \,\, \nonumber \\
\!\!\!\! && \times A(\{P_i,\Qi,V^*_i M_i/P^0_i\}) \nonumber \\
\!\!\!\! && = \sum_N P_N \Bigg[\prod_{i=1}^N \sum_{\Qi} \int \d^4 P_i \, \d V^*_i 
   \d \LL_i \Bigg] X(\{\LL_i\}) \,\, \nonumber \\
\!\!\!\! && \times u_*(\{\LL^{-1}_i(P_i),\Qi,V^*_i\}) \, A(\{P_i,\Qi,V^*_i M_i/P^0_i\})  
\end{eqnarray}   
where $P_N$ is the probability for $N$ clusters to be produced in an event,
(see Sect. 1). The variables defined in Eq.~(\ref{varia}) and 
Eqs.~(\ref{udis},\ref{ucon}) are now used to turn previous equation into:

\begin{eqnarray}\label{mean2}
 && \langle A \rangle = \sum_N P_N \Bigg[\prod_{i=1}^N \sum_{\Qi} \int \d^4 P'_i \, 
  \d V'_i \d \LL_i \Bigg] X(\{\LL_i\}) \, Y(\{V'_i\}) \, \nonumber \\
 && \times  v(\{P'_i,\Qi\}|\{V'_i\} ) \, 
   A(\{\LL_i(P'_i),\Qi,V'_i P'^0_i/\LL_i(P'_i)^0\})
\end{eqnarray} 
No weighting factor is implied in the $P'_i = \LL_i^{-1}(P_i)$ transformation as 
$| \det \LL_i | = 1$. If $A$ is a Lorentz invariant observable like $O$ in 
Eq.~(\ref{backb}), then it must be independent of $\LL_i$ and the integration 
of $X$ can be performed separately yielding 1 in accordance with Eq.~(\ref{norm2}). 
Hence:

\begin{eqnarray}\label{mean3}
 \langle O \rangle && = \sum_N P_N \Bigg[\prod_{i=1}^N \sum_{\Qi} \int \d^4 P'_i \, 
 \d V'_i \Bigg] v(\{P'_i,\Qi\}|\{V'_i\}) \, \nonumber \\
 && \times Y(\{V'_i\}) \, O(\{M'_i,\Qi,V'_i P'^0_i/M'_i\})   
\end{eqnarray}
The form of Eq.~(\ref{mean3}) suggests that one should set $v=w$. Indeed, 
if $O$ can be written as in Eq.~(\ref{linear}), the whole derivation in Subsect. 2.1 
for average primary multiplicities can be repeated, leading to: 

\begin{eqnarray}\label{mean4}
 \langle O \rangle && = \sum_N P_N \Bigg[\prod_{i=1}^N \sum_{\Qi} \int \d^4 P'_i \, 
  \d V'_i \Bigg] w(\{P'_i,\Qi\}|\{V'_i\})_{P'} \, \nonumber \\
 && \times Y(\{V'_i\}) \, O(\{M'_i,\Qi,V'_i P'^0_i/M'_i\}) \nonumber \\
 && = \Bigg[\prod_{i=1}^N \int \d V'_i \Bigg] Y(\{V'_i\}) \, 
     O(M',\Qz,P'^0\Sigma_i V'_i/M') \nonumber \\
 && = \int \d V' \; y(V') \, O(M',\Qz,V'P'^0/M')    
\end{eqnarray}
where $O$ in the last expression is meant to be the same observable calculated 
in an EGC with four-momentum $P'$, mass $M'=\sqrt{P'^2}$, quantum 
numbers $\Qz=\sum_i \Qi$, volume $V'=\sum_i V'_i$ (in the frame where four-momentum 
is $P'$) and:

\begin{equation}\label{ipsilon}
 y(V') \equiv \sum_N P_N \Bigg[\prod_{i=1}^N \int \d V'_i \Bigg] Y(\{V'_i\}) \,
\delta (V' -\Sigma_i V'_i)
\end{equation}
Now we can finally write the sought relationship between $u_*$ and $w$ on the
basis of Eqs.~(\ref{udis},\ref{ucon}) and the condition $v=w$:

\begin{eqnarray}\label{relation}
 && u_*(\{P_i,\Qi,V^*_i\}) =  \nonumber \\
 && = w(\{P_i,\Qi\}|\{V^*_i \frac{M_i}{P^0_i}\})_{P'} 
 Y(\{V^*_i \frac{M_i}{P^0_i}\}) \prod_{i=1}^N \frac{M_i}{P^0_i} 
\end{eqnarray}
Therefore, for any given $f_*$ configuration probability distribution, the
integral equation (\ref{eqint3}) for the reduction to the EGC finally becomes:

\begin{eqnarray}\label{eqint4}
 && \!\!\!\!\!\!\!\!\!\!\! 
  f_*(\{P_i,\Qi,V^*_i\}) = \Bigg[ \prod_{i=1}^N \int \d \LL_i 
 \frac{M_i}{\LL_i^{-1}(P_i)^0} \Bigg] X(\{\LL_i\}) \nonumber \\
 && \!\!\!\!\!\!\!\!\!\!\!\!\! \times 
 w(\{\LL_i^{-1}(P_i),\Qi\}|\{V^*_i \frac{M_i}{\LL_i^{-1}(P_i)^0}\})_{P'} 
 Y(\{V^*_i \frac{M_i}{\LL_i^{-1}(P_i)^0}\}) \nonumber \\  
 && 
\end{eqnarray}
with $X$ function to be determined. 
Whilst $w$ is given by Eq.~(\ref{w}), $Y$ is determined by integrating both
sides of above equation in the variables $P_i$'s and summing over the 
$\Qi$'s. After the change of variables $P_i \rightarrow P'_i$ in Eq.~(\ref{varia}), 
one obtains:

\begin{eqnarray}
 && \Bigg[ \prod_{i=1}^N \sum_{\Qi} \int \d^4 P_i \Bigg] 
 f_*(\{P_i,\Qi,V^*_i\}) = \nonumber \\
 && \!\!\!\! \Bigg[ \prod_{i=1}^N \sum_{\Qi} \!\! \int \d^4 P'_i 
 \frac{M_i}{P'^0_i} \Bigg] w(\{P'_i,\Qi\}|\{V^*_i \frac{M_i}{P'^0_i}\})_{P'} 
 Y(\{V^*_i \frac{M_i}{P'^0_i}\}) \nonumber \\
 &&
\end{eqnarray}
whose left hand side is apparently the actual marginal probability distribution 
for cluster proper volumes. Thus, the function $Y$ should satisfy the integral 
equation:

\begin{eqnarray}\label{eqintv}
 H_*(\{V^*_i\}) && \equiv \frac {\d P}{\d V^*_1\ldots\d V^*_N} =   
 \Bigg[ \prod_{i=1}^N \sum_{\Qi} \int \d^4 P'_i \frac{M'_i}{P'^0_i} \Bigg] 
 \nonumber \\
 && \times w(\{P'_i,\Qi\}|\{V^*_i \frac{M'_i}{P'^0_i}\})_{P'} 
 Y(\{V^*_i \frac{M'_i}{P'^0_i}\}) 
\end{eqnarray}  
Like for Eq.~(\ref{eqint3}), we will simply assume the existence of a positive 
definite solution for $Y$.
 
We are now in a position to answer the question whether one can make a reduction 
of Lorentz invariants calculation to the EGC starting from a $f_*$ distribution 
whatsoever by solving the two integral equations~(\ref{eqint4},\ref{eqintv}). 
According to the discussion with respect to Eq.~(\ref{mass}), this is not true in
general because the first integral transformation through $X$ does not affect 
the distribution of cluster masses. Likewise, Eq.~(\ref{eqintv}) only implements 
the accordance between the actual proper volumes distribution and that determined 
by the required insertion of $w$, hence it has no effect on masses. 
Therefore, there must be a definite constraint to be fulfilled by $f_*$ for the
reduction to the EGC to be possible and this is concerned with the 
conditional probability distribution of cluster masses and quantum numbers 
once their volumes are fixed. This is best seen by integrating both sides of 
Eq.~(\ref{eqint4}) with the usual change of variables $P_i \rightarrow P'_i$ 
in Eq.~(\ref{varia}) and taking into account the normalization condition in
Eq.~(\ref{norm2}):  

\begin{eqnarray}
 && \Bigg[ \prod_{i=1}^N \int \d^4 P_i \; \delta\left(M_i - \sqrt{P^2_i}\right) \Bigg]
 f_*(\{P_i,\Qi,V^*_i\}) \nonumber \\
 && \equiv f_{*M}(\{M_i,\Qi,V^*_i\}) \equiv g_{*M}(\{M_i,\Qi\}|\{V^*_i\}) 
 \, H_*(\{V^*_i\}) \nonumber \\ 
 && = \Bigg[ \prod_{i=1}^N \int \d^4 P'_i \frac{M_i}{P'^0_i} \,\, 
 \delta \left(M_i - \sqrt{P'^2_i}\right) \Bigg] \nonumber \\
 && \times w(\{P'_i,\Qi\}|\{V^*_i \frac{M_i}{P'^0_i}\})_{P'} \, 
 Y(\{V^*_i \frac{M_i}{P'^0_i}\}) 
\end{eqnarray}  
where $f_*$ has been decomposed into the product of a conditional probability 
distribution of masses and quantum numbers $g_{*M}$ for fixed proper volumes and the 
probability distribution of proper volumes $H_*$. The above equation can be solved 
to determine $g_{*M}$ by using the expression of $H_*$ in Eq.~(\ref{eqintv}). 
Indeed, the insertion of $H_*$ does but to set the due overall normalization of 
$g_{*M}$ to 1. Thus:

\begin{eqnarray}\label{condmass}
 && \!\!\!
 g_{*M}(\{M_i,\Qi\}|\{V^*_i\}) = \frac{1}{N_f} 
 \Bigg[ \prod_{i=1}^N \int \d^4 P'_i \frac{M'_i}{P'^0_i} \,\, \delta (M_i - M'_i) 
 \Bigg] \nonumber \\
 && \!\!\!
 \times w(\{P'_i,\Qi\}|\{V^*_i \frac{M'_i}{P'^0_i}\})_{P'} \,\,
 Y(\{V^*_i \frac{M'_i}{P'^0_i}\}) 
\end{eqnarray}
where $N_f$ is the normalization factor obtained by integrating the numerator
in the $M_i$'s and summing over $\Qi$.
Since $Y$ can be determined as a function of $w$ and $H_*$ through the integral
equation (\ref{eqintv}), EQ.~(\ref{condmass}) states that $g_{*M}$, the actual 
probability distribution of cluster masses and quantum numbers for fixed proper 
volumes, ought to fulfill an independent constraint. This is definitely not a 
trivial requirement and must be taken as a further assumption to make it 
possible the existence of an EGC. Most likely, the actual $g_{*M}$ shall 
not be equal to the right hand side of Eq.~(\ref{condmass}) but, hopefully, 
close enough to it to make the equivalence with the EGC a good approximation,  
thus justifying the surprising accuracy of multiplicity fits performed under that
assumption.

It is of utmost importance to stress that the four-vector $P'$ in 
Eq.~(\ref{mean4}), whose appearance is related to the particular form of 
$w$ (see Eq.~(\ref{w})), is not the initial four-momentum of the collision. 
Instead, $P'$ is the four-momentum of the EGC to be set as a result of cluster 
back-boosting and can be taken as a free parameter. For any chosen 
value of $P'$, there shall possibly be two corresponding solutions $X_{P'}$ 
and $Y_{P'}$ of the integral equations (\ref{eqint4},\ref{eqintv}). There 
are, however, some restrictions: $M' = \sqrt{P'^2}$ must certainly be smaller 
than centre-of-mass energy $\sqrt s$ and greater than the actual maximum value 
of $\sum_i M_i$, the sum of cluster masses. The latter constraint is the most 
relevant, for it might spoil at once our EGC reduction scheme falsifying the 
relationship (\ref{condmass}). For instance, a $g_{*M}$ distribution can be 
envisaged which is in agreement with that needed for an EGC with mass 
$M \ll \sqrt s$ except for a small tail or bump over which the sum of cluster 
masses is very close to $\sqrt s$. In this case, it is apparent that the 
equivalence with an EGC having one definite mass turns out to be impossible. 
Nevertheless, in principle, it should be possible to split collision events 
into suitable subsets having different values for $\max[\sum_i M_i]$ and link 
them to different EGC's four-momenta $P'$, just because $P'$ is a free parameter. 
Pushing along this idea, one is led to consider a continuous distribution 
$\Upsilon(P')$ to be chosen so as to restore the agreement between the 
actual and EGC-splitting distributions of $\sum_i M_i \equiv M_t$:

\begin{eqnarray}\label{totmass} 
&&  G (M_t) \equiv \Bigg[ \prod_{i=1}^N \sum_{\Qi} \int \d^4 P_i \, 
 \d V^*_i \Bigg] f_*(\{P_i,\Qi,V^*_i\}) \nonumber \\
&&  \times \delta (M_t - \Sigma_i M_i) \nonumber \\
&&  = \int \d^4 P' \; \Upsilon (P') \Bigg[ \prod_{i=1}^N \sum_{\Qi} 
 \int \d^4 P'_i \, \d V'_i \Bigg] \delta (M_t - \Sigma_i M'_i) \nonumber \\ 
&&  \times w(\{P'_i,\Qi\}|\{V'_i\})_{P'}\, Y(\{V'_i\})  
\end{eqnarray}   
where Eq.~(\ref{eqint4}) has been integrated by using the change of variables
~(\ref{varia}) and the normalization condition~(\ref{norm2}). Again, we will 
assume that a positive definite solution $\Upsilon$ of this integral equation 
exists. Of course, the introduction of the function $\Upsilon(P')$ requires 
Eqs.~(\ref{eqint4},\ref{eqintv}) to be changed accordingly: 

\begin{eqnarray}\label{eqint5}
 && f_*(\{P_i,\Qi,V^*_i\}) = \Bigg[ \prod_{i=1}^N \int \d \LL_i 
 \frac{M_i}{\LL_i^{-1}(P_i)^0} \Bigg] X(\{\LL_i\}) \nonumber \\
 && \times \int \d^4 P' \; \Upsilon(P') \, 
 w(\{\LL_i^{-1}(P_i),\Qi\}|\{V^*_i \frac{M_i}{\LL_i^{-1}(P_i)^0}\})_{P'} \nonumber \\
 && \times Y(\{V^*_i \frac{M_i}{\LL_i^{-1}(P_i)^0}\})  
\end{eqnarray}
and:

\begin{eqnarray}\label{eqintv2}
 && H_*(\{V^*_i\}) = \Bigg[ \prod_{i=1}^N \sum_{\Qi} \int \d^4 P'_i 
 \frac{M_i}{P'^0_i} \Bigg] \int \d^4 P' \; \Upsilon(P') \nonumber \\
 && \times w(\{P'_i,\Qi\}|\{V^*_i \frac{M_i}{P'^0_i}\})_{P'} 
 Y(\{V^*_i \frac{M_i}{P'^0_i}\}) 
\end{eqnarray}  
Altogether, the equations~(\ref{totmass},\ref{eqint5},\ref{eqintv2}) 
are a system of coupled integral equations for the unknown functions $X$, $Y$ and 
$\Upsilon$ to be solved at the same time in order to achieve the reduction
to a superposition of EGC's with different four-momentum and volume. 
Note that in Eqs.~(\ref{totmass},\ref{eqint5},\ref{eqintv2}) $X$ and $Y$ 
have been assumed to be independent of $P'$, yet this dependence can be introduced 
without affecting most of previous and following arguments.\\
As has been discussed with regard to Eq.~(\ref{condmass}), these equations are not 
sufficient to ensure that {\em any} configurational distribution $f_*$ 
is equivalent to such a superposition, as far as the calculation of Lorentz 
invariant quantities is concerned. In fact, after the introduction of $\Upsilon$, 
the constraint~(\ref{condmass}) is left almost unchanged:

\begin{eqnarray}\label{condmass2}
 && \!\!\! g_{*M}(\{M_i,\Qi\}|\{V^*_i\}) = \frac{1}{N_f} 
 \Bigg[ \prod_{i=1}^N \int \d^4 P'_i \; \frac{M'_i}{P'^0_i} \, 
 \delta (M_i - M'_i) \Bigg] \nonumber \\
 && \!\!\! \times \int \d^4 P' \; \Upsilon (P') \, 
 w(\{P'_i,\Qi\}|\{V^*_i \frac{M'_i}{P'^0_i}\})_{P'}\, Y(\{V^*_i \frac{M'_i}{P'^0_i}\})   
\end{eqnarray}    
and should be fulfilled independently once $Y$ and $\Upsilon$ have been 
determined as a function of the given $f_*$ and $w$. Finally, the average
value of a Lorentz invariant observable $O$ reads 
(see Eqs. (\ref{mean4},\ref{ipsilon})):

\begin{eqnarray}\label{mean5}
 \langle O \rangle && = \sum_N P_N \Bigg[\prod_{i=1}^N \int \d V'_i \Bigg] 
 \int \d^4 P' \; \Upsilon(P')\, Y(\{V'_i\}) \, \nonumber \\
 && \times O(M',\Qz, \Sigma_i V'_i P'^0/M') \nonumber \\
 && \!\!\!\!\!\!\!\!\!\!\!\!
 = \int \d V' \int \d^4 P' \; \Upsilon(P') \, y(V')\, O(M',\Qz,V' P'^0/M') \nonumber \\
 && \!\!\!\!\!\!\!\!\!\!\!\!
 = \int \d V^* \int \d^4 P' \; \frac{M'}{P'^0} \,\, \Upsilon(P') \, 
 y(V^*\frac{M'}{P'^0})\, O(M',\Qz,V^*) \nonumber \\
 && \!\!\!\!\!\!\!\!\!\!\!\!
 = \int \d V^* \int \d M' \; \chi(M',V^*) \, O(M',\Qz,V^*)
\end{eqnarray}   
where $V^*$ is the EGC's proper volume and: 

\begin{equation}
 \chi(M',V^*) \equiv \int \d^4 P' \; \delta \left (\sqrt{P'^2}-M'\right) 
 \, \Upsilon(P') \, y(V^*\frac{M'}{P'^0}) \nonumber 
\end{equation} 
If the distribution $\chi$ allows only large values of volume and mass, the 
canonical approximation of the last expression in Eq.~(\ref{mean5}) could be 
a satisfactory one. Specifically, the microcanonical average $O(M',\Qz,V^*)$ is to
be replaced by the canonical average $O_T(T(M',V^*),\Qz,V^*)$ calculated
for the suitable temperature $T$ dependent on mass and volume (see discussion 
at the end of Subsect. 2.1), so that:

\begin{eqnarray}\label{canmean}
 \langle O \rangle && = \int \d V^* \!\! \int \d M' \; \chi(M',V^*) \, 
  O(M',\Qz,V^*) \nonumber \\
 && \simeq \int \d V^* \!\! \int \d M' \; \chi(M',V^*) \, O_T(T(M',V^*),\Qz,V^*) 
  \nonumber \\
 && = \int \d V^* \!\! \int \d T \; \chi(M'(T,V^*),V^*) \, 
 \frac{\partial M'}{\partial T} \,\, O_T(T,\Qz,V^*) \nonumber \\
 && \equiv \int \d V^* \!\! \int \d T \; \zeta (T,V^*) \, O_T(T,\Qz,V^*) 
\end{eqnarray}
Thereby, the actual average $\langle O \rangle$ has been written as a
superposition of canonical averages at different temperatures and volumes.
The last integral in Eq.~(\ref{canmean}) can be replaced with the mean 
value of the observable calculated for average values of $T$ ($\equiv \overline T$)
and $V^*$ ($\equiv \overline V^*$), namely:

\begin{equation}\label{tmean}
 \int \d V^* \int \d T \; \zeta (T,V^*) \, O_T(T,\Qz,V^*) = O_T 
 (\overline T,\Qz,\overline V^*)
\end{equation}
In principle, $\overline T$ and $\overline V^*$ depend on the particular observable 
$O$ (e.g. on the hadron mass if $O$ is a multiplicity). Notwithstanding, in all 
previous studies \cite{beca1,beca2,beca3} on average multiplicities of hadrons
in the canonical ensemble, the single temperature and volume average in the right 
hand side of Eq.~(\ref{tmean}) has been used, tacitly assuming that $\overline T$ 
and $\overline V^*$ were independent of $O$. This is a reasonable assumption only if 
$\zeta$ is a strongly peaked function of $T$ and $V^*$ or, in other words, $T$ and $\oV$
have small fluctuations. On the other hand, one of the main results of such analyses 
is indeed the very good agreement between the data and single-temperature and volume 
fits at a fairly constant temperature value, which somehow {\em a posteriori} justifies  
the above assumption. That finding also provides a strong indication that hadronisation 
occurs at a critical value of cluster energy density \cite{beca3}.

Instead of deriving canonical formulae as a limiting case of microcanonical ones, 
it might be possible, in principle, to enforce the exact validity of the canonical
ensemble by a suitable choice of $\Upsilon$, that is of EGC's mass fluctuations.  
For this purpose, $\Upsilon$ must be set equal to the probability density $\sigma(P')$ 
of getting a four-momentum $P'$ in a canonical system with four-temperature $\beta$, 
volume $V' = \sum_i V'_i$ and quantum numbers $\Qz$:

\begin{eqnarray}\label{sigma}
 && \sigma(P')_{\beta,\Qz,V'} = \frac{\sum_{\rm states} \delta^4(P'-P_{\rm state}) 
 \, \e^{-\beta\cdot P_{\rm state}} \delta_{\Qz,{\bf Q}_{\rm state}}}{Z(\beta,\Qz,V')} 
 \nonumber \\
 && = \frac{\Omega(P',\Qz,V')}{Z(\beta,\Qz,V')} \, \e^{-\beta\cdot P'} 
\end{eqnarray}
where $Z$ is the canonical partition function. Note that in Eq.~(\ref{sigma}) the
EGC's four-momentum distribution function depends on $V'=\sum_i V'_i$ besides $P'$, 
but this does not affect any of the previous and forthcoming arguments provided 
that integrations are performed in the appropriate order. 
An integral equation corresponding to equation~(\ref{totmass}) in the canonical 
ensemble can be obtained by replacing $\Upsilon(P')$ with a superposition of 
$\sigma(P')$'s for different temperatures, with $\beta=(1/T,{\bf 0})$:

\begin{eqnarray}\label{superp}
 && \Upsilon(P') = \int_0^\infty \!\!\! \d T \; \Psi(T) \, \sigma(P')_{\beta,\Qz,V'} = 
 \nonumber \\
 && = \Omega(P',\Qz,V') \int_0^\infty \d T \; \frac{\Psi(T)}{Z(T,\Qz,V')} 
 \, \e^{-P'^0/T} 
\end{eqnarray}
where $\Psi(T)$ is an unknown temperature distribution. By plugging 
Eq.~(\ref{superp}) into Eq.~(\ref{totmass}) one obtains again a Fredholm integral 
equation of the first kind for $\Psi$:

\begin{eqnarray}\label{totmasscan} 
&& \!\!\! G (M_t) = \!\! \int_{0}^{\infty} \!\!\!\! \d T \; \Psi(T) 
  \Bigg[ \prod_{i=1}^N \sum_{\Qi} \int \d^4 P'_i \, \d V'_i \Bigg] \, 
  \delta (M_t - \!\! \Sigma_i M'_i) \nonumber \\
&& \!\!\! \times \int \d^4 P' \; \frac{\Omega(P',\Qz,\Sigma_i V'_i)}
 {Z(T,\Qz,\Sigma_i V'_i)} \,\e^{-P'^0/T} \, w(\{P'_i,\Qi\}|\{V'_i\})_{P'} 
  \nonumber \\
&& \!\!\! \times \, Y(\{V'_i\}) 
\end{eqnarray}      
which, however, may not always have a solution for the actual $G(M_t)$ might vanish 
if $M_t$ is less than a finite threshold value whereas the right hand side does not, 
regardless of $\Psi$'s shape. This is indeed one of the key properties of the 
canonical ensemble: for any temperature, the distribution of EGC's invariant mass 
(and of $\Sigma M_i$ too) always has non-vanishing tails down to zero and up to 
infinity whilst a superposition of EGCs with definite masses may not. Therefore, 
the sought exact equivalence between the actual 
system and a canonical superposition of global clusters with finite temperature 
and volume has one more difficulty with respect to the microcanonical superposition. 
It is interesting to derive the average value of the Lorentz invariant observable 
$O$ in this case similarly to what has been done before in the microcanonical 
case. This can be done by replacing $\Upsilon(M)$ in Eq.~(\ref{mean5}) with 
Eq.~(\ref{superp}) :

\begin{eqnarray}\label{canmean2}
&& \!\!\!\!\! \langle O \rangle = \sum_N P_N \int_0^{\infty} \!\!\! \d T \, \Psi(T)
 \Bigg[\prod_{i=1}^N \int \d V'_i \Bigg] Y(\{V'_i\}) \nonumber \\
&& \!\!\!\!\! \times \int \d^4 P' \; \e^{-P'^0/T} \, 
 \frac{\Omega(P',\Qz,\Sigma_i V'_i)}{Z(T,\Qz,\Sigma_i V'_i)} \,\,  
  O(M',\Qz,\frac{P'^0}{M'} \Sigma_i V'_i) \nonumber \\
&& \!\!\!\!\! = \int_0^\infty \!\!\! \d T \, \Psi(T) \int \d V' \; y(V') 
 \int \d^4 P' \; \e^{-P'^0/T} \nonumber \\
&& \!\!\!\!\! \times \frac{\Omega(P',\Qz,V')}{Z(T,\Qz,V')} \; O(M',\Qz,V'P'^0/M') 
  \nonumber \\
&& \!\!\!\!\! = \int_0^\infty \!\!\! \d T \, \Psi(T) \int \d V' \; y(V') \, 
 O_T(T,\Qz,V')   
\end{eqnarray}  
where $y(V')$ is the same function defined in Eq.~(\ref{ipsilon}) and:

\begin{eqnarray}\label{tmean2}
 && \!\!\!\!\!\!\!\!\!\!\! O_T(T,\Qz,V') = \nonumber \\
 && \!\!\!\!\!\!\!\!\!\!\! = \int \d^4 P' \, \frac{\e^{-P'^0/T}\Omega(P',\Qz,V')}
 {Z(T,\Qz,V')} \, O(M',\Qz,V'P'^0/M')
\end{eqnarray}
is apparently the average value of the observable $O$ for an equivalent global
cluster at temperature $T$, volume $V'$ and quantum numbers $\Qz$. 

\subsection{Summary} 

Before moving to the central topic of this paper, i.e. tranverse momentum 
spectra, it is worth summarizing the main points of this Section. 
\begin{itemize}
\item{} The statistical hadronisation model assumes hadrons to be 
produced from a set of clusters with dynamically generated configurations 
in terms of quantum numbers, volumes, mass and momenta; within each cluster,
all hadronic states are equally likely.
\item{} Since cluster configurations are dynamically generated, the statistical
{\it ansatz} needs a supplementary dynamical information. However, Lorentz 
invariant observables, such as hadron multiplicities, are unaffected by a 
change of cluster momenta. Then, in order to reduce the number of free 
parameters to a minimum, cluster momenta distribution can be modified so as 
to achieve an equivalence with the calculation for one global cluster (EGC); 
this is done through a system of integral equations. 
\item{} Provided that positive definite solutions of the aforementioned integral
equations exist, the reduction to the EGC ultimately relies on a particular form 
(see Eqs. (\ref{condmass},\ref{condmass2})) of cluster mass and charges 
fluctuations at fixed volumes. If this is not the case, the equivalence is 
spoiled. Still, the actual distribution could be close enough to that 
particular form to ensure a sufficiently accurate approximate equivalence.  
\item{} If mass and volume of the EGC are large, the basic microcanonical treatment 
can be replaced by a more manageable canonical one. Thanks to the equivalence
with the EGC, this should be possible even though individual physical clusters are
too small to be treated canonically. Therefore, temperature could be a well 
defined quantity only in a global sense and not at locally for each single 
cluster.
\end{itemize}    
            
\section{Transverse momentum spectra}

We can now proceed the main subject of this paper, that is transverse momentum 
spectra. Unlike average multiplicities, this observable 
is not a Lorentz invariant one and a special treatment is necessary. To start
with, let us consider the spectrum of the $j^{\rm th}$ hadron species relevant 
to the $i^{\rm th}$ cluster:

\begin{equation}
  \frac{\d n_{ji}}{\d p_T}(p_T)\vl_{P_i,\Qi,V^*_i}
\end{equation}
The possible dependences of this spectrum on the four-momenta $P_i$ are crucial
for forthcoming arguments. If a Lorentz transformation ${\sf \Lambda}$ is performed
and the new transverse momentum is $p'_T ={\sf \Lambda}(p)_T$, the corresponding 
spectrum must have the same functional dependence on $p'_T$ as the transverse
momentum spectrum associated with the cluster whose four-momentum is Lorentz-transformed 
with the same ${\sf \Lambda}$. Hence ($\Qi$ and $V^*_i$ have been omitted):
  
\begin{equation}
  \frac{\d n_{ji}}{\d p_T}(\lala^{-1}(p)_T)\vl_{P_i} \Big| 
 \frac{\d \lala^{-1}(p)_T}{\d p_T}\Big| = \frac{\d n_{ji}}{\d p_T}(p_T)\vl_{\lala(P_i)}
\end{equation}
A rotation around $z$ axis does not change $p_T$; thus, according to the above equation, 
the spectrum must depend on $P_i^x$ and $P_i^y$ only through $P_{iT}$ and not
on the azimuthal angle. Furthermore, $p_T$ is not changed by a boost along $z$ 
axis either, so $\rho_{ji}$ must depend on $P^0_i$ and $P_i^z$ only through 
the combination $P_i^{0 2}-P_i^{z 2} = M_i^{2}+P_{iT}^2$. Therefore, the spectrum 
ought to have a dependence on $M_i$ and $P_{iT}$ only, so that:

\begin{equation}
 \frac{\d n_{ji}}{\d p_T}(p_T)\vl_{P_i,\Qi,V^*_i} \rightarrow 
 \frac{\d n_{ji}}{\d p_T}(p_T)\vl_{M_i,P_{iT},\Qi,V^*_i}
\end{equation}    
Indeed, what we are really interested in is the overall average transverse momentum 
spectrum of a given hadron species (labelled by $j$). To obtain it, the distribution 
of mass, volume and quantum numbers of the clusters must be folded with the sum 
of the single cluster spectra for a given configuration, yielding (see 
Eq.~(\ref{genaver})): 

\begin{eqnarray}\label{ptsp1}
&& \Big\langle \frac{\d n_{j}}{\d p_T} \Big\rangle = \sum_N P_N 
 \Bigg[ \prod_{i=1}^N \sum_{\Qi} \int \d^4 P_i \, \d V^*_i \Bigg] 
 f_*(\{P_i,\Qi,V^*_i\}) \nonumber \\ 
&& \times \sum_{i=1}^N \frac{\d n_{ji}}{\d p_T}(p_T)\vl_{M_i,P_{iT},\Qi,V^*_i}
\end{eqnarray}
The distribution $f_*$ can be replaced with the right hand side of Eq.~(\ref{eqint5}) 
and, by using the change of variables (\ref{varia}), Eq.~(\ref{ptsp1}) becomes:

\begin{eqnarray}\label{ptsp2}
 && \Big\langle \frac{\d n_{j}}{\d p_T} \Big\rangle = \sum_N P_N 
 \Bigg[ \prod_{i=1}^N  \sum_{\Qi} \int \d^4 P'_i \d V'_i \d \LL_i \Bigg] 
 X(\{\LL_i\}) \nonumber \\
 && \times \int \d^4 P' \, \Upsilon(P') \, w(\{P'_i,\Qi\}|\{V'_i\})_{P'}
  Y(\{V'_i\}) \nonumber \\
 && \times \sum_{i=1}^N \frac{\d n_{ji}}{\d p_T}(p_T)\vl_{M_i,\LL_i(P'_i)_T,
 \Qi,V'_i\frac{P'^0_i}{M_i}}
\end{eqnarray}
The physical meaning of this expression is essentially that the average
transverse spectrum of the hadron $j$ is the convolution of a distribution
$X$ of Lorentz transformation on clusters (assumed to be positive definite, 
see Sect.~2) with the spectra obtained from EGC splitting into $N$ sub-clusters. 
The integration over $\LL_i$ can be considerably simplified by 
using an alternative decomposition of the Lorentz transformation ${\sf L}$ 
defined in Eq.~(\ref{boost}) (see Appendix B):

\begin{equation}\label{boost2}
  {\sf L} = {\sf \hat L}_3 (\eta) \, {\sf \hat R}_3 (\phi) \, 
  {\sf \hat L}_1 (\zeta) 
\end{equation}
where ${\sf \hat L}_k$ and ${\sf \hat R}_k$ are to be understood as in 
Eq.~(\ref{boost}), $\phi \in [0,2\pi)$ is an angle and $\eta \in (-\infty,+\infty)$ 
and $\zeta \in [0,\infty)$ are hyperbolic sectors. This particular decomposition, 
along with its associated measure (see again Appendix B):

\begin{equation}
    \d{\sf L} = \frac{1}{2} \, \sinh 2 \zeta \, \d \eta \, \d \varphi \, 
    \d \zeta
\end{equation}    
is advantegeous in that it allows to integrate away at once two parameters in 
Eq.~(\ref{ptsp2}). In fact:

\begin{eqnarray}
&& \LL_i(P'_i)_T = ({\sf \hat L}_3(\eta_i)({\sf \hat R}_3(\phi_i)
 ({\sf \hat L}_1(\zeta_i)(P'_i))))_T  
  = \nonumber \\
&& = ({\sf \hat R}_3(\phi_i)({\sf \hat L}_1(\zeta_i)(P'_i)))_T = 
  {\sf \hat L}_1(\zeta_i)(P'_i)_T
\end{eqnarray}   
because both ${\sf \hat L}_3$ and ${\sf \hat R}_3$ leave the transverse component 
of their argument unchanged. Thereby, the transverse component of the four-momentum 
$P'_i$ undergoing a Lorentz transformation like (\ref{boost}) or (\ref{boost2}), 
is actually the same obtained by applying one suitable Lorentz boost along $x$ 
axis. The parameter $\zeta_i > 0$ is related to the four-velocity associated with 
this transverse boost, namely:

\begin{equation}
\!\!\!\! \sinh \zeta_i = \beta_{Ti} \gamma_{Ti}= \ui \qquad \cosh \zeta_i 
 = \gamma_{Ti} = \sqrt{1+\ui^2} 
\end{equation}
while the associated measure reads:

\begin{equation}   
  \frac{1}{2} \, \sinh 2 \zeta_i \, \d \zeta_i = \ui \, \d \ui 
\end{equation}   
Therefore, Eq.~(\ref{ptsp2}) becomes:

\begin{eqnarray}\label{ptsp3}
 \Big\langle \frac{\d n_{j}}{\d p_T} \Big\rangle && = \sum_N P_N 
 \Bigg[ \prod_{i=1}^N  \sum_{\Qi} \int \d^4 P'_i \d V'_i \;
 \frac{\d \uqi}{2} \Bigg] X_T(\{\uqi\}) \nonumber \\
 && \times \int \d^4 P' \; \Upsilon(P') \, w(\{P'_i,\Qi\}|\{V'_i\})_{P'} Y(\{V'_i\}) 
 \nonumber \\
 && \times \sum_{i=1}^N 
 \frac{\d n_{ji}}{\d p_T}(p_T)\vl_{M_i,\hat \LL_1(\ui)(P'_i)_T,\Qi,V'_i \frac{P'^0_i}{M_i}}
\end{eqnarray}     
where $X_T(\{\uqi\})$ is defined as:

\begin{equation}
  X_T(\{\uqi\}) \equiv \Bigg[\prod_{i=1}^N \int_{-\infty}^{+\infty} \!\!\! 
  \d \eta_i \int_0^{2\pi} \!\!\! \d \phi_i \Bigg] 
  X(\{\LL_i(\eta_i,\phi_i,\ui)\})
\end{equation}
Before doing anything else, it is advantageous to work out the integral:

\begin{equation}\label{utint}
  \Bigg[ \prod_{i=1}^N \int_0^{\infty} \!\! \frac{\d \uqi}{2} \Bigg] 
  X_T(\{\uqi\}) \sum_{i=1}^N \frac{\d n_{ji}}{\d p_T}(p_T)\vl_
  {\hat \LL_1(\ui)(P'_i)_T} 
\end{equation}
in which the other arguments of $\d n_{ji}/ \d p_T$ have been omitted. In fact, at
this stage, we can take advantage of a physical information concerning the 
range over which the distribution $X_T$ is expected to be significantly different 
from zero. Indeed, it is well known that in high energy collisions hadrons are 
mostly emitted with a limited transverse momentum (with respect to event axis 
in \ee or beam line in hadronic collisions) of the order of few hundreds MeV,
slowly increasing with centre-of-mass energy. Consequently, in the framework
of a cluster hadronisation model, most clusters must have a little transverse 
momentum and, by taking as educated guess their mass values between 1 and 3 GeV, 
transverse four-velocities generally $< 1$. In other words, the distribution 
$X_T(u^2_{T1},\ldots,u^2_{TN})$ is expected to be significantly different from 
zero only in the region $\uqi < 1$ $i=1,\ldots,N$ over a reasonably large 
centre-of-mass energy range. Certainly, at very high energy, the increase of 
transverse phase space (for instance, the increase of radiated gluon $p_T$ 
in \ee collisions) is likely to extend the effective transverse four-velocity 
domain beyond 1. Nevertheless, in the present work, we will confine our attention
to a centre-of-mass energy range where the condition $\uqi \ll 1$ applies.   
Under this circumstance, one is allowed to expand all $\d n_{ji}/ \d p_T$'s in
powers of $\uqi$ starting from a suitable point $\utms$ common to all of the 
clusters:

\begin{eqnarray}\label{expansion}
&& \frac{\d n_{ji}}{\d p_T}(p_T)\vl_{\hat \LL_1(\ui)(P'_i)_T} \equiv 
  \tau_{ji}(\uqi) = \tau_{ji}(\utms) \nonumber \\
&& + \frac{\partial \tau_{ji}}{\partial \uqi}(\utms) \, (\uqi-\utms) + 
   {\cal O}((\uqi-\utms)^2) 
\end{eqnarray}
so that the integral~(\ref{utint}) becomes:

\begin{eqnarray}\label{utintex}
 && \Bigg[ \prod_{i=1}^N \int_0^{\infty} \!\! \frac{\d \ui^2}{2} \Bigg] 
  X_T(\{\ui\}) \Bigg[ \sum_{i=1}^N \tau_{ji}(\utms) \nonumber \\ 
 && + \frac{\partial \tau_{ji}}{\partial \uqi}(\utms) \, (\uqi-\utms) + 
  {\cal O}((\uqi-\utms)^2) \Bigg] 
\end{eqnarray}
The $X_T$ distribution will be assumed to be such that even the first order
term in the expansion~(\ref{expansion}) can be neglected. Then, only
the zeroth order term will be retained and Eq.~(\ref{ptsp3}) turns into:

\begin{eqnarray}\label{ptsp4}
 && \Big\langle \frac{\d n_{j}}{\d p_T} \Big\rangle = \sum_N P_N 
 \Bigg[ \prod_{i=1}^N  \sum_{\Qi} \int \d^4 P'_i \d V'_i \Bigg] 
 \int \d^4 P' \; \Upsilon(P') \,\nonumber \\
 && \times w(\{P'_i,\Qi\}|\{V'_i\})_{P'} Y(\{V'_i\}) \nonumber \\
 && \times \sum_{i=1}^N \frac{\d n_{ji}}{\d p_T}(p_T)
 \vl_{M_i,\hat \LL_1(\utm)(P'_i)_T,\Qi,V'_i \frac{P'^0_i}{M_i}}
\end{eqnarray}
where $\utm \equiv \sqrt{\utms}$.\\ 
Now the reduction to the EGC may apply due to the fact that all clusters in the 
above equation are boosted by the {\em same} transverse four-velocity $\utm$ along 
$x$ axis. In other words, the sum of transverse momentum spectra of all clusters 
for a given configuration of four-momenta and volumes, with subsequent integration 
over them weighted by $w$, must yield the transverse momentum spectrum of the given 
hadron in an EGC boosted with four-velocity $\utm$. This statement can be proved not 
only for transverse momentum spectrum but, more generally, for the invariant 
four-momentum spectrum. For this purpose, the convolution equation 
Eq.~(\ref{ptsp4}) is generalised to four-momentum spectra:

\begin{eqnarray}\label{fmsp}
 && \Big\langle \frac{\d n_{j}}{\d^4 p} \Big\rangle = \sum_N P_N 
 \Bigg[ \prod_{i=1}^N  \sum_{\Qi} \int \d^4 P'_i \d V'_i \Bigg] 
 \int \d^4 P' \; \Upsilon(P') \,\nonumber \\
 && \times w(\{P'_i,\Qi\}|\{V'_i\})_{P'} Y(\{V'_i\}) \nonumber \\
 && \times \sum_{i=1}^N 
 \frac{\d n_{ji}}{\d^4 p}(p)\vl_{\hat \LL_1(\utm)(P'_i),\Qi,V'_i 
 \frac{P'^0_i}{M_i}}
\end{eqnarray} 
To complete the proof, four-momentum spectrum for a given configuration must be 
expressed as in Eq.~(\ref{linear}), according to what has been established 
in Sect.~1. Since:

\begin{equation}\label{spetran}
 \frac{\d n_{ji}}{\d^4 p}(p)\vl_{\hat \LL_1(\utm)(P'_i),\Qi,V^*_i}=
 \frac{\d n_{ji}}{\d^4 p}(\hat \LL^{-1}_1(\utm)(p))\vl_{P'_i,\Qi,V^*_i} 
\end{equation} 
so that Eq.~(\ref{fmsp}) can be rewritten as:

\begin{eqnarray}\label{fmsp2}
 && \Big\langle \frac{\d n_{j}}{\d^4 p} \Big\rangle = \sum_N P_N 
 \Bigg[ \prod_{i=1}^N  \sum_{\Qi} \int \d^4 P'_i \d V'_i \Bigg] 
 \int \d^4 P' \; \Upsilon(P') \,\nonumber \\
 && \times w(\{P'_i,\Qi\}|\{V'_i\})_{P'} Y(\{V'_i\}) \nonumber \\
 && \times \sum_{i=1}^N 
 \frac{\d n_{ji}}{\d^4 p}(\hat \LL^{-1}_1(\utm)(p))\vl_{P'_i,\Qi,V'_i \frac{P'^0_i}{M_i}}
\end{eqnarray} 
Now, the contributions of primary hadrons, i.e. directly emitted from hadronising 
clusters, are separated from that of decay products of heavier hadrons:

\begin{equation}
 \Big\langle \frac{\d n_{ji}}{\d^4 p} \Big\rangle = 
 \Big\langle \frac{\d n_{ji}}{\d^4 p} \Big\rangle\Big|^{\rm primary} 
 + \sum_k \Big\langle \frac{\d n_{ji}}{\d^4 p} \Big\rangle\Big|^{k \rightarrow j}
\end{equation} 
where $\d n_{ji}/\d^4 p |^{k \rightarrow j}$ is the four-momentum density
of hadrons $j$ stemming from the decay of the {\em primary} hadron $k$ either 
directly or through intermediate steps (i.e. $k \rightarrow h \rightarrow \ldots 
\rightarrow j$). The primary part of the spectrum can be calculated by using a 
technique similar to that for average primary multiplicities, introduced in Eq.~(\ref{mult1}). 
Instead of taking the derivative of $\log \Oi$ with respect to a fictitious 
fugacity $\lambda_j$, one has to take the functional derivative of $\log \Oi$ 
with respect to a fictitious 'fugacity' function $\lambda_j(p)$ depending on 
four-momentum $p$ and set to 1 thereafter:

\begin{equation}
 \frac{\d n_{ji}}{\d^4 p}\Big|^{\rm primary}= 
 \frac{\delta \log \Oi[\lambda_j]}{\delta \lambda_j(p)}\Bigg|_
 {\lambda_j(p)=1}
\end{equation}
The proof is similar to that for average primary multiplicities and can be conveniently
carried out by using the expression of $\Oi$ in~(\ref{Omega2}) with the obviously
needed replacement, for single-mass-valued particles (i.e. not resonances):

\begin{equation}
 \int \d^3 {\rm p} = \int \d^4 p \; \delta (p^0-\sqrt{{\rm p}^2+m^2})
\end{equation}
It is worth emphasizing that, unlike the derivative with respect to $\lambda_j$, 
the functional derivative does not return a Lorentz invariant quantity. 
The secondary spectra $\d n_{ji}/\d^4 p|^{k \rightarrow j}$ can be expressed as 
a convolution of a cluster-independent kernel function $K_{jk}(p,t)$ (i.e. the
four-momentum density of hadron $j$ stemming from the decays of primary 
hadrons $k$ with four-momentum $t$) either directly or through intermediate steps,
with the four-momentum density of primary hadrons $k$ expressed again as a functional 
derivative:

\begin{equation}  
 \frac{\d n_{ji}}{\d^4 p}\Big|^{k \rightarrow j} = \int \d^4 t \, K_{jk}(p,t)\, 
 \frac{\delta \log \Oi[\lambda_k]}{\delta \lambda_k(t)}\Bigg|_{\lambda_k(t)=1}
\end{equation}    
Hence, the four-momentum spectrum of the hadron $j$ for a given configuration of 
clusters, i.e. the last factor on the right hand side of Eq.~(\ref{fmsp2}), reads:  

\begin{eqnarray}\label{reduct}
\!\!\!\!\! && \sum_{i=1}^N \frac{\d n_{j}}{\d^4 p}_{ji}(\hat \LL_1^{-1}(\utm)(p))\vl_
  {P'_i,\Qi,V^*_i} = \sum_{i=1}^N \frac{\delta \log \Oi[\lambda_j]}{\delta 
  \lambda_j(\hat \LL_1^{-1}(\utm)(p))} \nonumber \\
\!\!\!\!\! && + \sum_k \int \d^4 t \, K_{jk}(\hat \LL^{-1}_1(\utm)(p),t)\, 
 \frac{\delta \log \Oi[\lambda_k]}{\delta \lambda_k(t)}  \nonumber \\
\!\!\!\!\! && = \frac{1}{\prod_i \Oi} 
  \Bigg[\frac{\delta}{\delta \lambda_j(\hat \LL^{-1}_1(\utm)(p))}+ \sum_k \nonumber \\
\!\!\!\!\! && \int \d^4 t \, K_{jk}(\hat \LL^{-1}_1(\utm)(p),t) \, 
 \frac{\delta}{\delta \lambda_k(t)} \Bigg] \left( \prod_{i=1}^N \Oi[\{\lambda_k\}] \right) 
  \nonumber \\
\!\!\!\!\! && = \frac{{\cal L}_{\hat \LL^{-1}_1(\utm)(p)} \left(\prod_{i=1}^N 
  \Oi(P_i,\Qi,V_i)\right)}{\prod_{i=1}^N \Oi(P_i,\Qi,V_i)}   
\end{eqnarray}   
where the final setting $\lambda_j=1$, $\lambda_k=1$ are implied.
The operator between square brackets in the above equation, defined as 
${\cal L}_{\hat \LL^{-1}_1(\utm)(p)}$, is a linear one, so that the four-momentum spectrum
of the hadron $j$ for a given configuration of clusters has the form required in 
Eq.~(\ref{linear}) for the reduction to the EGC to apply. Therefore, using 
Eq.~(\ref{reduct}) and proceeding like for Eq.~(\ref{mean4}), the four-momentum 
spectrum in Eq.~(\ref{fmsp}) can also be written as:

\begin{eqnarray}\label{fmsp3}
 && \!\!\! \Big\langle \frac{\d n_{j}}{\d^4 p} \Big\rangle = \sum_N P_N 
 \Bigg[ \prod_{i=1}^N \int \d V'_i \Bigg] Y(\{V'_i\}) \int \d^4 P' \; 
 \Upsilon(P') \, \nonumber \\
&& \!\!\! \times \frac{{\cal L}_{\hat \LL^{-1}_1(\utm)(p)} \left(\Omega(P',\Qz,\Sigma_i V'_i) 
 \right)}{\Omega(P',\Qz,\Sigma_i V'_i)} = \nonumber \\
&& \!\!\! = \int \d V' \!\! \int \d^4 P' \; \Upsilon(P') \, y(V') \, 
 \frac{{\cal L}_{\hat \LL^{-1}_1(\utm)(p)} \left(\Omega(P',\Qz,V') \right)}
 {\Omega(P',\Qz,V')} \nonumber \\    
\end{eqnarray} 
 
From now on, calculations will be carried out in the canonical ensemble only.
Formally, this amounts to take the $\Upsilon$ function as in Eq.~(\ref{superp}); 
we have mentioned in Sect.~1 that this choice might not be an appropriate one 
if a lower bound on the sum of cluster exists, yet we will neglect this 
possibility. Moreover, we will assume a temperature distribution function $\Psi$ 
(see again Eq.~(\ref{superp})) equal to, or at least very close, to a Dirac's delta 
or, in other words, a unique temperature (see discussion at the end of Sect.~2). 
These assumptions amount to enforce an EGC's four-momentum distribution 
function $Y$ in Eq.~(\ref{fmsp3}) equal to $\sigma$ in Eq.~(\ref{sigma}) with 
$\beta=(1/T,{\bf 0})$. The main justification of it is a very strong indication 
in favour of a critical come out from the analysis of multiplicities. Whether the same 
critical value is retrieved in the analysis of transverse momentum spectra, it is 
just the main issue to be studied in this paper. Hence: 

\begin{eqnarray}\label{fmspcan}
\!\!\! && \Big\langle \frac{\d n_{j}}{\d^4 p} \Big\rangle = \int \d V' \int \d^4 P' 
 \; y(V') \,\frac{\Omega(P',\Qz,V')}{Z(T,\Qz,V')}\,\e^{-P'^0/T}\, \nonumber \\ 
\!\!\! && \times \frac{{\cal L}_{\hat \LL^{-1}_1(\utm)(p)} \left( \Omega(P',\Qz,V') \right)}
 {\Omega(P',\Qz,V')} = \nonumber \\
\!\!\! && = \int \d V \; y(V') \, \frac{{\cal L}_{\hat \LL^{-1}_1(\utm)(p)}
  \left( \int \d^4 P'\; \e^{-P'^0/T} \, \Omega(P',\Qz,V') \right)}{Z(T,\Qz,V')} 
  \nonumber \\
\!\!\! && = \int \d V \; y(V') \, \frac{{\cal L}_{\hat \LL^{-1}_1(\utm)(p)}
  \left( Z(T,\Qz,V') \right)}{Z(T,\Qz,V')} 
\end{eqnarray}
where advantage has been taken of the linearity of ${\cal L}$ and Eq.~(\ref{zpart})
has been used. The integration over volumes $V'$ weighted by the $y(V')$ distribution 
is now replaced by a mean volume $\oV$ which is assumed to be independent of hadron 
species; this would be an exact operation were not for the dependence of chemical 
factors on the volume (see Eq.~(\ref{primary}) below). Thus, the final result is 
the four-momentum invariant spectrum of hadron $j$ in a canonical EGC with volume 
$\oV$ and temperature $T$ evaluated at the four-momentum $\hat \LL^{-1}_1(\utm)(p)$, 
that is the same spectrum, evaluated at the four-momentum $p$, in a canonical EGC 
boosted with $\hat \LL_1(\utm)$ (see Eq.~(\ref{spetran})):  

\begin{eqnarray}\label{fmspcan2}
 \Big\langle \frac{\d n_{j}}{\d^4 p} \Big\rangle && = 
 \frac{{\cal L}_{\hat \LL^{-1}_1(\utm)(p)}\left( Z(T,\Qz,\oV)\right)}{Z(T,\Qz,\oV)}
 \nonumber \\ 
 && = \frac{{\cal L}_{p}\left( Z(\beta_{\utm},\Qz,\oV /\gmt)\right)}{Z(T,\Qz,\oV)}
\end{eqnarray}
where $\beta_{\utm} \equiv \hat \LL_1(\utm)((1/T,{\bf 0}))=(\gmt/T,\gmt\btt/T,0,0) 
=(\sqrt{1+\utms}/T,\utm/T,0,0)$.

The transverse momentum spectrum can be obtained by integrating the above spectrum
in mass, $p_z$ and $\varphi$ in cylindrical coordinates. In fact, it can be shown
that this operation, from a formal point of view, amounts to a redefinition of the 
functional derivative and integral operator in 
Eqs.~(\ref{reduct},\ref{fmsp3},\ref{fmspcan},\ref{fmspcan2}):  

\begin{eqnarray}\label{ptspcan}
\!\!\!\! && \Big\langle \frac{\d n_{j}}{\d p_T} \Big\rangle = \frac{\delta 
 \log Z([\lambda_j],\beta_{\utm},\Qz,\oV/\gmt)}{\delta \lambda_j(p_T)}
 \Bigg|_{\lambda_j=1} \nonumber \\
\!\!\!\! && + \sum_k \int \d^4 t \, K_{Tjk}(p_T,t) \, \frac{\delta \log 
 Z([\lambda_k],\beta_{\utm},\Qz,\oV/\gmt)}{\delta \lambda_k(t)}\Bigg|_{\lambda_k=1}
\end{eqnarray}
where $K_{Tjk}$ is the transverse momentum density of hadron $j$ stemming from the 
decays of hadron $k$ with four-momentum $t$. 
We are now going to show how to perform a detailed calculation of the above 
expression for different temperatures and transverse velocities. 

\section{Calculation of primary and secondary spectra in the canonical ensemble}    
    
Before working out Eq.~(\ref{ptspcan}), it is worth summarizing the key 
assumptions that have led to that expression: 
\begin{enumerate}
\item{}     
the reducibility of the actual set of clusters to an EGC as far as the calculation 
of mean values of Lorentz invariant observables is concerned; this has been 
extensively discussed in Sect.~2 and requires that integral equations (\ref{eqint4},
\ref{eqintv}) for fixed EGC mass, or integral equations 
(\ref{totmass},\ref{eqint5},\ref{eqintv2}) for variable EGC mass, have positive 
definite solutions;  
\item{}
an EGC large enough to allow the use of canonical ensemble; 
\item{}
transverse four-velocities $u_T= \beta_T \gamma_T$ of the clusters much smaller 
than 1, making the replacement of a transverse four-velocity distribution with an 
average value $\utm$ a good approximation.        
\end{enumerate}    
Eq.~(\ref{ptspcan}) has two terms which are the contributions to the final 
spectrum of the primary and secondary (i.e. produced by hadronic decays) hadrons 
respectively. Taking into account that the canonical partition function reads 
\cite{beca3}:

\begin{equation}\label{partfunc}
 Z([\{\lambda_k\}],\beta_{\utm},\Qz,\frac{\oV}{\gmt}) = 
 \frac{1}{(2 \pi)^n} \!\!\! \int_{-\piv}^{\piv} \!\!\!\! \d^n \phi_i \;
 \e^{ \i \Qi \cdot \phivs_i + F_{\rm c}([\{\lambda_k\}],\phivs_i) } 
\end{equation}    
where:

\begin{eqnarray}
&& \!\!\!\!\!\!\!\!\!\!\!\!\! F_{\rm c}([\{\lambda_j\}],\phiv_i) = \frac{\oV}{\gmt (2\pi)^3} 
   \sum_j (2J_j + 1) \nonumber \\
&& \!\!\!\!\!\!\!\!\!\!\!\!\! \times \int \d^3{\rm p} \; 
\log \, [1\pm \lambda(p_T) \, \exp (-\beta_{\utm} \cdot p_j + \i \qj\cdot\phiv)]^{\pm 1} 
\end{eqnarray}
the first functional derivative in the right hand side of Eq.~(\ref{ptspcan}) 
yields:

\begin{eqnarray}\label{primary} 
 && \!\!\!\!\!\!\!\!\!\Big\langle \frac{\d n_{j}}{\d p_T} \Big\rangle^{\rm primary} 
 = \frac{\oV (2J_j+1)}{(2\pi)^3 \sqrt{1+\utms}} \sum_{n=1}^\infty (\mp 1)^{n+1} 
  \nonumber \\
 && \!\!\!\!\!\!\!\!\! \times \int_{-\infty}^{+\infty} \!\!\!\! \d p_z 
 \int_{-\pi}^{\pi} \!\!\! \d \varphi \; p_T \, e^{-n \beta_{\utm} \cdot p_j} 
 \frac{Z(T,\Qz-n\qj,\oV)}{Z(T,\Qz,\oV)}
\end{eqnarray} 
with $\gmt=\sqrt{1+\utms}$, as defined below Eq.~(\ref{fmspcan2}).
The ratios of partition functions in the above equations have been defined as
{\em chemical factors} in refs.~\cite{beca2,beca3} as they contain the dependence 
of the production rate of the considered hadron on its quantum numbers. The 
integration in $p_z$ and $\varphi$ can be done analytically, yielding:

\begin{eqnarray}\label{primary2} 
 && \!\!\!\Big\langle \frac{\d n_{j}}{\d p_T} \Big\rangle^{\rm primary} 
 = \frac{\oV (2J_j+1)}
 {2\pi^2 \sqrt{1+\utms}} \sum_{n=1}^\infty (\mp 1)^{n+1} \, m_T \, p_T \nonumber \\ 
 && \!\!\!\times {\rm K}_1 \left( \frac{n \sqrt{1+\utms}\,m_T}{T}\right) 
 {\rm I}_0 \left( \frac{n \utm p_T}{T}\right) \frac{Z(T,\Qz-n\qj,\oV)}{Z(T,\Qz,\oV)}
 \nonumber \\
 && 
\end{eqnarray}
where $m_T = \sqrt{p_T^2+m_j^2}$ is the transverse mass and $\rm K_1$, $\rm I_0$
are modified Bessel functions.
By integrating the above spectrum, the formulae for primary multiplicities quoted in 
refs.~\cite{beca2,beca3} can be recovered. At a constant temperature of about 
160 MeV found in the analysis of multiplicities, the effect of quantum statistics 
can be neglected for all hadrons except pions and only the first term of the 
series in Eq.~(\ref{primary}) can be retained. This approximation makes the 
spectrum shape unaffected by the chemical factor $Z(\Qz-\qj)/Z(\Qz)$ which can
be embodied in an overall normalization factor. Thus, in the Boltzmann limit:

\begin{eqnarray}\label{primboltz} 
 && \!\!\!\Big\langle \frac{\d n_{j}}{\d p_T} \Big\rangle^{\rm primary} 
 = \frac{\langle n_j \rangle^{\rm prim}}{m^2_j T {\rm K_2}(m_j/T)\sqrt{1+\utms}} 
 \, m_T \, p_T \nonumber \\ 
 && \!\!\!\times {\rm K}_1 \left( \frac{\sqrt{1+\utms}\,m_T}{T}\right) 
 {\rm I}_0 \left( \frac{\utm p_T}{T}\right)
\end{eqnarray}
where $\langle n_j \rangle^{\rm prim}$ is the primary multiplicity. For resonances with 
width $\Gamma > 1$ MeV, the single mass value $m_j$ in Eq.~(\ref{primary2}) 
is replaced with a relativistic Breit-Wigner distribution over an interval 
$[m_0-\Delta m,m_0+\Delta m]$, where $\Delta m = \min [m-m_{\rm threshold},2\Gamma]$ 
and $m_{\rm threshold}$ is the minimal required mass to open all decay modes.

The calculation of the second term in Eq.~(\ref{ptspcan}) is much more involved
as a further four-dimensional integration and up to hundreds of terms are 
implied in the sum for each hadron species. This troublesome problem has been 
dealt with in refs.~\cite{sollfran,sollfran2} and worked out by an analytical
calculation of the kernels $K_{Tjk}$ for two and three body decays and performing
integrations either analytically (whenever possible) or numerically for a 
reduced set of hadrons feeding the examined $j^{\rm th}$ hadron species. So far,
this method has not allowed in practice an exhaustive computation of
the resonance decay effect onto observed spectra. Therefore, we have devised 
a new method based on a mixing of Monte-Carlo and analytical calculations that
is described in the following.

Let us first consider one term in the sum in the right hand side of 
Eq.~(\ref{ptspcan}); the Boltzmann limit is appropriate for all $k$'s as pions, 
the lightest hadrons, are certainly not involved and one can write: 

\begin{eqnarray}\label{second}
 && \!\!\!\Big\langle \frac{\d n_{j}}{\d p_T} \Big\rangle^{k \rightarrow j}
  \simeq \int \d^4 t \; K_{Tjk}(p_T,t) \, \frac{\oV (2J_k+1)}{(2\pi)^3} 
 \e^{-\beta_{\utm} \cdot p_j} \, \nonumber \\
 && \times \frac{Z(T,\Qz-\qj,\oV)}{Z(T,\Qz,\oV)} 
\end{eqnarray} 
The above expression is in fact the transverse momentum spectrum of hadron $j$ 
stemming from the decays of hadron $k$ emitted from a cluster at temperature $T$ 
boosted along $x$ axis. Instead of trying to calculate all kernels $K_{Tjk}$, our 
starting point is to express this spectrum as a function of the same spectrum in the 
EGC's rest frame. Let us then consider the invariant spectrum:

\begin{equation}
  \epsilon \Big\langle \frac{\d n}{\d^3 {\rm p}}\Big\rangle^{k\rightarrow j} = 
  \epsilon^*(p)\Big\langle\frac{\d n}{\d^3 {\rm p^*}}\Big\rangle^{k \rightarrow j}
  \!\!\!\!\!\!(p^*(p))
\end{equation}
where the starred quantities pertain to the cluster's rest frame,
$p=(\epsilon,{\bf p})$ is the four-momentum and $p^*$ is linked to $p$ through 
the Lorentz transformation $\hat \LL_1(\utm)^{-1}$. The transverse momentum 
spectrum can be obtained by integrating the cylindrical coordinates $p_z$ and 
$\varphi$, namely:

\begin{equation}\label{spectransf}
 \Big\langle\frac{\d n}{\d p_T}\Big\rangle^{k\rightarrow j} = 
 \int_{-\infty}^{+\infty} \!\!\!\! \d p_z \int_{-\pi}^{\pi} \!\!\! \d \varphi 
 \; p_T \,\frac{\epsilon^*(p)}{\epsilon} \Big\langle\frac{\d n}{\d^3 {\rm p^*}}
 \Big\rangle^{k\rightarrow j}\!\!\!\!\!\!(p^*(p))
\end{equation}              
The rest-frame spectrum $\Big\langle \d n /\d^3 {\rm p^*}\Big\rangle^{k\rightarrow j}$,
expressed as a function of $p^*$, has the remarkable feature of being isotropic.
In fact, no dependence on direction is possible for the momentum of a hadron
($j$) produced by the decay of a hadron ($k$) whose primary spectrum is supposed
to be, according to the prediction of the statistical model, isotropic. Therefore:

\begin{equation}\label{isotropy}
 \Big\langle\frac{\d n}{\d^3 {\rm p^*}}\Big\rangle^{k\rightarrow j}\!\!\!\!\!\!(p^*)= 
 \frac{\Big\langle\frac{\d n}{\d {\rm p^*}}\Big\rangle^{k\rightarrow j}
 \!\!\!\!\!\!({\rm p^*})}{4 \pi {\rm p^{*2}}}= 
 \frac{\Big\langle\frac{\d n}{\d \epsilon^*}\Big\rangle^{k\rightarrow j}\!\!\!\!\!\!
 ({\epsilon^*})}{4 \pi {\rm p^*} \epsilon^*}
\end{equation}
By using Eq.~(\ref{isotropy}), Eq.~(\ref{spectransf}) can be written as an integral 
transform:

\begin{eqnarray}\label{spectransf2}
 && \Big\langle\frac{\d n}{\d p_T}\Big\rangle^{k\rightarrow j} = \nonumber \\
 && = \int \d x \int_{-\infty}^{+\infty} \!\!\!\! \d p_z \int_{-\pi}^{\pi} 
 \!\!\! \d \varphi \; \frac{p_T}{4 \pi {\rm p^*} \epsilon} \,
 \Big\langle\frac{\d n}{\d \epsilon^*}\Big\rangle^{k\rightarrow j}\!\!\!\!\!\!
 (x) \, \delta(x-\epsilon^*(p)) \nonumber \\
 &&
\end{eqnarray} 
The energy in the cluster rest frame can be calculated by using the formula
for a Lorentz transformation along $x$ axis:

\begin{equation}\label{epsstar}
  \epsilon^*(p) = \gmt \epsilon - \btt \gmt p_T \cos \varphi = 
  \sqrt{1+\utms} \epsilon - \utm p_T \cos \varphi  
\end{equation}  
The integration on the variables $p_z$ and $\varphi$ can be performed analytically
and the result is (see Appendix C):

\begin{eqnarray}\label{second2}
&& \Big\langle \frac{\d n}{\d p_T} \Big\rangle^{k \rightarrow j} =  
\frac{4 p_T}{\sqrt{1+\utms}\,m_T} \int_0^{+\infty} \!\!\!\! \d {\rm p}^* 
\Big\langle\frac{\d n}{\d {\rm p}^*}\Big\rangle^{k \rightarrow j} \nonumber \\
&& \times \; \frac{1}{2 \pi {\rm p}^* \sqrt{(z_+ - z_{\rm min})(z_{\rm max}+1)}} 
\; {\rm F}\left( \frac{\pi}{2},r \right)
\end{eqnarray}
where ${\rm F}$ is the elliptic integral of the first kind and:

\begin{eqnarray}\label{zetas} 
 && r = \sqrt{\frac{(z_+ - z_{\rm max})(z_{\rm min}+1)}{(z_+ - z_{\rm min})(z_{\rm max}+1)}}
  \nonumber \\ 
 && z_{\rm max} = {\rm max}(1,z_-) \qquad z_{\rm min} = {\rm min}(1,z_-) 
  \nonumber \\ 
 && z_{\pm}= \frac{\epsilon^* \pm \utm p_T}{\sqrt{1+\utms}\, m_T} \;\;
 \epsilon^* = \sqrt{{\rm p}^{*2}+m_j^2}
\end{eqnarray} 
What have we gained by expressing the secondary spectrum as in Eq.~(\ref{second2})? 
In fact, we have unfolded its dependence on $\utm$ in an explicit form, unlike in 
Eq.~(\ref{second}), and this has a strong impact on the problem overall, since we
have now to calculate only the momentum spectrum $\Big\langle \d n/\d {\rm p}^*
\Big\rangle^{k \rightarrow j}$ in the cluster rest frame, whose shape depends 
only on $T$. Instead of doing this analytically, which is much too complicated 
for three or more body decays, we have adopted a Monte-Carlo procedure which may 
be summarized as follows:
\begin{enumerate}
 \item{} at a fixed temperature, for each light flavoured hadron $k$ with mass 
$< 1.8$ GeV, we have simulated 200,000 decays distributed among the known decays 
modes according to the known branching ratios. If the decay products of $k$ were 
unstable particles, the decay chain has been continued until 'stable' states 
(where 'stable' is an experiment-dependent definition) were reached. The kinematic 
distributions of hadronic decays have been calculated according to relativistic 
phase space \cite{jetset}. Resonances with a width $\Gamma > 1$ MeV 
have been given a distributed mass according to a relativistic Breit-Wigner (see
discussion about Eq.~(\ref{primboltz})). 
The considered hadron species, their masses and widths, their decay modes and 
relevant branching ratios have been taken from ref.~\cite{pdg}. For the special 
case of \ee collisions, where heavy quarks are abundantly produced, also heavy 
flavoured hadron decays have been simulated by using known branching ratios 
\cite{pdg} along with the predictions of the Lund model \cite{jetset} for unmeasured 
channels. The set of heavy flavoured states included all measured ones \cite{pdg} 
and all other states quoted in JETSET program tables \cite{jetset} predicted by HQET 
 \item{} the obtained momentum spectra of the various hadrons emerging from 
the decay chain initiated by the hadron $k$ have been stored in 200 bins histograms 
ranging from 0 to 3 GeV for hadronic collisions and in 333 bins histograms ranging
from 0 to 5 GeV for \ee collisions, owing to the hardness of heavy flavoured hadron decay
products. Thereafter, they have been smoothed according to a quadratic interpolation 
procedure \cite{hbook}, normalized to 1 and stored into readable files     
 \item{} the procedure described in 1) and 2) has been repeated for temperatures
ranging from 140 to 190 MeV in steps of 1 MeV. 
\end{enumerate}

\begin{table*}[htb]\label{tab1}
\caption{Parameters of the statistical hadronisation model fitted with average 
multiplicities in hadronic and \ee collisions. Also quoted the Wroblewski factor 
$\lambda_S = 2\ssb/(\uub+\ddb)$ calculated with the newly produced valence quark 
pairs from fitted primary hadron multiplicities. The very large value of $\chi^2$ 
in pp collisions is owing to the lack of systematic errors \cite{chlia} for the 
measurements quoted in the main reference~\cite{agui91} (see also Table~2).}      
\begin{tabular}{lllllll}
\hline\noalign{\smallskip}
 Collision & $\sqrt s$ (GeV) & $T$ (MeV) & $VT^3$  & $\ssb$ & $\chi^2/{\rm dof}$      &$\lambda_S$\\
\noalign{\smallskip}\hline\noalign{\smallskip}
 K$^+$p    & 11.5 &  176.9 $\pm$ 2.6 & 5.85 $\pm$ 0.39  & 0.347 $\pm$ 0.020 & 68.0/14  & 0.203$\pm$0.012 \\
 K$^+$p    & 21.7 &  175.8 $\pm$ 5.6 & 8.5 $\pm$ 1.1    & 0.578 $\pm$ 0.056 & 38.0/9   & 0.227$\pm$0.028 \\
 $\pi^+$p  & 21.7 &  170.5 $\pm$ 5.2 & 10.8 $\pm$ 1.2   & 0.734 $\pm$ 0.049 & 39.7/7   & 0.282$\pm$0.032 \\
 pp        & 27.4 &  162.4 $\pm$ 1.6 & 14.24 $\pm$ 0.66 & 0.653 $\pm$ 0.017 & 315.2/29 & 0.200$\pm$0.005 \\
\noalign{\smallskip}\hline
\noalign{\smallskip}
 Collision & $\sqrt s$ (GeV) & $T$ (MeV) & $VT^3$  & $\gamma_S$ & $\chi^2/{\rm dof}$  &$\lambda_S$ \\
\noalign{\smallskip}\hline\noalign{\smallskip}
 \ee       & 14   &  167.4 $\pm$ 6.5 & 9.7  $\pm$ 1.6   & 0.795 $\pm$ 0.088 & 1.5/3   & 0.243$\pm$0.036 \\
 \ee       & 22   &  172.5 $\pm$ 6.7 & 10.6 $\pm$ 2.2   & 0.767 $\pm$ 0.094 & 1.0/3   & 0.263$\pm$0.042 \\ 
 \ee       & 29   &  159.0 $\pm$ 2.6 & 17.3 $\pm$ 1.5   & 0.710 $\pm$ 0.047 & 29.3/12 & 0.228$\pm$0.015 \\ 
 \ee       & 35   &  158.7 $\pm$ 3.4 & 17.5 $\pm$ 1.8   & 0.746 $\pm$ 0.040 & 8.8/7   & 0.242$\pm$0.017 \\
 \ee       & 43   &  162.5 $\pm$ 8.1 & 16.2 $\pm$ 3.1   & 0.768 $\pm$ 0.065 & 3.0/3   & 0.261$\pm$0.030 \\ 
\noalign{\smallskip}\hline 
\end{tabular}
\end{table*}     
Overall, for hadronic collisions, the Monte-Carlo calculation has involved the 
simulation of 200,000 decays of 144 light flavoured hadronic states for 51 temperature 
steps, yielding a grand-total of 59,721 spectra and $\approx 1.2\,10^7$ stored 
real numbers. For \ee collisions, also the decays of 83 heavy flavoured hadron decays 
have been simulated yielding a grand-total of 336,447 spectra and $\approx 1.1\,10^8$
stored real numbers.\\
If $\nu_{kj}$ is the momentum spectrum for the decay $k \rightarrow \ldots \rightarrow j$, 
normalized to 1, and $f_{kj}$ is the mean number of hadrons $j$ produced by the
decay chain started by the primary hadron $k$, the final formula obtained by 
adding primary (\ref{primboltz}) and secondary (\ref{second2}) contributions reads:

\begin{eqnarray}\label{final} 
 && \! \Big\langle \frac{\d n_{j}}{\d p_T} \Big\rangle = 
 \frac{\langle n_j \rangle^{\rm prim}}{m^2_j T {\rm K_2}(\frac{m_j}{T})} \, 
 \frac{m_T \, p_T}{\sqrt{1+\utms}} \, {\rm K}_1 \left( \frac{\sqrt{1+\utms}\,m_T}{T}\right) 
 \nonumber \\ 
 && \! \times {\rm I}_0 \left( \frac{\utm p_T}{T}\right)
 + \sum_k \frac{4 \langle n_k \rangle^{\rm prim} f_{kj} p_T}{\sqrt{1+\utms}\,m_T} \nonumber \\ 
 && \! \times \int_0^{+\infty} \!\!\!\!\!\!\! \d {\rm p}^* \; \frac{\nu_{kj}({\rm p}^*)\,
 {\rm F}(\pi/2,r)}{2 \pi {\rm p}^* \sqrt{(z_+ - z_{\rm min})(z_{\rm max}+1)}}
 \nonumber \\
 &&  
\end{eqnarray}
for all hadrons except pions, for which Bose-Einstein statistics cannot be 
neglected and Eq.~(\ref{primary2}) for the primary spectrum must be used. 
The $\langle n \rangle^{\rm prim}$'s are the primary multiplicities. The $f_{kj}$ have 
been computed on the basis of experimentally known branching ratios. The 
momentum integral in Eq.~(\ref{final}) has been computed numerically by taking into
account that $\nu_{jk}$ are stepwise functions vanishing for ${\rm p}^* > 3$
GeV (5 GeV for \ee collisions). This truncation introduces a further numerical 
error in the computation which is found to be negligible.    

\section{Data analysis}    

The main goal of data analysis is the assessment of the consistency of 
experimental spectra with the prediction of the statistical hadronisation
model. This has a twofold implication: firstly, spectrum shapes of different hadrons 
at a given centre-of-mass energy should be described by essentially the same 
parameters $T$ and $\utm$, as long as $\utms \ll 1$, as discussed before; secondly, 
the best-fit $T$'s should be in agreement with the temperature fitted with hadron 
multiplicities. The latter statement holds provided that hadrons do not undergo
significant elastic rescattering after their formation (or chemical freeze-out), a 
generally accepted belief in elementary collision. In view of this objective, a
natural requirement for the data set for a given collision and centre-of-mass energy, 
is the existence of a considerably large sample of both measured transverse momentum 
spectra and integrated multiplicities of different hadron species. Furthermore,
centre-of-mass energy must be high enough to allow the use of a canonical formalism,
what is expected to occur above roughly $\sqrt s \approx 10$ GeV (see discussion in Sect.~1). 
Essentially, four collision systems fulfilling these requirementshave been found: 
K$^+$p at $\sqrt s = 11.5$ and $\sqrt s = 21.7$ GeV, $\pi^+$p at $\sqrt s = 21.7$ 
GeV and pp at $\sqrt s = 27.4$ GeV. Though the covered energy range is not large,  
these points should allow to settle both of the previous issues.

\begin{table}[!t]\label{tab2}
\begin{tiny}
\caption{Particle multiplicities in hadronic collisions.} 
\begin{tabular}{llll}
\hline\noalign{\smallskip} 
  {\scriptsize Particle}&{\scriptsize Measured}&{\scriptsize Fitted}&
  {\scriptsize Reference}\\ 
\noalign{\smallskip}\hline
\noalign{\smallskip}
\multicolumn{4}{l}{\scriptsize K$^+$p $\sqrt s$ = 11.5 GeV}  \\                    
\noalign{\smallskip}\hline
  $\pi^0$          &   2.50 $\pm$ 0.12      &  2.724	  &  \cite{barth84}	  \\
  $\pi^+$	   &   2.860 $\pm$ 0.087    &  2.802	  &  \cite{barth81}$^{ab}$  \\
  $\pi^-$	   &   1.960 $\pm$ 0.025    &  1.973	  &  \cite{barth81}$^{ab}$  \\   
  K$^+$	           &   0.685 $\pm$ 0.062    &  0.6563	  &  \cite{barth81}$^{ab}$  \\
  K$^-$	           &   0.087 $\pm$ 0.025    &  0.09420    &  \cite{barth81}$^{ab}$  \\ 
  $\rho^0$	   &   0.254 $\pm$ 0.020    &  0.3468	  &  \cite{barth82}$^{b}$  \\    
  K$^{*0}$	   &   0.249 $\pm$ 0.023    &  0.2150	  &  \cite{barth82}$^{b}$  \\
$\bar{\rm K}^{*0}$ &   0.0685 $\pm$ 0.019   &  0.03225    &  \cite{barth82}$^{ab}$  \\  
  K$^{*+}$	   &   0.253 $\pm$ 0.019    &  0.2749	  &  \cite{barth82}$^{b}$  \\
  K$^{*-}$	   &   0.0180 $\pm$ 0.0093  &  0.02757    &  \cite{barth82}$^{ab}$  \\ 
  f$_2(1270)$	   &   0.0734 $\pm$ 0.0140  &  0.05044    &  \cite{barth82}$^{ab}$  \\   
  K$_2^{*0}$       &   0.0616 $\pm$ 0.0068  &  0.02266    &  \cite{barth82}$^{ab}$  \\
  K$_2^{*+}$       &   0.0653 $\pm$ 0.016   &  0.03292    &  \cite{barth82}$^{ab}$  \\
  $\Lambda$        &   0.0678 $\pm$ 0.0031  &  0.06828    &  \cite{barth81_2}$^{bc}$ \\
  $\bar\Lambda$    &   0.0404 $\pm$ 0.0025  &  0.04266    &  \cite{barth81_2}$^{bc}$ \\  
  $\Sigma^{*+}$    &   0.0124 $\pm$ 0.0037  &  0.01089    &  \cite{barth81_2}$^{bc}$ \\
  $\Sigma^{*-}$    &   0.00436 $\pm$ 0.00310&  0.005099   &  \cite{barth81_2}$^{bc}$ \\
\hline				       
\noalign{\smallskip}
\multicolumn{4}{l}{\scriptsize K$^+$p $\sqrt s$ = 21.7 GeV}  \\  
\noalign{\smallskip}\hline				      
   charged           &    8.21 $\pm$ 0.12    &   8.228       & \cite{adam86}$^{c}$ \\
   $\pi^0$	     &    2.93 $\pm$ 0.47    &   3.717       & \cite{atay92}$^{d}$ \\
   K$^0_S$           &    0.390 $\pm$ 0.045  &	 0.4166      & \cite{azhi90} \\ 
   $\rho^0$          &    0.308 $\pm$ 0.037  &	 0.4820      & \cite{agab89} \\     
   K$^{*0}$	     &    0.290 $\pm$ 0.028  &	 0.2486      & \cite{agab89}$^{a}$ \\
$\bar{\rm K}^{*0}$   &    0.142 $\pm$ 0.027  &	 0.06152     & \cite{agab89}$^{a}$ \\   
   K$^{*+}$	     &    0.319 $\pm$ 0.042  &	 0.2980      & \cite{agab89}$^{a}$ \\
   f$_2(1270)$       &    0.095 $\pm$ 0.017  &	 0.06931     & \cite{agab89}$^{a}$ \\
   $\Lambda$         &    0.092 $\pm$ 0.008  &	 0.1009      & \cite{azhi89}$^{a}$    \\
   $\bar\Lambda$     &    0.061 $\pm$ 0.009  &	 0.06688     & \cite{azhi89}$^{a}$    \\   
   $\Delta^{++}$     &    0.214 $\pm$ 0.034  &	 0.1596      & \cite{azhi89}          \\
   $\Sigma^{*+}$     &    0.021 $\pm$ 0.009  &	 0.01519     & \cite{azhi89}$^{a}$    \\   
\hline
\noalign{\smallskip}
\multicolumn{4}{l}{\scriptsize $\pi^+$p $\sqrt s$ = 21.7 GeV} \\
\noalign{\smallskip}\hline
   charged           &   8.31 $\pm$ 0.10   &  8.434        & \cite{adam86}$^{c}$\\
   $\pi^0$	     &   3.092 $\pm$ 0.47  &  4.003        & \cite{atay92}$^{d}$\\
   K$^0_S$           &   0.236 $\pm$ 0.025 &  0.2673       & \cite{azhi90}      \\
   $\rho^0$          &   0.460 $\pm$ 0.020 &  0.5322       & \cite{agab90}$^{a}$\\   
   K$^{*0}$	     &   0.129 $\pm$ 0.014 &  0.09909      & \cite{agab90}$^{a}$\\
$\bar{\rm K}^{*0}$   &   0.148 $\pm$ 0.014 &  0.08879      & \cite{agab90}$^{a}$\\
   $\Lambda$         &   0.103 $\pm$ 0.008 &  0.1195       & \cite{azhi89}$^{a}$\\
   $\bar\Lambda$     &   0.032 $\pm$ 0.005 &  0.03259      & \cite{azhi89}$^{a}$\\
   $\Delta^{++}$     &   0.221 $\pm$ 0.025 &  0.1707       & \cite{azhi89}\\
   $\Sigma^{*+}$     &   0.015 $\pm$ 0.007 &  0.01984      & \cite{azhi89}$^{a}$\\
\hline
\noalign{\smallskip}
\multicolumn{4}{l}{\scriptsize pp $\sqrt s$ = 27.4 GeV} \\
\noalign{\smallskip}\hline
  $\pi^0$	    &   3.87 $\pm$ 0.12     &  4.594      & \cite{agui91}$^{a}$\\			      
  $\pi^+$           &   4.10 $\pm$ 0.11     &  4.479      & \cite{agui91}$^{a}$\\
  $\pi^-$	    &   3.34 $\pm$ 0.08     &  3.612      & \cite{agui91}$^{a}$\\
  K$^+$	            &   0.331 $\pm$ 0.016   &  0.3085     & \cite{agui91}$^{a}$\\
  K$^-$	    	    &   0.224 $\pm$ 0.011   &  0.1852     & \cite{agui91}$^{a}$\\
  $K^0_S$	    &	0.225 $\pm$  0.014  &  0.2377     & \cite{kich79}$^{a}$, \cite{kass79}$^{c}$\\
  $\eta$	    &   0.30 $\pm$ 0.02     &  0.4046     & \cite{agui91}$^{a}$\\  
  $\rho^0$	    &   0.384 $\pm$ 0.018   &  0.5830     & \cite{agui91}$^{a}$\\
  $\rho^+$	    &   0.552 $\pm$ 0.082   &  0.6236     & \cite{agui91}$^{a}$\\
  $\rho^-$	    &   0.354 $\pm$ 0.058   &  0.4698     & \cite{agui91}$^{a}$\\ 
  $\omega$     	    &   0.390 $\pm$ 0.024   &  0.4798     & \cite{agui91}$^{a}$\\ 
  K$^{*0}$	    &   0.120 $\pm$ 0.021   &  0.09458    & \cite{agui91}$^{a}$\\
$\bar{\rm K}^{*0}$  &   0.0902 $\pm$ 0.016  &  0.06278    & \cite{agui91}$^{a}$\\
  K$^{*+}$	    &   0.132 $\pm$ 0.016   &  0.1080     & \cite{agui91}$^{a}$\\
  K$^{*-}$	    &   0.0875 $\pm$ 0.012  &  0.05710    & \cite{agui91}$^{a}$\\    
  f$_0(980)$	    &   0.0226 $\pm$ 0.0079 &  0.03876    & \cite{agui91}$^{a}$\\
  $\phi$	    &	0.0189 $\pm$ 0.0018 &  0.02401    & \cite{agui91}$^{a}$\\  
  f$_2(1270)$	    &   0.0921 $\pm$ 0.012  &  0.06623    & \cite{agui91}$^{a}$\\  
  $\rho_3(1690)$    &   0.078 $\pm$ 0.049   &  0.009045   & \cite{suzu80}$^{a}$\\  
  p	    	    &   1.200 $\pm$ 0.097   &  1.054      & \cite{agui91}$^{a}$\\
$\bar{\rm p}$	    &   0.063 $\pm$ 0.0020  &  0.05277    & \cite{agui91}$^{a}$\\ 
  $\Lambda$         &   0.1230 $\pm$ 0.0062 &  0.1461     & \cite{kich79}$^{a}$, \cite{kass79}$^{c}$\\
  $\bar\Lambda$     &	0.0155 $\pm$ 0.0034 &  0.01669    & \cite{kich79}$^{a}$, \cite{kass79}$^{c}$\\ 
  $\Sigma^+$        &	0.0479 $\pm$ 0.015  &  0.04369    & \cite{agui91}$^{a}$\\
  $\Sigma^-$        &	0.0128 $\pm$ 0.0061 &  0.03252    & \cite{agui91}$^{a}$\\
  $\Delta^{++}$     &	0.218 $\pm$ 0.003   &  0.2514     & \cite{agui91}$^{a}$\\
  $\Delta^0$        &	0.1410 $\pm$ 0.0079 &  0.2057     & \cite{agui91}$^{a}$\\
  $\bar\Delta^{--}$ &   0.0128 $\pm$ 0.0049 &  0.009645   & \cite{agui91}$^{a}$\\
  $\bar\Delta^0$    & 	0.0335 $\pm$ 0.0098 &  0.01426    & \cite{agui91}$^{a}$\\
  $\Sigma^{*+}$     &	0.0204 $\pm$ 0.0024 &  0.02060    & \cite{kich79}$^{ad}$\\
  $\Sigma^{*-}$     &	0.0101 $\pm$ 0.0018 &  0.01396    & \cite{kich79}$^{ad}$\\
  $\Lambda(1520)$   &	0.0171 $\pm$ 0.003  &  0.01054    & \cite{agui91}$^{a}$\\
\hline 
\noalign{\smallskip} 
\multicolumn{4}{l}{\tiny $a$ - Only statistical error}\\
\multicolumn{4}{l}{\tiny $b$ - The multiplicity has been calculated by dividing the quoted cross section}\\ 
\multicolumn{4}{l}{\tiny by the inelastic cross section $\sigma_{\rm in} = 16.07 \pm 0.11$ mb measured 
                   by the same}\\
\multicolumn{4}{l}{\tiny collaboration \cite{barth82_2}}\\
\multicolumn{4}{l}{\tiny $c$ - Systematic error maybe not included in the quoted experimental error}\\
\multicolumn{4}{l}{\tiny $d$ - The multiplicity has been calculated by dividing the quoted cross section}\\ 
\multicolumn{4}{l}{\tiny by the inelastic cross section $\sigma_{\rm in} = 17.53$ mb for
                   K$^+$p, $\sigma_{\rm in} = 20.71$ mb for $\pi^+$p}\\
\multicolumn{4}{l}{\tiny and $\sigma_{\rm in} = 32.80$ mb for pp collisions, inferred from combined 
                   quotations of}\\
\multicolumn{4}{l}{\tiny multiplicities and cross sections of other particles.}\\
\noalign{\smallskip}\hline    
\end{tabular}
\end{tiny}
\end{table}    
        
\subsection{Fit to average multiplicities} 
     
The first step of the analysis is a fit to measured average multiplicities of
the various hadron species. The fit procedure is very much alike that in 
refs.~\cite{beca2,beca3}, though with some significant improvement. The used formula for 
the primary multiplicities is the integral of the spectrum (\ref{primary2}), 
in which only the first term of the series is retained for all hadrons except pions.
Hadron multiplicities can be reproduced only if a further suppression of hadrons 
with valence strange quarks is introduced. This extra strangeness suppression 
has been implemented in refs.~\cite{beca1,beca2,beca3} by means of a phenomenological 
parameter $\gamma_S < 1$ reducing the average primary multiplicities by $\gamma_S^{n_S}$,
$n_S$ being the number of valence strange quarks, with respect to the full 
equilibrium values. However, in view of the constancy of the ratio between newly 
produced strange quarks with respect to u, d quarks \cite{becah}, the extra 
strangeness suppression has been parametrised differently here. Indeed, the 
number of s+${\rm \bar s}$ quarks has been considered as an additional
charge $N_S$ to be conserved into final hadrons along with electric charge $Q$, 
baryon number $N$ and strangeness $S$. This means that quantum number vectors
actually have four components for light flavoured hadrons:

\begin{eqnarray}\label{vectors}
 && \Qz = (Q,N,S,N_S)  \nonumber \\
 && \qj = (Q_j,N_j,S_j,N_{Sj}) 
\end{eqnarray} 
Unlike $Q$, $N$ and $S$, the number of strange quarks to be shared among the 
primary hadrons is not fixed by the initial conditions and may fluctuate. 
Hence, it is assumed that the \ss pairs are independently produced from the
vacuum with fluctuations governed by Poisson distribution. The mean number of \ss pairs, 
$\ssb$, has been taken as a free parameter replacing $\gamma_S$. Therefore, the 
actual formula for average primary multiplicities reads:

\begin{eqnarray}\label{mult}
 && \langle n_j\rangle^{\rm primary} = \frac{\oV T(2J_j+1)}{2\pi^2} \sum_{K=0}^\infty
 \frac{\e^{-\ssb}\ssb^K}{K!} \nonumber \\
 && \times \sum_{n=1}^\infty (\mp 1)^{n+1}\;\frac{m^2}{n}\;
 {\rm K}_2\left(\frac{n m}{T}\right)\, \frac{Z(\Qz-\qj)}{Z(\Qz)}
\end{eqnarray} 
with $\Qz$ and $\qj$ like in Eq.~(\ref{vectors}) and $N_S = 2K + N_S^0$ where $N_S^0$
is the number of valence strange quarks in the colliding particles. The numerical 
computation of partition functions with four fixed quantum numbers, such as in 
Eq.~(\ref{mult}), requires the implementation of a quite involved numerical algorithm 
(described in Appendix D) to keep it within practical times. Notwithstanding, it 
becomes extremely
slow even for relatively small number of \ss pairs, say 5. Therefore, the sum over
\ss pairs in Eq.~(\ref{mult}) has been truncated to $\min[3,\ssb+3\sqrt\ssb]$ and
the Poisson distribution renormalized accordingly. As $\ssb$ turns out to be 
$< 1$ in all examined collisions (see Table 1), the truncation is essentially 
harmless.
    
Once primary average multiplicities of all (light-flavoured) hadron species up 
to a mass of 1.8 GeV are calculated for a given set of free parameters $T$, $\oV$ 
and $\ssb$, final multiplicities to be compared with experimental data are calculated 
by using the known decay modes and branching ratios. The hadronic decay chain is 
continued until particles considered stable by the experiments are reached. For the 
presently examined hadronic collisions, measured at fixed target, all weakly decaying 
light flavoured hadrons are considered as stable. The various hadron species, 
their masses and widths, their decay modes and relevant branching ratios have been 
taken from ref.~\cite{pdg}.
   
The free parameters $T$, $\oV$ and $\ssb$ are determined by means of a $\chi^2$
fit to measured average multiplicities. For each experiment, the most recent 
measurements have been considered. Multiple measurements from different experiments
have been averaged according to the PDG's \cite{pdg} method with error rescaling
in case of discrepancy. The minimisation of the $\chi^2$ is 
performed in two steps in order to also take into account the uncertainties on 
input parameters such as hadron masses, widths and branching ratios, according
to the procedure described in ref.~\cite{becah}, which is summarized hereafter. 
Firstly a $\chi^2$ with only experimental errors is minimised and preliminary 
best-fit model parameters are determined. Then, keeping the preliminarly 
fitted parameters fixed, the variations $\Delta n_j^{l \, {\rm theo}}$ of the 
multiplicities corresponding to the variations of the $l^{\rm th}$ input parameter 
by one standard deviation are calculated. Such variations are considered as 
additional systematic uncertainties on the multiplicities and the following 
covariance matrix is formed:

\begin{equation}
   {\sf C}_{ij}^{\rm sys} = \sum_l \Delta n_i^l  \Delta n_j^l 
\end{equation}
to be added to the experimental covariance matrix ${\sf C}^{\rm exp}$. Finally a 
new $\chi^2$ is minimised with covariance matrix $\sf C^{\rm exp}+{\sf C}^{\rm sys}$
from which the best-fit estimates of the parameters and their errors are 
obtained. Actually more than 130 among the most relevant or poorly known input 
parameters have been varied.  This fit technique upgrades that used in a previous 
analysis of multiplicities in \ee, pp and \ppb collisions \cite{beca1,beca2,beca3} 
in that also the off-diagonal elements of ${\sf C}^{\rm sys}$ are included.

The results of the fit are shown in Tables~1, 2. The $\chi^2$'s are not as 
small as it would have been expected in a statistically consistent analysis. 
Nevertheless, it should be taken into account that many measurements only include 
statistical errors and that the large $\chi^2$ value is often owing to a single 
large deviation. Moreover, it should be emphasized once more that the assumptions
the existence of an EGC relies on, may not be fully realised.
 
The found values of temperatures and Wroblewski parameters $\lambda_S$ are
very close to those found in previous analyses \cite{beca2,beca3,becah}; the 
difference between old \cite{beca2,beca3} and present $T$ value in pp collisions 
at $\sqrt s = 27.4$ GeV (and \ee collisions as well, see Subsect. 5.3) is mainly 
owing to the new parametrisation ($\ssb$ instead of $\gamma_S$), to the upgraded 
fitting procedure taking into account correlations in the systematic errors, to 
the extension of hadron mass spectrum cut-off (from 1.7 to 1.8 GeV) and the use 
of an updated set of hadronic data \cite{pdg}. A noteworthy feature is the 
sizeable increase of temperature as centre-of-mass energy decreases (see Table~1). 
Due to uncertainties relevant to the assumed physical picture and the lack of 
systematic errors in several measurements, it cannot be established, for the present, 
whether this increase is a genuine physical effect rather than a numerical artefact 
in the fit. However, it should be remarked that such an effect has been been 
advocated as a possible explanation of the constancy of ${\rm \bar p}/\pi$ ratio 
\cite{gore}. 

\begin{figure}
 \resizebox{0.48\textwidth}{!}{%
 \includegraphics{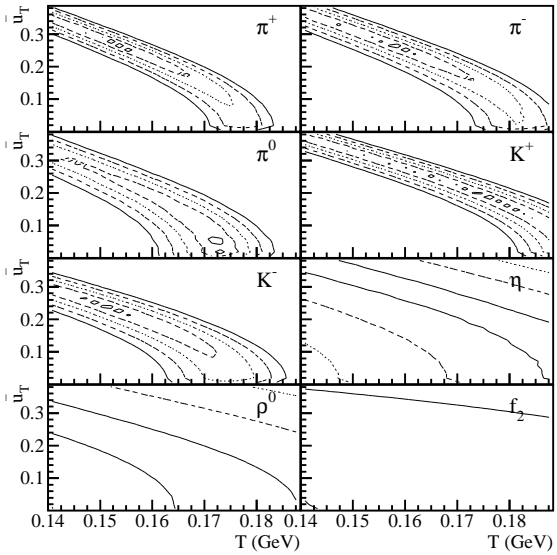}}
\caption{$\chi^2$ contour lines in the $T-\utm$ plane for the fits to 
identified particle transverse momentum spectra in pp collisions at 
$\sqrt s = 27.4$ GeV. The contours range from $\chi^2_{\rm min}+1$ 
(1 $\sigma$) to $\chi^2_{\rm min}+81$ (9 $\sigma$'s).}
\label{contpp}
\end{figure}
      
\subsection{Fit to transverse momentum spectra} 
 
The measured transverse momentum spectra are fitted to the Eq.~(\ref{final})
for all hadron species except pions; for the latter, a special treatment was
necessary, which is discussed later. A peculiar feature of the statistical
model is the strong relation between multiplicities and spectra as both
have a chief dependence on the same parameter $T$. However, as Eq.~(\ref{final}) 
shows, final hadron spectra have both a direct and an indirect dependence 
on temperature through primary multiplicities. Since we want to test the 
consistency between the {\em chemical temperature} $T_{ch}$ obtained from fits 
to multiplicities and that - denoted as $T$ - obtained from spectrum shape 
analysis, we have fixed the primary multiplicities $\langle n_k \rangle^{\rm prim}$ and 
$\langle n_j \rangle^{\rm prim}$ in Eq.~(\ref{final}) to the values obtained from the 
independent multiplicity fits as shown in Tables~1 and 2. This position also 
allows to disentangle the dependence on $T$ of the shapes of all spectrum components 
from that of their overall normalization. As far as pions are concerned, Bose-
Einstein statistics cannot be neglected and the first term on the right
hand side of Eq.~(\ref{final}) must be replaced with the series on the right 
hand side of Eq.~(\ref{primary2}), actually truncated at $n=5$. This is quite 
a special case compared with all other hadrons because chemical factors also 
affect the shape of the spectrum besides its normalization. Therefore, both 
integrated primary multiplicity and chemical factors of pions have been fixed 
to those calculated with the parameters $\oV,T_{ch},\ssb$ of the multiplicity
fit.    

\begin{figure}
 \resizebox{0.48\textwidth}{!}{%
 \includegraphics{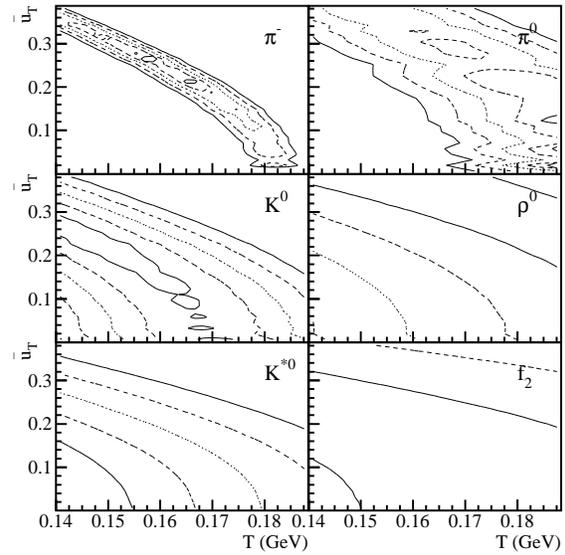}}
\caption{$\chi^2$ contour lines in the $T-\utm$ plane for the fits to 
identified particle transverse momentum spectra in K$^+$p collisions at 
$\sqrt s = 21.7$ GeV. The contours range from $\chi^2_{\rm min}+1$ 
(1 $\sigma$) to $\chi^2_{\rm min}+25$ (5 $\sigma$'s).}
\label{contkp}
\end{figure}
\begin{figure}
 \resizebox{0.48\textwidth}{!}{%
 \includegraphics{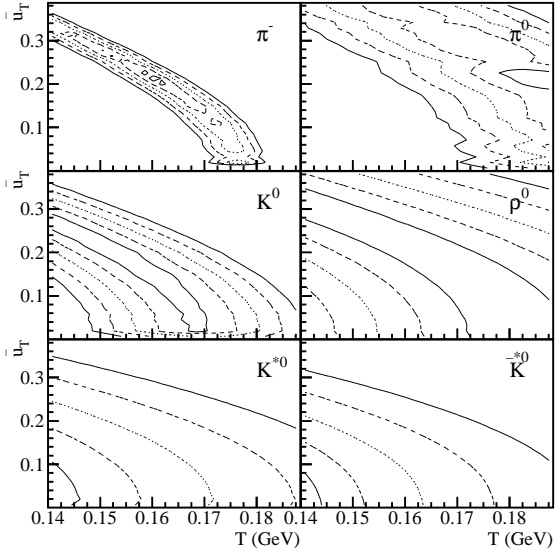}}
\caption{$\chi^2$ contour lines in the $T-\utm$ plane for the fits to 
identified particle transverse momentum spectra in $\pi^+$p collisions at 
$\sqrt s = 21.7$ GeV. The contours range from $\chi^2_{\rm min}+1$ 
(1 $\sigma$) to $\chi^2_{\rm min}+25$ (5 $\sigma$'s).}
\label{contgp}
\end{figure}
Each measured spectrum is fitted with three free parameters: $T$, $\utm$ and 
an overall normalization factor $A$ multiplying the right hand side
of Eq.~(\ref{final}). In principle, this factor would not be needed if the
average multiplicities $\langle n_k \rangle$ and $\langle n_j \rangle$ were
in perfect agreement with experimental measurements. However, this is never 
the case as more or less pronounced deviations from measured values usually 
occur and, also, transverse momentum spectrum and multiplicity may have been measured 
in different experiments. Therefore, the normalization parameter $A$ has to 
be introduced in order to ultimately decouple the dependence of spectrum shape 
on $T$ and $\utm$ from that of its integral.

\begin{figure*}
 \resizebox{0.99\textwidth}{!}{%
 \includegraphics{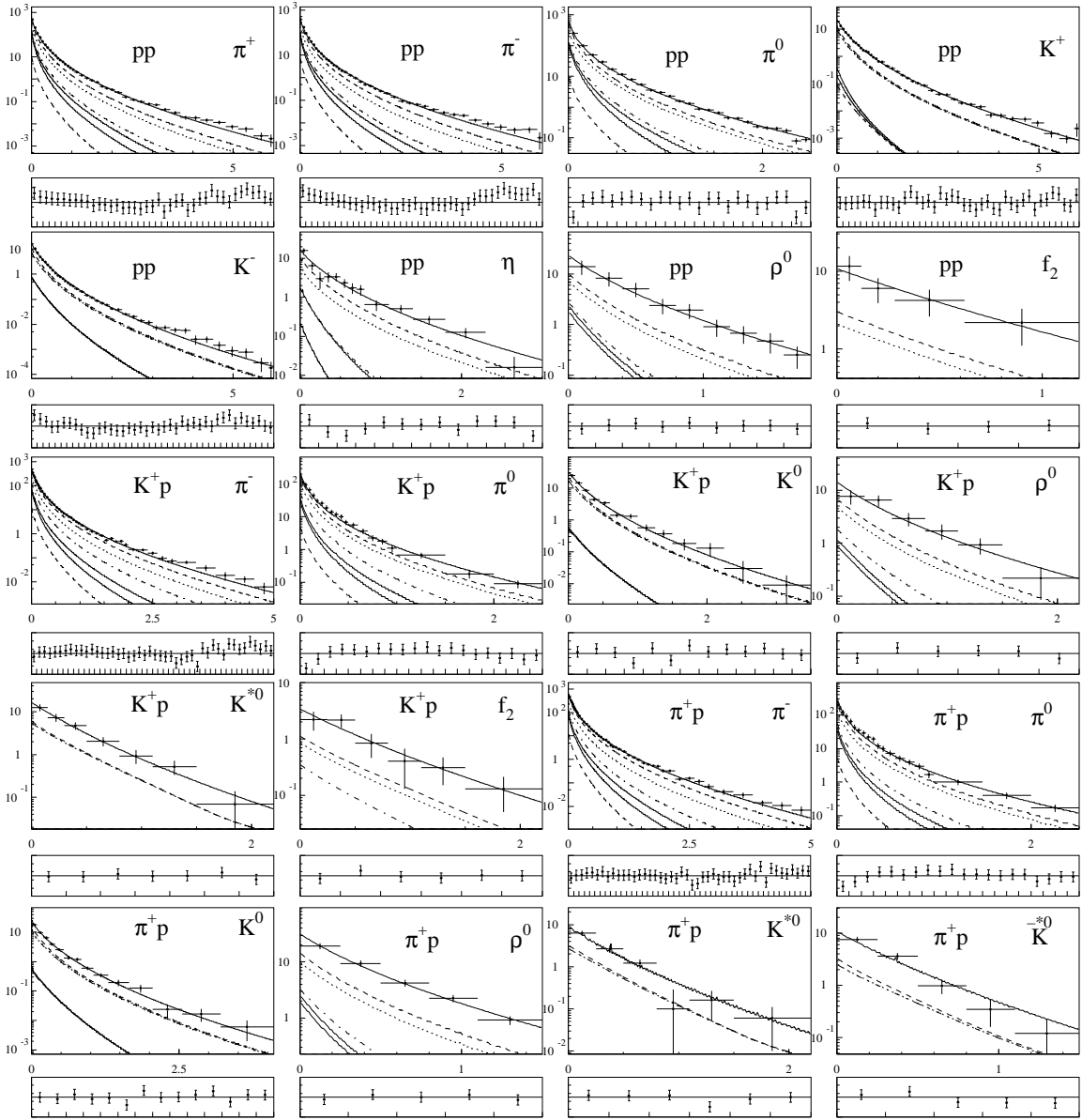}}
\caption{Transverse momentum spectra of identified particles in pp, K$^+$p and
$\pi^+$p collisions at $\sqrt s = 27.4, 21.7$ and 21.7 GeV respectively. For each
particle, the upper plot is $\d \sigma/ \d p_T^2$ (mb/GeV$^2$) versus $p_T^2$ (GeV$^2$) in 
logarithmic scale, the lower plot is the fit residuals distribution in linear scale
in standard deviation units. In the spectrum plots, the upper solid lines are the 
best-fit ones while data are shown as full dots with error bars. Also shown as dashed 
or solid lines the cumulative contributions due to secondary decays for six classes 
of parent hadrons (from bottom to top: strange baryons, delta-like baryons, N-like 
baryons, strange mesons, non-strange charged mesons and completely neutral mesons).}
\label{spectra}      
\end{figure*}

The fit consists of a minimisation of a $\chi^2$ for each measured hadron $j$:
 
\begin{equation}\label{chi2}
  \chi^2_j= \sum_{i=1}^M \frac{\left( \Delta n_{ji \; {\rm theo}}
   - \Delta n_{ji \; {\rm exp}}\right)^2}
   {\sigma_{i \; {\rm exp}}^2+\sigma_{i \; {\rm sys}}^2}
\end{equation}      
where the sum runs over the $M$ bins of the experimental spectrum and $\Delta n_{ji}$
are the integrals over the $i^{\rm th}$ bin of the measured spectrum: 

\begin{equation}
  \Delta n_{ji} = \int_{i^{\rm th} {\rm bin}} \d p_T^k \;
   \Big\langle \frac{\d n_{j}}{\d p_T^k} \Big\rangle
\end{equation}
The exponent $k$ stands for the different variables used for the spectra, 
essentially $p_T$ or $p_T^2$.  
As far as the experimental value is concerned, the above integral is simply
the product of the quoted differential spectrum times the bin width. 
Besides the quoted experimental error $\sigma_{i \; {\rm exp}}$, we have also 
included in the $\chi^2$ (\ref{chi2}) the systematic error (added in quadrature) 
due to the uncertainty on masses, widths and branching ratios of involved hadrons. 
This is calculated by using the estimated systematic uncertainties $\varepsilon_j$ 
on the total multiplicity with the method described in the previous subsection:

\begin{equation}
  \varepsilon_j = \sqrt{\sum_l (\Delta n_j^l)^2} \nonumber
\end{equation}     
This error is assumed to contribute to the errors of all bins independently 
and proportionally to the spectrum height:

\begin{equation}
  \sigma_{i \; {\rm sys}} = \varepsilon_j \frac{(\Delta n_{ji \; {\rm exp}})^2}
  {\sqrt{\sum_{i=1}^M (\Delta n_{ji \; {\rm exp}})^2}}
\end{equation}
so that the relative systematic error on the spectrum height in each bin is 
constant. With such an assumption, the correlations between different bins
stemming from the correlations between uncertainties on primary multiplicities 
(pointed out in Subsect. 5.1) in Eq.~(\ref{final}) are neglected and this should 
lead to somewhat optimistic $\chi^2$ values. This effect can be seen in the
residuals distributions of fits to pion spectra in Fig.~\ref{spectra} which
show some coherent structure indicating a bin-to-bin correlation.   
\begin{figure}
 \resizebox{0.48\textwidth}{!}{%
 \includegraphics{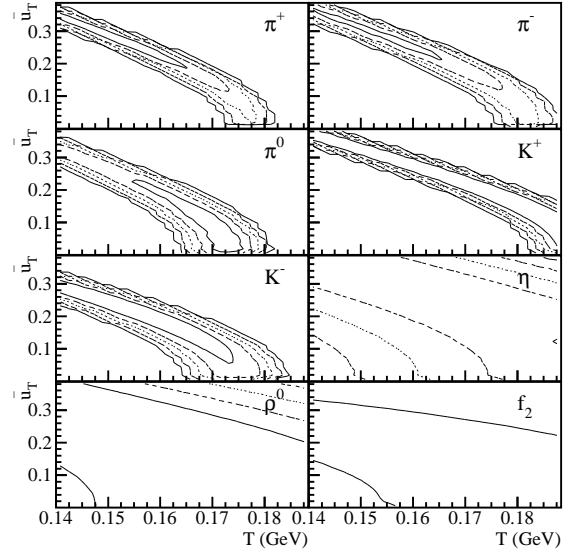}}
\caption{Confidence level contour lines relevant to the $\chi^2$ test for 
spectra in pp collisions at $\sqrt s = 27.4$ GeV. The contours range from
a confidence level of 31.7\% (1 $\sigma$) to 5.7 10$^{-5}$ (5 $\sigma$'s)
in steps of 1 $\sigma$, except K$^-$ for which the innermost contour 
corresponds to a confidence level of 21.1\%.}
\label{confpp}
\end{figure}

The $\chi^2$ minimisation yields preliminary $T$, $\utm$ and $A$ values for each
hadron. Since both $T$ and $\utm$ parameters determine the slope of the spectrum,
they are expected to be correlated variables in the fit. Hence,
in order to check the fit outcome quality and assess the degree of correlation
between $T$ and $\utm$, $\chi^2$ contour plots in the $T-\utm$ plane are calculated 
by keeping $A$ fixed equal to its preliminary best-fit value. It must be remarked 
that best-fit $A$ is almost entirely independent of $T$ and $\utm$ as the full 
spectrum integral in Eq.~(\ref{final}) depend only on the multiplicities 
$\langle n_j \rangle$ and $\langle n_k \rangle$, which are fixed in the spectrum
fit:

\begin{equation}
 \langle n_j\rangle = A \left[ \langle n_j\rangle^{\rm prim} + 
 \langle n_k \rangle f_{kj} \right] 
\end{equation}
Indeed, a residual correlation between $A$ and $T-\utm$ is brought about by 
the unmeasured portion of the spectrum at high $p_T$ which, however, gives a 
very small contribution to the overall integral. Thus, the quasi-independence 
of $A$ on $T$ and $\utm$ makes the pattern of local minima of the $\chi^2$ in 
the $T-\utm$ plane an (almost) absolute one.
 
Most $\chi^2$ contour plots exhibit a pattern of several shallow local 
minima along a valley in the $T-\utm$ plane (see Figs.~\ref{contpp}, \ref{contkp},
\ref{contgp}) demostrating a strong anticorrelation between $T$ and $\utm$ in the fit. 
This feature makes the definition of a best fit quite difficult; there are indeed 
many local minima with sufficiently low $\chi^2$ or, equivalently, providing 
a satisfactory confidence level for the $\chi^2$ test (see Fig.~\ref{confpp}). 

\begin{figure}
 \resizebox{0.48\textwidth}{!}{%
 \includegraphics{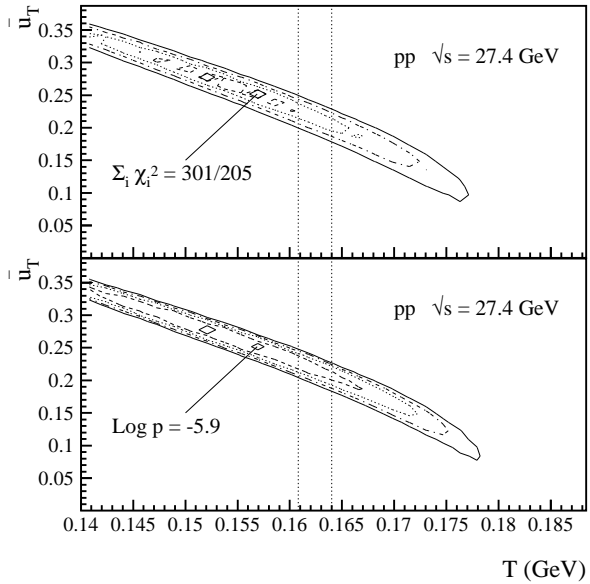}}
\caption{Global $\chi^2$ contour lines (top) and confidence level contours
for the test variable $p$ in Eq.~(\ref{ptest}) (bottom) in the $T-\utm$ 
plane for the fits to transverse momentum spectra in pp 
collisions at $\sqrt s = 27.4$ GeV. The $\chi^2$ contours range from 
$\chi^2_{\rm min}+1$ (1 $\sigma$) to $\chi^2_{\rm min}+81$ (9 $\sigma$'s)
whereas the confidence level contours range from $10^{-6}$ to $10^{-14}$. 
Also shown the absolute minimum (top) and maximum (bottom) and the vertical 
1 $\sigma$ temperature band determined by the fit to multiplicities. The 
local $\chi^2$ minimum  chosen as starting point of single particle fits is 
the small one beside the vertical band on the left hand side.}
\label{globpp}
\end{figure}
\begin{figure}
 \resizebox{0.48\textwidth}{!}{%
 \includegraphics{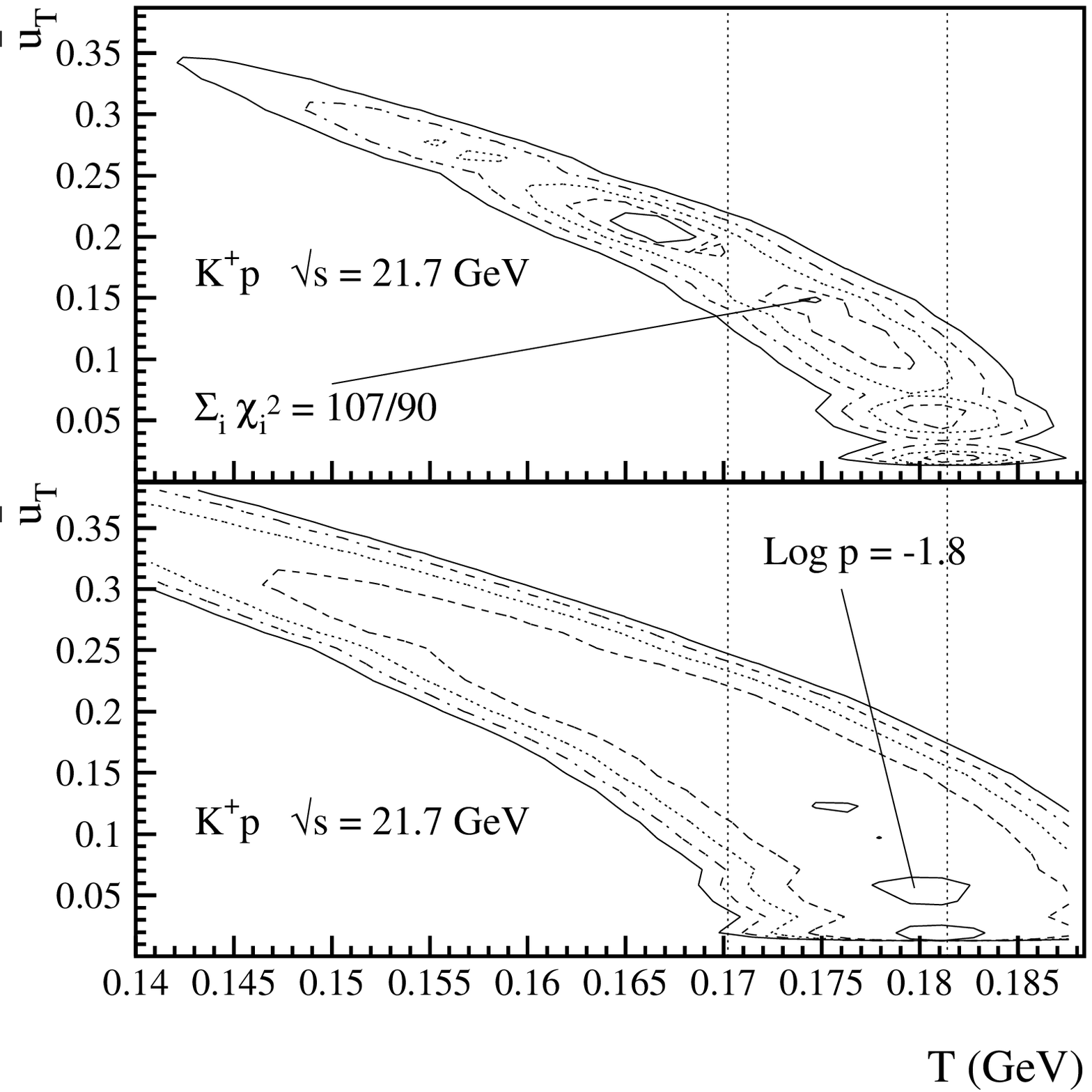}}
\caption{Global $\chi^2$ contour lines (top) and confidence level contours
for the test variable $p$ in Eq.~(\ref{ptest}) (bottom) in the $T-\utm$ 
plane for the fits to transverse momentum spectra in 
K$^+$p collisions at $\sqrt s = 21.7$ GeV. The $\chi^2$ contours range from 
$\chi^2_{\rm min}+1$ (1 $\sigma$) to $\chi^2_{\rm min}+25$ (5 $\sigma$'s)
whereas the confidence level contours range from $10^{-2}$ to $10^{-10}$. 
Also shown the absolute minimum (top) and maximum (bottom) and the vertical 
1 $\sigma$ temperature band determined by the fit to multiplicities. The 
local $\chi^2$ minimum 
chosen as starting point of single particle fits is the absolute minimum.}
\label{globkp}
\end{figure}
\begin{figure}
 \resizebox{0.48\textwidth}{!}{%
 \includegraphics{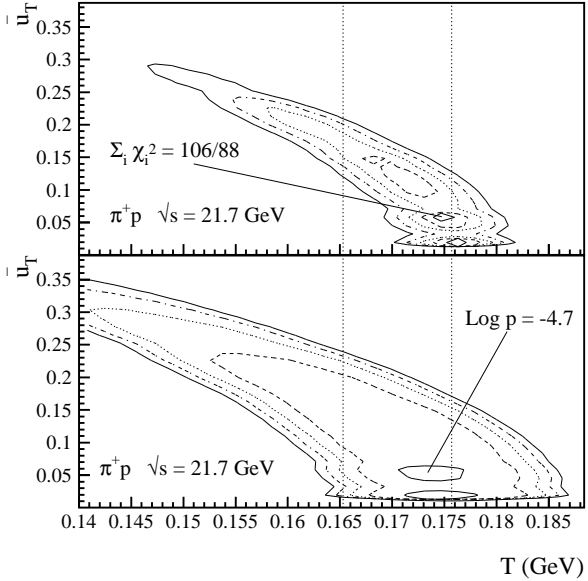}}
\caption{Global $\chi^2$ contour lines (top) and confidence level contours
for the test variable $p$ in Eq.~(\ref{ptest}) (bottom) in the $T-\utm$ 
plane for the fits to transverse momentum spectra in $\pi^+$p collisions 
at $\sqrt s = 21.7$ GeV. The $\chi^2$ contours range from $\chi^2_{\rm min}+1$ 
(1 $\sigma$) to $\chi^2_{\rm min}+25$ (5 $\sigma$'s) whereas the confidence 
level contours range from $10^{-5}$ to $10^{-13}$. Also shown the absolute 
minimum (top) and maximum (bottom) and the vertical 1 $\sigma$ temperature 
band determined by the fit to multiplicities. The local $\chi^2$ minimum 
chosen as starting point of single particle fits is the absolute minimum.}
\label{globgp}
\end{figure}
In order to better define the best-fit values, the search of local minima is 
performed in the $T-\utm$ plane for the sum of all $\chi^2_j$'s at a given 
centre-of-mass energy, thereby assuming the universality of the slope parameters 
$T$ and $\utm$ for different hadron species. Due to the quasi-independence 
between normalization parameter $A$ and $(T, \utm)$, this procedure amounts 
to fit the spectra of all measured hadrons at the same time, by using one $T$, 
one $\utm$ and different normalization parameters $A_1, A_2, \ldots$ for each 
hadron species. A suitable minimum for $\sum_j \chi^2_j$ can be determined by 
intersecting the valley of local minima with the $1\sigma$ band of the 
temperature $T_{ch}$ extracted from the fit to average multiplicities for the 
same colliding system. This is shown in Figs.~\ref{globpp}, \ref{globkp}, \ref{globgp} 
for pp collisions at $\sqrt s= 27.4$ GeV, K$^+$p and $\pi^+$p collisions at 
$\sqrt s= 21.7$ GeV. It should also be emphasized that for K$^+$p and 
$\pi^+$p collision the local minimum located within the $T_{ch}$ band coincides
with the absolute minimum and, moreover, the local minima have apparently moved 
towards higher $T$ with respect to pp collisions, in agreement with the same 
observed trend of $T_{ch}$ (see Table~1); this is a desired indication of a correlation 
between $T$ and $T_{ch}$. In Figs.~\ref{globpp}, \ref{globkp}, \ref{globgp}
are also shown the confidence level contour lines for the test variable $p$
\cite{roe}:

\begin{eqnarray}\label{ptest}
 && p = \prod_j P(\chi^2_j(T,\utm),n_j) \\
 && P(\chi^2_j(T,\utm),n_j) = \int_0^{\chi^2_j(T,\utm)} \!\!\!\!\!\!\!\!\!\!\!\!\! 
 f(\chi^2,n_j)\; d\chi^2 \nonumber 
\end{eqnarray}    
where $f(\chi^2,n_j)$ is the $\chi^2-$distribution for $n_j$ degrees of freedom.
This variable can be used as a suitable test for the hypothesis of consistency,
with one $T$ and one $\utm$ value, among the spectrum slopes of different species,
in that it combines symmetrically the confidence levels of all particles. Conversely, 
the sum of $\chi^2$'s, albeit the most appropriate quantity to estimate the
{\em assumed} common $T$ and $\utm$, cannot be used as a consistency test variable 
because it favours particle spectra with more, and more accurately, measured points. 
The maximum confidence levels for the test variable $p$ (whose location does not
need to be the same as for the global $\chi^2$) in Figs.~\ref{globpp}, 
\ref{globkp}, \ref{globgp} are quite poor. However, it should be reminded 
that the used fit model, for hadron multiplicities as well as for transverse momentum 
spectra, rests on several additional assumptions and approximations, besides the basic 
postulate of local statistical equilibrium, essentially those listed at the beginning
of Sect.~4, which make theoretical formulae expected to be valid only to a certain
degree of accuracy. Moreover, the use of fixed fitted relative multiplicities of 
parent hadrons in transverse momentum spectrum fits involves a further systematic 
uncertainty, which is very difficult to estimate without resorting to a global fit 
of spectra and multiplicities at the same time, which is beyond the scope of the present
work. In view of such considerations, it should not be too surprising that rigorous 
statistical tests do not turn out to be fully satisfactory, especially for very accurately
measured spectra. Nonetheless, it would be desirable to have a more direct and
intuitive feeling whether the examined spectra are in agreement with each other 
within the expected accuracy of the model, whereas the measure of consistency 
by means of a maximum confidence level can be too criptic in this regard.
This and related issues will be discussed in detail in the next section by 
addressing the $m_T$ scaling property.

\begin{table}[tb]\label{tab3}
\caption{Best-fit temperatures and mean transverse four-velocities $\utm$ for each
identified hadron transverse momentum spectrum in hadronic collisions.} 
\begin{tabular}{lllll}
\hline\noalign{\smallskip} 
    Particle  & $T$ (MeV)  & $\utm$  &  $\chi^2/{\rm dof}$ & Reference \\ 
\noalign{\smallskip}\hline
\noalign{\smallskip}
\multicolumn{5}{l}{K$^+$p $\sqrt s$ = 21.7 GeV}  \\                    
\noalign{\smallskip}\hline
\noalign{\smallskip}
  $\pi^0$	   &   181.6       &   0.2101   &  19.2/17   & \cite{azhi87} \\  
  $\pi^-$	   &   175.0       &   0.1485   &  39.3/43   & \cite{adam88} \\
  K$^0_S$	   &   158.0       &   0.1631   &  11.2/10   & \cite{azhi90} \\
  $\rho^0$	   &   174.7       &   0.3155   &  3.55/3    & \cite{agab89} \\
  K$^{*0}$	   &   141.4       &   0.0218   &  1.23/4    & \cite{agab89} \\
  f$_2(1270)$	   &   161.7       &   0.1553   &  1.61/3    & \cite{agab89} \\
\noalign{\smallskip}\hline				       
\noalign{\smallskip}
\multicolumn{5}{l}{$\pi^+$p $\sqrt s$ = 21.7 GeV}  \\  
\noalign{\smallskip}\hline				      
\noalign{\smallskip}					      
  $\pi^0$	   &   174.0	   &   0.3021	&  13.2/17   & \cite{azhi87}\\
  $\pi^-$	   &   171.4	   &   0.1178	&  25.3/43   & \cite{adam88}\\
  K$^0_S$	   &   162.9	   &   0.1251	&  8.00/11   & \cite{azhi90}\\
  $\rho^0$	   &   174.5	   &   0.1252	&  1.05/2    & \cite{agab90}\\
  K$^{*0}$	   &   140.0	   &   0.0018	&  4.31/3    & \cite{agab90}\\
$\bar{\rm K}^{*0}$ &   140.0	   &   0.0031	&  5.36/2    & \cite{agab90}\\
\noalign{\smallskip}\hline				       
\noalign{\smallskip}
\multicolumn{5}{l}{pp $\sqrt s$ = 27.4 GeV}  \\  
\noalign{\smallskip}\hline				      
\noalign{\smallskip}					      
  $\pi^0$	   &   159.6	   &   0.1948	&  22.5/22   & \cite{agui91}\\
  $\pi^+$	   &   161.0	   &   0.2140	&  37.5/38   & \cite{agui91}\\
  $\pi^-$	   &   165.0	   &   0.2179	&  40.2/38   & \cite{agui91}\\
  K$^+$	           &   168.5	   &   0.2324	&  26.9/38   & \cite{agui91}\\
  K$^-$  	   &   158.3	   &   0.1960	&  31.2/38   & \cite{agui91}\\
  $\eta$	   &   174.4	   &   0.2260	&  13.3/10   & \cite{agui91}\\
  $\rho^0$	   &   158.9	   &   0.1971	&  1.29/6    & \cite{agui91}\\
  f$_2(1270)$	   &   156.0	   &   0.2004	&  0.571/1   & \cite{suzu79}\\
\noalign{\smallskip}\hline 
\noalign{\smallskip}     
\end{tabular}
\end{table}    
Once that a local minimum is found through the intersection of the $T_{ch}$ band
with the best-fit valley of the global $\chi^2$, all single spectra are refitted
in turn by instructing the minimisation algorithm \cite{minuit} to start from
that very point with a small initial search step. This allows to determine the
single-particle $\chi^2$'s local minimum closest to the selected initial point.    
The refit results are summarized in Table~3. The fitted spectra are shown
in Fig.~\ref{spectra} along with the calculated cumulative contributions of 
several classes of parent hadrons and residuals distribution. Because of the 
closely spaced local minima in the $\chi^2$ graphs, the minimisation 
algorithm has not been able to produce a reliable error estimate. Indeed, in most 
cases, the found local minima are so shallow that it is not even possible to define 
an error on the basis of a nearby $\chi^2 = \chi^2_{\rm min}+1$ closed contour 
line. Thus, we have decided not to quote any error on the best-fit parameters 
in Table~3.
  
\begin{figure}
 \resizebox{0.48\textwidth}{!}{%
 \includegraphics{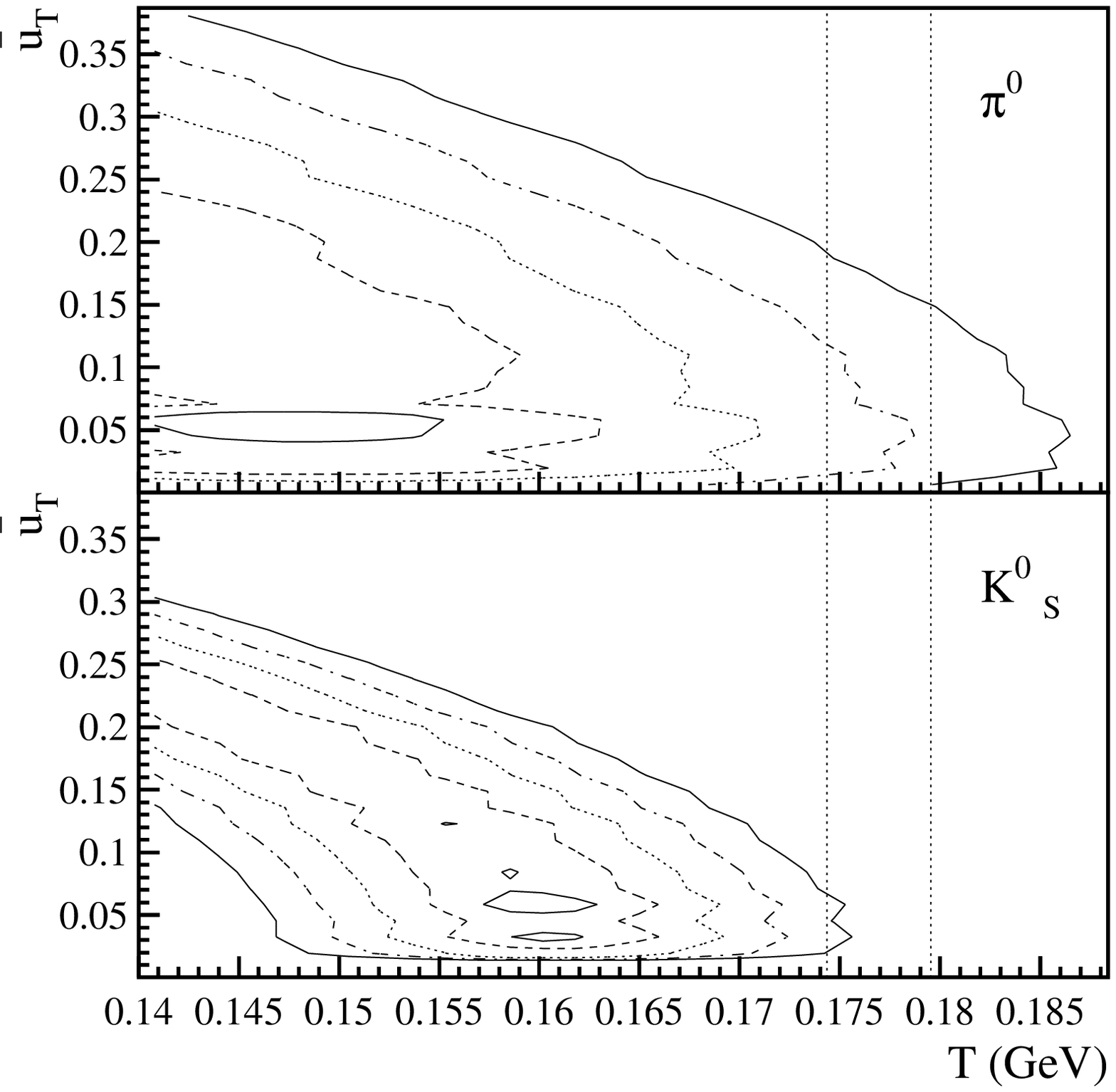}}
\caption{$\chi^2$ contour lines in the $T-\utm$ plane for the fits to 
identified particle transverse momentum spectra in K$^+$p collisions at 
$\sqrt s = 11.5$ GeV. The contours range from $\chi^2_{\rm min}+1$ 
(1 $\sigma$) to $\chi^2_{\rm min}+25$ (5 $\sigma$'s).}
\label{cont70}
\end{figure}

We have also analysed transverse momentum spectra of hadrons in K$^+$p collisions
at $\sqrt s$ = 11.5 GeV; the relevant contour plots of $\pi^0$ \cite{kp70_1} and 
K$^0_S$ \cite{kp70_2} are shown in Fig.~\ref{cont70}. It can be seen that in both 
cases the best-fit valley does not intersect the $1\sigma$ $T_{ch}$ band. Therefore, 
it is not possible to 
find a suitable value of temperature to reproduce both integrated yields and 
transverse momentum spectra in this collision system. This failure of can be 
possibly explained by the inadequacy of the canonical framework at low centre-of-mass 
energy, where microcanonical effects are expected to show up. Particularly, exact 
total transverse momentum 
conservation should entail a steepening of the single inclusive transverse momentum 
spectra (a high $p_T$ suppression analogous to chemical canonical suppression), an 
effect which is in agreement with the observed discrepancy in Fig.~\ref{cont70}. 
However, this explanation is still to be thoroughly tested by carrying out a 
detailed microcanonical calculation.
\begin{figure}
 \resizebox{0.48\textwidth}{!}{%
 \includegraphics{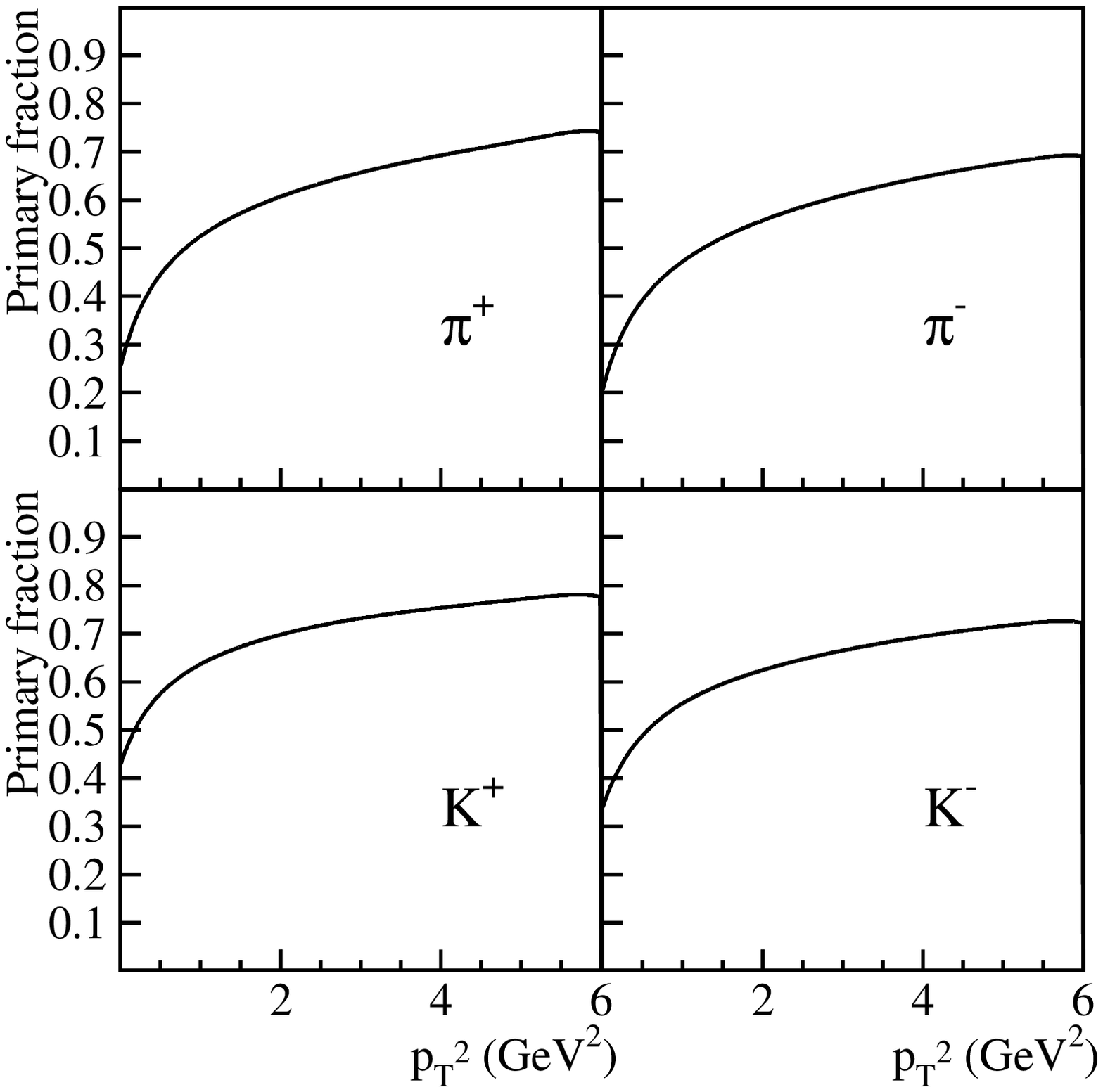}}
\caption{Fraction of primary pions and kaons in pp collisions at $\sqrt s = 27.4$ 
GeV estimated by using the fitted parameters in Table~3. The relative primary 
component increases as a function of $p_T^2$ but does not reach a value higher
than 70-80\% even at $p_T^2$ as high as 6 GeV$^2$.}
\label{prima}
\end{figure}
\begin{table}[!t]\label{tab4}
\begin{tiny}
\caption{Particle multiplicities in \ee collisions.} 
\begin{tabular}{llll}
\hline\noalign{\smallskip} 
  {\scriptsize Particle}&{\scriptsize Measured}&{\scriptsize Fitted}&
  {\scriptsize Reference}\\ 
\noalign{\smallskip}\hline
\noalign{\smallskip}
\multicolumn{4}{l}{\scriptsize $\sqrt s$ = 14 GeV}  \\                    
\noalign{\smallskip}\hline
  $\pi^0$    &  4.69  $\pm$ 0.20     &	  4.640  	 & \cite{jade}   \\
  $\pi^+$    &  3.6   $\pm$ 0.3	     &    3.831  	 & \cite{tass}  \\				      
  K$^+$      &  0.60  $\pm$ 0.07     &    0.5872	 & \cite{tass}   \\
  K$^0_S$    &  0.563 $\pm$ 0.045    &    0.5557	 & \cite{tass,jade} \\
  p	     &  0.21  $\pm$ 0.03     &	  0.1943	 & \cite{tass}   \\
  $\Lambda$  &  0.065 $\pm$ 0.020    &	  0.07972	 & \cite{tass}   \\
\hline				       
\noalign{\smallskip}
\multicolumn{4}{l}{\scriptsize $\sqrt s$ = 22 GeV}  \\  
\noalign{\smallskip}\hline				      
  $\pi^0$    &  5.5    $\pm$ 0.4   &  5.488	     &  \cite{jade}      \\
  $\pi^+$    &  4.4    $\pm$ 0.5   &  4.611	     &  \cite{tass}      \\
  K$^+$      &  0.75   $\pm$ 0.10  &  0.6802	     &  \cite{tass}      \\
  K$^0_S$    &  0.638  $\pm$ 0.057 &  0.6473	     &  \cite{tass,jade}     \\
  p	     &  0.31   $\pm$ 0.03  &  0.3012	     &  \cite{tass}      \\
  $\Lambda$  &  0.110  $\pm$ 0.025 &  0.1227	     &  \cite{tass}      \\   
\hline
\noalign{\smallskip}
\multicolumn{4}{l}{\scriptsize $\sqrt s$ = 29 GeV} \\
\noalign{\smallskip}\hline
 $\pi^0$      &  5.3    $\pm$ 0.7  	 & 6.395 	 & \cite{tpc} \\
 $\pi^+$      &  5.35   $\pm$ 0.25	 & 5.417  	 & \cite{tpc} \\
 K$^+$        &  0.70  $\pm$ 0.05	 & 0.7405 	 & \cite{tpc} \\
 K$^0_S$      &  0.691 $\pm$ 0.029	 & 0.7072 	 & \cite{jade,tpc,mark2,pluto,hrs} \\
 $\eta$       &  0.584  $\pm$ 0.075	 & 0.5636 	 & \cite{mark2,hrs} \\
 $\rho^0$     &  0.90   $\pm$ 0.05	 & 0.7604   	 & \cite{hrs} \\
 K$^{*0}$     &  0.281 $\pm$ 0.022	 & 0.2309  	 & \cite{tpc,mark2,pluto,hrs} \\
 K$^{*+}$     &  0.310 $\pm$ 0.030	 & 0.2338  	 & \cite{tpc} \\
 $\eta'$      &  0.26   $\pm$ 0.10	 & 0.05988  	 & \cite{mark2} \\				   
 $\phi$       &  0.084  $\pm$ 0.022	 & 0.08672  	 & \cite{tpc} \\
 p            &  0.30   $\pm$ 0.05	 & 0.2812  	 & \cite{tpc} \\
 $\Lambda$    &  0.0983 $\pm$ 0.006	 & 0.1023  	 & \cite{tpc,mark2,hrs} \\
 $\Xi^-$      &  0.0083 $\pm$ 0.0020	 & 0.006844  	 & \cite{mark2,hrs}  \\ 	     
 $\Sigma^{*+}$&  0.0083 $\pm$ 0.0024	 & 0.01030  	 & \cite{hrs} \\
 $\Omega$     &  0.0070 $\pm$ 0.0036	 & 0.0004667 	 & \cite{mark2} \\ 
\hline
\noalign{\smallskip}
\multicolumn{4}{l}{\scriptsize $\sqrt s$ = 35 GeV} \\
\noalign{\smallskip}\hline
 $\pi^0$   &  6.31 $\pm$ 0.35      &  6.4786 	     &	\cite{jade35,cello}     \\
 $\pi^+$   &  5.45 $\pm$ 0.25      &  5.494  	     &	\cite{tass35}    \\
 K$^+$     &  0.88 $\pm$  0.10     &  0.7789	     &	\cite{tass35}     \\
 K$^0_S$   &  0.740 $\pm$ 0.017    &  0.7444	     &	\cite{jade,cello,tass}     \\
 $\eta$    &  0.636 $\pm$ 0.080    &  0.5791	     &	\cite{jade35,cello}     \\
 $\rho^0$  &  0.756 $\pm$ 0.077    &  0.7660	     &	\cite{tass,jade35}     \\
 K$^{*+}$  &  0.361 $\pm$ 0.046    &  0.2461	     &	\cite{tass,jade35,cello}     \\
 p	   &  0.303 $\pm$ 0.033    &  0.2900	     &	\cite{jade35,tass35} \\ 
 $\Lambda$ &  0.108 $\pm$ 0.010    &  0.1103	     &	\cite{cello,tass35}     \\
 $\Xi^-$   &  0.0060 $\pm$ 0.0021  &  0.007783 	     &	\cite{tass35}  \\
\hline 
\noalign{\smallskip}
\multicolumn{4}{l}{\scriptsize $\sqrt s$ = 43 GeV} \\
\noalign{\smallskip}\hline
  $\pi^0$   &	6.66 $\pm$ 0.65      & 6.561	   & \cite{jade35,tass35} \\
  $\pi^+$   &	5.55 $\pm$ 0.25      & 5.565	   & \cite{tass35}	 \\
  K$^+$     &   0.97 $\pm$ 0.15      & 0.8089	   & \cite{tass43}	 \\
  K$^0_S$   &	0.760 $\pm$ 0.035    & 0.7735      & \cite{tass}	 \\					      
  K$^{*+}$  &	0.385 $\pm$ 0.094    & 0.2613      & \cite{tass}	 \\
  $\Lambda$ &	0.128 $\pm$ 0.024    & 0.1310      & \cite{tass35}	 \\
\noalign{\smallskip}\hline    
\end{tabular}
\end{tiny}
\end{table}    
 
\subsection{Transverse momentum spectra in \ee collisions} 
 
The very definition of transverse momentum spectrum considerably changes from  
hadronic to \ee collisions. In the latter, the reference projection axis for
$p_T$ is no longer the beam line but a suitably defined event or thrust axis, 
which is the best approximation to the direction of the primarily created 
q$\bar{\rm q}$ pair. 
However, this axis has to be determined experimentally on an event by event basis 
and this is generally done by minimising the sum of tranverse projections of particle 
momenta. Hence, the transverse momentum of a particle with respect to the event 
axis is a biased variable in that the same quantity has been used to determine 
the axis itself. Only when number of particle per event becomes very large, those 
spurious correlations should be negligible.

Being aware of this difficulty, we have fitted charged particle transverse 
momentum spectra in \ee collisions at four different centre-of-
mass energies: 14, 22, 29, and 35 GeV \cite{eept}. Unlike in hadronic collisions, 
transverse spectra have not been measured for several identified particles and this 
compelled us to consider only charged tracks. The fit procedure and the treatment
of the data was essentially the same as for hadronic collisions except for the preliminary 
multiplicity fit which has been performed with the old $\gamma_S$ parametrisation 
\cite{beca2} instead of the aforementioned (see Subsect.~5.1) new method. Moreover, 
the hadronic decay 
chain has been extended to include weak decay products of K$^0_S$ and hyperons in
order to match the experimental definition of final hadrons in \ee experiments. 
The inclusion of heavy flavoured events has been implemented according to the 
procedure described in ref.~\cite{beca2} and the branching ratios 
$R_q = \sigma({\rm e}^+{\rm e^-} \rightarrow q \bar q)/\sigma_{\rm had}$ have been 
taken as the lowest order ones $R_q \propto Q^2_q$. Fit results are quoted in 
Tables~1 and 4.

\begin{figure}
 \resizebox{0.48\textwidth}{!}{%
 \includegraphics{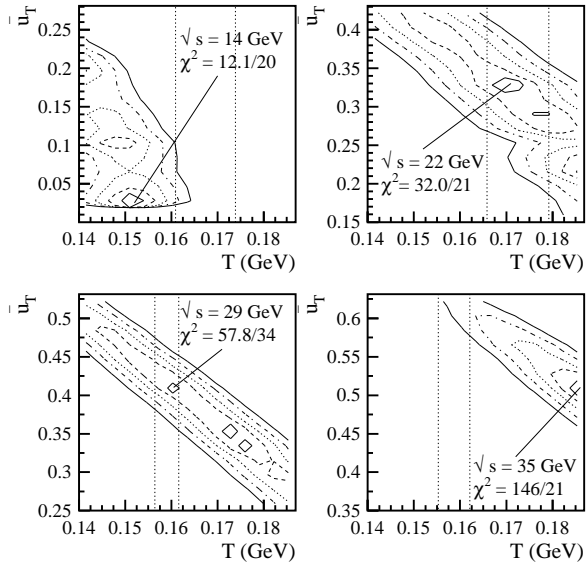}}
\caption{$\chi^2$ contour lines in the $T-\utm$ plane for the fits to 
charged particle transverse momentum spectra in \ee collisions at 
$\sqrt s =14, 22, 29$ and 35 GeV. The contours range from $\chi^2_{\rm min}+1$ 
(1 $\sigma$) to $\chi^2_{\rm min}+25$ (5 $\sigma$'s). Also shown the 
absolute minima and the vertical 1 $\sigma$ temperature band determined by 
the fit to multiplicities.}
\label{contee}      
\end{figure}
The $\chi^2$ contour plots in the $T-\utm$ plane are shown in Fig.~\ref{contee}
along with the relevant $1\sigma$ $T_{ch}$ bands. Whilst there is no intersection
between best-fit valley and $T_{ch}$ band for \ee collisions at $\sqrt s = 14$ GeV, 
as it was found in K$^+$p collisions at $\sqrt s = 11.5$ GeV, fairly consistent 
local minima can be determined up to $\sqrt s = 29$ GeV with sufficiently low
$\utm$ values. However, already at $\sqrt s= 29$ GeV and chiefly at $\sqrt s = 35$
GeV, the fitted $\utm$ is maybe too large for the assumed zero-order approximation, 
described in Sect.~3, to hold. This is confirmed by the increasingly high $\chi^2$ values
demonstrating the inability to fit experimental transverse momentum spectrum to 
the shape (\ref{final}). Otherwise stated, as the centre-of-mass energy increases, 
transverse momentum spectra of hadrons become more and more influenced by the 
shape (which can no longer be parametrised with an average $\utm$ only) of the
transverse four-velocity distribution of hadronising clusters, which is related in
turn to the shape of $p_T$ distribution of gluons radiated in the perturbative 
QCD shower. Conversely, the primordial $T$-dependent hadronisation spectrum 
becomes less and less important; at high energy, its trace is only a small 
sumperimposed noise smearing out original partonic distributions.   

\section{Discussion and $m_T$ scaling}    
 
As has been discussed at the beginning of previous section, two main predictions 
of the statistical hadronisation model have been provided concerning transverse 
momentum spectra: the parameters describing the shape of 
different species should be approximately the same and the extracted temperature 
should be in agreement with that obtained from fits to average multiplicities.
 
As far as the latter statement is concerned, a definite answer cannot be given 
because we have in fact used multiplicity fits to pin down the temperature 
in spectrum fits. Yet, an independent indication of consistency is provided by 
the correlation between the temperature location of the global $\chi^2$ minima 
in K$^+$p and $\pi^+$p collisions at $\sqrt s = 21.7$ GeV and the corresponding 
$1\sigma$ band of $T_{ch}$ (see Figs.~\ref{globkp}, \ref{globgp}). A clearcut 
test would be the analysis of spectra with $\utm \simeq 0$, a situation that,
in view of the present analysis, is foreseen to occur in a small energy window 
below $\sqrt s \approx 20$ GeV. Focussing solely on the most accurately measured 
spectra, namely charged pions, 
it can be seen that $\utm$ appreciably increases from $\sqrt s= 21.7$ GeV to 27.4 
GeV (from 0.12-0.15 to 0.21) whilst $T$ decreases like $T_{ch}$ does, not a 
surprising fact because the $T_{ch}$ band has been actually used to locate a 
suitable best-fit value for $T$. On the other hand, if pion spectra are singled 
out in all three examined collisions, their mimimum $\chi^2$ valleys in 
Figs.~\ref{contpp}, \ref{contkp}, \ref{contgp} nicely overlap and one might 
be led to argue that, in fact, spectra do no show any significant change of 
shape from 21.7 to 27.4 GeV. In the assumed statistical formalism, this lack 
of an observable broadening is due to an accidental compensation between a 
decrease in temperature and an increase in transverse four-velocity. This 
effect would be essentially related to a decrease in the chemical temperature 
$T_{ch}$ from 21.7 to 27.4 GeV and an indication of its physical genuiness 
would be the apparent correlation between the best-fit global $T$ and $T_{ch}$ 
band in Figs.~\ref{globkp}, \ref{globgp}. However, for the present, such 
behaviour of temperatures and transverse four-velocities and the possible 
reason of it cannot be firmly established, as discussed in Subsect. 5.1
with regard to chemical temperature. It might well happen that, resorting to a more 
fundamental microcanonical description (with hadronisation occurring, for instance, 
at constant cluster energy density $M/V =$ const), a monotonic evolution of the 
parameters governing $p_T$ and mass slopes as a function of centre-of-mass 
energy is retrieved while the compensation effect between $T$ and $\utm$ would 
be only the effect of enforcing a canonical approximation. In general,
this topic needs a more detailed study. 

Turning back to the first issue, whether statistical parameters describing 
spectra are sufficiently universal for different hadron species, we have seen in 
Sect.~5 that rigorous statistical tests would rule out this possibility,
though a fully affirmative answer would have indeed been unlikely because of the 
various approximations inherent to the theoretical formulae. However, the 
implementation of a single statistical test does not tell us the main sources
of the discrepancy between model and data and a further investigation is necessary.
As far as pp collisions are concerned, it can be realised by looking at Table~3 
and the location of $\chi^2$ minimum valleys in Fig.~\ref{contpp}, \ref{confpp}, 
that the different fitted temperatures and tranverse four-velocities 
are indeed in good agreement with each other with the exception of K$^+$ whose
spectrum is seemingly harder than expected and significantly different from
K$^-$'s. The $T$ and $\utm$ central values extracted for $\eta$, $\rho^0$ and 
$f_2$ are affected by large uncertainties ($\chi^2$ minimum valleys are much
wider, see Fig.~\ref{contpp}) so they are essentially in agreement with 
the more accurate ones. The situation is slightly worse in the other 
two examined hadronic collisions, K$^+$p and $\pi^+$p at $\sqrt s = 21.7$ GeV. 
By inspecting Table~3 and Figs.~\ref{contkp}, \ref{contgp}, several discrepancies
can be noticed: $\pi^0$'s spectrum is harder than $\pi^-$'s, K$^0_S$'s is softer 
and K$^*$'s much too soft; these features are common to both collisions.
\begin{figure}
 \resizebox{0.48\textwidth}{!}{%
 \includegraphics{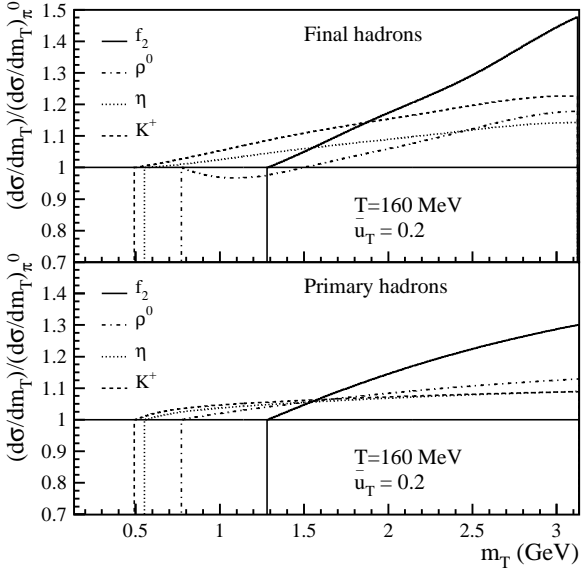}}
\caption{Ratio between $m_T$ spectra of quoted particles and $\pi^0$ spectrum 
taken as reference for primary (bottom) and final observable hadrons (top) calculated
for $T=160$ MeV and $\utm=0.2$ in pp collisions. The spectra have been normalized 
so as to have the same value at the threshold $m_T=m$.}
\label{mtscal}
\end{figure}
\begin{figure}
 \resizebox{0.48\textwidth}{!}{%
 \includegraphics{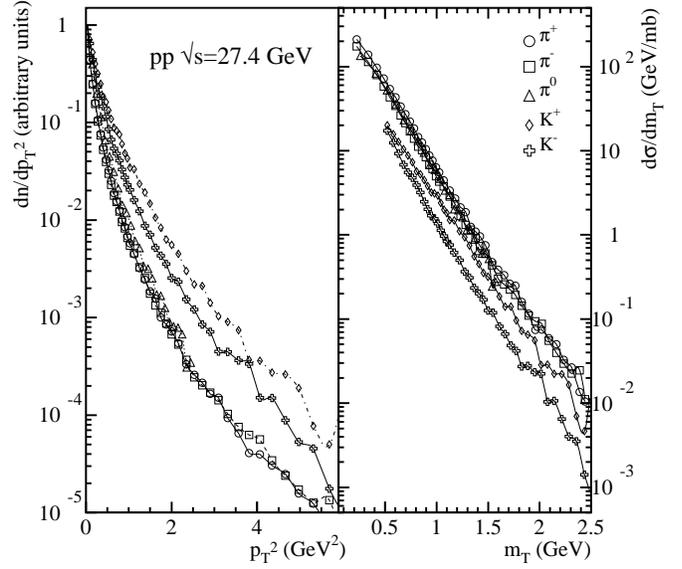}}
\caption{Comparison between experimentally measured $p_T^2$ and corresponding 
$m_T$ spectra of pions and kaons in pp collisions at $\sqrt s = 27.4$ GeV. 
The $p_T^2$ spectra have been normalized so as to have the same value in the 
first bin. Lines connecting data points are drawn to guide the eye.}
\label{mtslopepp}
\end{figure}

In order to assess the universality as a function of different hadron species,
a very useful property is the so-called $m_T$-scaling, a well known feature 
of statistical and thermal models \cite{hage,hage4}: if $\utm=0$, the shape of the 
{\em primary} spectrum of any hadron (except for charged pions at very low $p_T$, 
owing to quantum statistics corrections) depends on its mass only through the 
variable $m_T=\sqrt{p_T^2+m^2}$, as shown in Eq.~(\ref{final}).
The $m_T$-scaling is broken by a non-vanishing transverse velocity and by the
hadronic decay chain but both are small effects for the $\utm$ values found in
the present analysis, as it can be seen in Fig.~\ref{mtscal}: the scaling
violation is limited to 25\% at the top $m_T$ value ($\simeq 2.4$ GeV) in
the analysed pp collisions.  
The general trend of the scaling violation can be roughly summarized with a 
relative softening of $m_T$ spectra of lighter particles with respect to 
heavier particles.

\begin{figure}
 \resizebox{0.48\textwidth}{!}{%
 \includegraphics{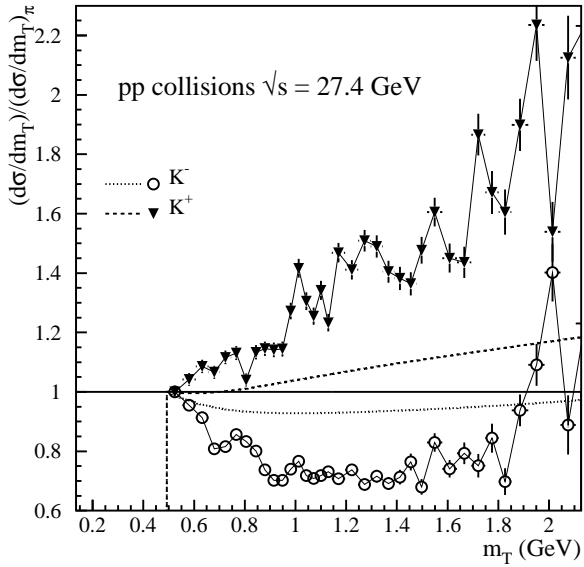}}
\caption{Ratio between $m_T$ spectra of kaons and $\pi^+$ spectrum 
taken as reference. The experimental ratios (dots with error bars) are calculated
by using the measured kaon spectra and the fitted $\pi^+$ spectrum (parameters quoted 
in Table~3), which is a very good approximation to the measured one. The theoretical 
ratios are calculated by taking common values for the relevant parameters, namely 
$T=160$ MeV and $\utm=0.2$, which are a fair average of values quoted in Table~3.
The spectra have been normalized so as to have the same value at the threshold 
$m_T=m_{\rm K}$. Lines connecting data points are drawn to guide the eye.}
\label{mtratio}
\end{figure}
The approximate $m_T$ scaling apparently holds in pp collisions, for the most 
accurately measured spectra of pions and kaons. That this is a non trivial fact,
it is proved in Fig.~\ref{mtslopepp}, where the actually measured and published 
\cite{agui91} $p_T^2$ spectra are compared with the corresponding spectra 
transformed in the variable $m_T$. While $p_T^2$ slopes are markedly different 
between pions and kaons, their $m_T$ slopes are much closer (heavier particle 
spectra are affected by too large errors to allow any conclusion). 
To a closer look, also the aforementioned discrepancy between K$^+$ 
and K$^-$ and the hardness of K$^+$ spectrum show up; this is better seen 
in Fig.~\ref{mtratio} where predicted and measured ratios between $m_T$ spectra
of K$^+$ and K$^-$ with respect to $\pi^+$'s spectrum are shown. Although 
the model succeeds in predicting a slight difference between the slopes of 
K$^+$ and K$^-$ (in the right direction) owing to different resonance feeding 
(the colliding system is not isospin symmetric), the observed
ratios with $\pi^+$ spectrum are steeper in shape, particularly for K$^+$, 
and the difference between K$^+$ and K$^-$ is larger. It is very difficult
to understand the reason of this discrepancy; at present, it can attributed
to deviations from the model scheme (i.e. from the assumptions listed at the 
beginning of Sect.~4) or to a systematic error in the estimation of relative 
abundances of parent hadrons.

\begin{figure}
 \resizebox{0.48\textwidth}{!}{%
 \includegraphics{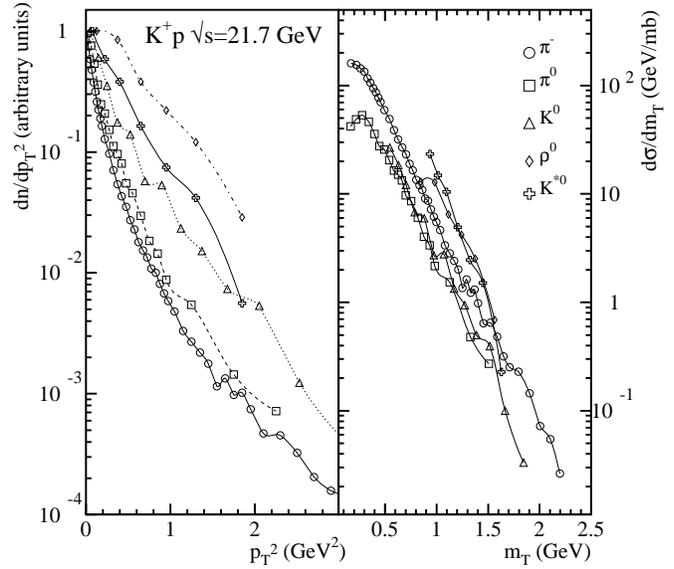}}
\caption{Comparison between experimentally measured $p_T^2$ and 
corresponding $m_T$ spectra of identified particles in K$^+$p collisions at 
$\sqrt s = 21.7$ GeV. The $p_T^2$ spectra have been normalized so as to have 
the same value in the first bin. Lines connecting data points are drawn to 
guide the eye.}
\label{mtslopekp}
\end{figure}
\begin{figure}
 \resizebox{0.48\textwidth}{!}{%
 \includegraphics{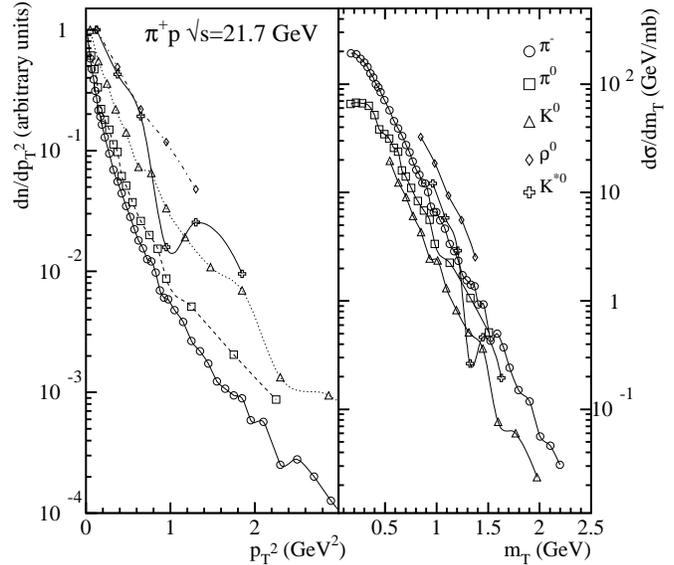}}
\caption{Comparison between experimentally measured $p_T^2$ and 
corresponding $m_T$ spectra of identified particles in $\pi^+$p collisions at 
$\sqrt s = 21.7$ GeV. The $p_T^2$ spectra have been normalized so as to have 
the same value in the first bin. Lines connecting data points are drawn to 
guide the eye.}
\label{mtslopegp}
\end{figure}
A similar behaviour is observed in K$^+$p and $\pi^+$p collisions: as shown
in Figs.~\ref{mtslopekp}, \ref{mtslopegp}, the slopes of $m_T$ spectra are 
definitely much closer than those of $p_T^2$ spectra. However, in agreement
with the previously observed contour plots in Figs.~\ref{contkp}, \ref{contgp}
and to the values quoted in Table~3, $\pi^0$ spectrum has a remarkably different 
shape at low $m_T$ with respect to $\pi^-$'s and K$^*$'s spectra are definitely
steeper than expected. The softness of K$^*$'s spectrum could be related
to the presence of a valence strange quark in the initial state for K$^+$p 
collisions; yet it is there in $\pi^+$p collisions too. Unfortunately, the 
spectra in both K$^+$p and $\pi^+$p collisions have been measured by the same 
experiment, NA22, and this prevents a cross-check on the data to assess the 
genuiness of such similarities. Furthermore, there is little overlap between 
the particle sample in pp and K$^+$p, $\pi^+$p with the exception of $\pi^0$ 
and $\pi^-$. In this regard, whilst in pp 
collisions very good agreement is found between the measured slopes of the 
three pion species, a considerable discrepancy is observed in in both K$^+$p and 
$\pi^+$p collisions with different initial colliding particles. Indeed, 
this apparent violation of a nice scaling observed at a centre-of-mass energy 
only 6 GeV higher is quite odd.    
  
\section{Summary and conclusions}    

Transverse momentum spectra of identified hadrons in several high energy 
collisions have been compared to the predictions of statistical hadronisation 
model, which has been described in detail starting from a microcanonical 
formulation, with emphasis on the assumptions needed to introduce the usual
concepts of single-cluster temperature and volume. The distortion effect due 
to secondary hadronic decays has been accurately and exhaustively taken into 
account by a newly proposed numerical method. A good agreement is found between 
model and data and, on top of that, the temperatures estimated by fits to average 
particle multiplicities and those extracted by fitting transverse momentum 
spectra are fully compatible with each other. 
This is an indication in favour of one of the key predictions of the 
statistical hadronisation model, namely the existence of a close relationship 
between the production of particles as a function of their mass and, for each 
particle species, the production as a function of momentum (measured in the 
rest frame of the cluster they belong to) at the hadronisation.
 
The model calculations have been performed by taking several assumptions.
Firstly, the canonical framework has been used, which is expected to fail at low 
centre-of-mass energies where the effect of exact transverse momentum conservation 
must be significant and, in fact, in \ee collisions at a centre-of-mass energy of 14 
GeV and K$^+$p collisions at 11.5 GeV, a clear discrepancy between data and
calculations shows up. Secondly, the approximation of very small average transverse 
velocity of the clusters has been introduced, which should break down at high 
centre-of-mass energy because of the increasingly large $p_T$ parton emission (giving 
rise to clusters with large tranverse velocity). At very high energy, this effect 
is predominant in determining the shape of transverse momentum spectra whereas 
hadronisation plays the role of a small superimposed smearing noise. Indeed, it 
is found that even at moderately high energy \ee collisions, at $\sqrt s =$ 35 GeV, 
the approximate formulae are no longer able to reproduce the shape of experimental 
charged particle spectrum and the resulting transverse four-velocity is not consistent 
with the initial requirement. In summary, the presently used parametrisation of 
$p_T$ spectra is found to work well only in a limited centre-of-mass energy range 
(roughly between 20 and 30 GeV). 

In order to further investigate statistical features of hadronisation, the study 
of $p_T$ spectrum at lower centre-of-mass energy would be greatly interesting but 
full complex microcanonical calculations are required.

\section*{Acknowledgements}    
    
We are grateful to U. Heinz and M. Gazdizcki for useful discussions.\\
Certainly, this work would not have been possible without the great 
Durham reaction data database \cite{durham}. We are glad to express our 
gratitude to their editors, compilers and curators for a superb piece 
of work.        
    
\section*{Appendix}    
\appendix
\renewcommand{\theequation}{\thesection.\arabic{equation}}

\setcounter{equation}{0}
\section{Proof of the reduction to the EGC}    

We want to prove by direct transformation that :

\begin{equation}\label{iniz}
 \Omega = \Big[ \prod_{i=1}^N \sum_{\Qi} \int \d^4 P_i \; \theta(P^0_i) 
 \, \Oi \Big] \delta^4(P - \Sigma_i P_i) \, \delta_{\Qz,\Sigma_i\Qi}
\end{equation}
is the density of states of an EGC with four-momentum $P = \sum_i P_i$, 
quantum numbers $\Qz = \sum_i \Qi$ and volume (in the reference frame where 
four-momentum is $P$) $V = \sum_i V_i$. Let us consider the integral representation 
of Dirac's and Kronecker's $\delta$:
 
\begin{eqnarray}\label{intrep}
&& \!\!\!\!\!\! \delta^4(P - \Sigma_i P_i) = \frac{1}{(2\pi)^4} 
 \int \d^4 x \; \e^{\i \,(P -\Sigma_i P_i)\cdot x} \nonumber \\
&& \!\!\!\!\!\! \delta_{\Qz,\Sigma_i\Qi} = \lim_{\etavs \to 0} \frac{1}{(2\pi)^n}  
 \int_{-\piv}^{\piv} \!\!\! \d^n \phi \; \e^{\i (\Qz -\Sigma_i\Qi) \cdot 
 \phivs -\Sigma_i \etavs_i \cdot \Qi} \nonumber \\ 
&& 
\end{eqnarray}
where $\etav=(\eta_1,\ldots,\eta_n)$ is a set of real positive numbers which
are introduced in order to regularize infinite sums for quantum numbers running
from 0 to $+\infty$, such as the absolute number of strange quarks. By
using Eq.~(\ref{intrep}) we can rewrite Eq.~(\ref{iniz}) as:

\begin{eqnarray}\label{extend}
&& \!\!\!\!\!\!\!\!\!\!
\Omega = \frac{1}{(2\pi)^{4+n}} \int \d^4 x \int_{-\piv}^{\piv} \!\!\! 
\d^n \phi \; \e^{\i P \cdot x + \i \Qz \cdot \phivs} \nonumber \\	 
&& \!\!\!\!\!\!\!\!\!\! \Big[ \prod_{i=1}^N \lim_{\etavs \to 0} 
\sum_{\Qi} \int \d^4 P_i \; \e^{-\i P_i\cdot x - 
\i \Qi \cdot \phivs -\etavs_i \cdot \Qi} \, \theta(P^0_i) \, \Oi \Big]
\end{eqnarray} 
Let us focus on the single $i^{\rm th}$ factor in the above product. 
If one plugs in it the expression (\ref{omega3}) of $\Oi$ \footnote{the 
integration over the line $(-\infty-\i\varepsilon,+\infty-\i\varepsilon)$,
with $\varepsilon \to 0$ in the complex plane for all $x^0_i$'s, as well as 
for $x^0$, is implied.} and the $\theta$ function's integral representation:

\begin{equation}\label{theta} 
\theta(P^0_i) = \frac{1}{2\pi\i} \int_{-\infty-\i \varepsilon}^{+\infty-\i 
\varepsilon} \!\!\!\!\!\!\!\!\!\! \d \alpha \;\frac{\e^{\i \alpha P^0_i}}{\alpha}   
\end{equation}
the following expression is obtained:

\begin{eqnarray}\label{ifactor}
&& \lim_{\etavs \to 0} \frac{1}{2\pi\i} \sum_{\Qi} \int \d^4 P_i  
\int_{-\infty-\i \varepsilon}^{+\infty-\i \varepsilon} \!\!\!\!\!\!\!\!\!\! 
\d \alpha \int \d^4 x_i \int_{-\piv}^{\piv} \!\!\! \d^n \phi_i \; \nonumber \\
&& \frac{\e^{\i \alpha P^0_i
-\i P_i \cdot (x - x_i) -\i \Qi \cdot (\phivs -\phivs_i) -\etavs_i \cdot \Qi}}
{\alpha} \, \e^{F (x_i,\phivs_i)} 
\end{eqnarray}
where the limit $\etav \rightarrow 0$ is implied. The integration in $P_i$ in 
Eq.~(\ref{ifactor}) formally (that is, in a distributional sense) yields $(2\pi)^4 
\delta(x^0_i-x^0+\alpha)\delta^3 ({\bf x}_i-{\bf x})$. On the other hand, the 
sum over the quantum number vectors $\Qi$ yields, for its generic $k^{\rm th}$ 
component:

\begin{eqnarray}\label{sumq}
&& \!\!\!\!\!\!\!\!\!\!\!\! 
2\pi \delta(\phi_k-\phi_{ik}) \qquad {\rm if} \;\; Q_{ik} = -\infty,\ldots,+\infty
\nonumber \\
&& \!\!\!\!\!\!\!\!\!\!\!\!
\lim_{\eta_k \to 0} \frac{1}{1-\exp[\i(\phi_{ik}-\phi_k)-\eta_k]} 
\qquad {\rm if} \;\; Q_{ik} = 0,\ldots,+\infty \nonumber \\
&& 
\end{eqnarray}
In fact, the second expression in the above equation is also equivalent 
to a $\delta$ distribution. This can be shown through a transformation
of the integration in $\phi_i$ over the interval $(-\pi,\pi)$ onto the 
unitary circle in the complex plane by setting $w=\exp[-\i\phi_{ik}]$:

\begin{eqnarray}\label{cmplint}
 &&\lim_{\eta_k \to 0} \int_{-\pi}^{\pi} \!\!\! \d \phi_{ik} \;
 \frac{1}{1-\exp[\i(\phi_{ik}-\phi_k)-\eta_k]} \, \e^{F (x_i,\phivs_i)} 
 \nonumber \\
 && = \lim_{\eta_k \to 0} \frac{1}{\i} \oint \d w \;
 \frac{\exp[F_w (x_i,\phi_1,\ldots,w,\ldots,\phi_n)]}{w-\e^{-\i\phi_k-\eta_k}} 
\end{eqnarray}
The function $F_w$ (see Eq.~(\ref{f2})) is analytic with 
respect to the complex variable $w$ as only positive integer powers of 
$w$ are involved (the $q_{kj}$'s are positive integer numbers by 
assumption) and the pole $\e^{-\i\phi_k-\eta_k}$ lies within the unitary 
circle. Thus, the result of the integral in Eq.~(\ref{cmplint}) is simply:

\begin{eqnarray}
 && \lim_{\eta_k \to 0} 
 2\pi \exp[F_w(x_i,\phi_{i1},\ldots,\e^{-\i\phi_k-\eta_k},\ldots,\phi_{in})]
 \nonumber \\
 = && 2\pi \exp[F(x_i,\phi_{i1},\ldots,\phi_k,\ldots,\phi_{in})]
\end{eqnarray}
which is the same result that would have been obtained by integrating 
with a $\delta$ distribution. Of course, the same procedure may be 
repeated for all components other than $k$. Eventually, the integrations in 
$x_i$ and $\phiv_i$ in (\ref{ifactor}) yield:

\begin{equation}\label{ifactor2}
\lim_{\varepsilon \to 0} \frac{1}{2\pi\i} \int_{-\infty-\i \varepsilon}^
{+\infty-\i \varepsilon} \!\!\!\!\!\!\!\!\!\! \d \alpha \; 
\frac{\exp[F(x^0-\alpha,{\bf x},\phiv)]}{\alpha}
\end{equation}
This integral can be calculated by closing the $z=-\i\varepsilon$ line 
with a semicircle lying in the upper half of the complex plane and taking the limit
for the circle radius going to infinity. Since the function $F$ is analytic 
with respect to $\alpha$ (see Eq.~(\ref{f2})), the result is simply:

\begin{equation}
 \exp[F(x^0,{\bf x},\phiv)] = \exp[F(x,\phiv)]
\end{equation} 
which is the final form of expression (\ref{ifactor}). Now the function $F$ 
depends on the particular $i^{\rm th}$ cluster only through the volume $V_i$ 
with a simple linear relationship, namely $F(x,\phiv) \equiv V_i f(x,\phiv)$ 
(see Eq.~(\ref{f2})). Therefore, the product of all the factors (\ref{ifactor}) 
yields: 

\begin{equation}
 \exp[\sum_i V_i f(x,\phiv)]
\end{equation}  
and this is the final form of the expression between square brackets in 
Eq.~(\ref{extend}). Hence, the $\Omega$ function reads:

\begin{equation}\label{omegafin}
\Omega = \frac{1}{(2\pi)^{4+n}} \int \d^4 x \int_{-\piv}^{\piv} \!\!\! 
\d^n \phi \; \e^{\i P \cdot x + \i \Qz \cdot \phivs} \, 
\e^{(\Sigma_i V_i) f(x,\phivs)}
\end{equation} 
QED.
  
\setcounter{equation}{0}
\section{Decomposition of Lorentz transformations}    

A general Lorentz transformation ${\sf \Lambda}$ can be uniquely decomposed 
as the product of a transformation like in Eq.~(\ref{boost}) and a general 
rotation \cite{group}:

\begin{equation}\label{deco}
 {\sf \Lambda} = \LL {\sf R} \qquad \LL = {\sf \hat R}_3(\varphi)\,{\sf \hat R}_2(\theta) 
   \,{\sf \hat L}_3 (\xi)
\end{equation} 
where $\varphi \in [0,2\pi)$, $\theta \in [0,\pi]$ and $\xi \ge 0$ (see 
Eq.~(\ref{boost})).\\ 
Any timelike vector can be uniquely written as $\LL(\hat t)$ where $\hat t=(1,0,0,0)$ 
is the time axis unit vector and $\LL$ like in Eqs.~(\ref{boost},\ref{deco}) 
\cite{group}. 
Hence, given two timelike vectors $P' = \LL'(\hat t)$ and $P = \LL(\hat t)$ there 
is only one Lorentz transformation of the type~(\ref{boost},\ref{deco}) 
transforming $P'$ into $P$, namely $\LL''$ such that $\LL''(\hat t) = 
\LL \LL'^{-1} (\hat t)$.
 
In order to prove the alternative decomposition of $\LL$ in Eq.~(\ref{boost2}), 
it is sufficient to show that the equation:

\begin{equation}\label{biuniv}
 {\sf \hat R}_3(\varphi)\,{\sf \hat R}_2(\theta) \, {\sf \hat L}_3 (\xi)(\hat t)=
 {\sf \hat L}_3(\eta)\,{\sf \hat R}_3(\phi)\,{\sf \hat L}_1(\zeta)(\hat t)
\end{equation} 
with $\phi \in [0,2\pi)$, $\eta \in (-\infty,+\infty)$ and $\zeta \ge 0$,
establishes a biunivocal correspondance between the two sets of coordinates. 
Eq.~(\ref{biuniv}) leads to the system of equations:

\begin{eqnarray}\label{coordchange}
  \cosh \zeta \cosh \eta & = & \cosh \xi \nonumber \\
  \sinh \zeta \cos \phi & = & \sinh \xi \sin \theta \cos \varphi \nonumber \\
  \sinh \zeta \sin \phi & = & \sinh \xi \sin \theta \sin \varphi \nonumber \\
  \cosh \zeta \sinh \eta & = & \sinh \xi \cos \theta
\end{eqnarray} 
whose solution is: 

\begin{eqnarray}\label{direct}
  \cosh \xi   & = & \cosh \zeta \cosh \eta \nonumber \\
  \tan \theta & = & \frac{\tanh \zeta}{\sinh \eta} \nonumber \\ 
  \varphi & = & \phi
\end{eqnarray} 
and its inverse:

\begin{eqnarray}\label{inverse}
  \cosh \eta  & = & \frac{\cosh \xi}{\sqrt{1+\sinh^2 \xi \sin^2 \theta}} \nonumber \\
  \sinh \zeta & = & \sinh \xi \sin \theta \nonumber \\ 
  \phi & = & \varphi
\end{eqnarray}
Therefore, if Eq.~(\ref{boost}) is a valid decomposition, Eq.~(\ref{boost2})
holds as well, within the relevant variable domains.

The associated invariant measure in the variables $\zeta$, $\eta$ and $\phi$
can be obtained by calculating the jacobian of the transformation (\ref{direct})
and taking into account that $\d {\sf L} = \sinh^2 \xi \, \sin \theta \, \d \xi 
\, \d \theta \, \d \varphi$ \cite{group2}. It is advantegeous to rewrite the above measure as 
$\d {\sf L} = \sinh \xi \, \d \cosh \xi \, \d \cos \theta \, \d \varphi$ and use 
the following transformation formulae derived by Eqs. (\ref{direct}) and 
(\ref{coordchange}):

\begin{eqnarray*}
\cosh \xi   & = & \cosh \zeta \cosh \eta \\
\cos \theta & = & \frac{\cosh \zeta \sinh \eta}{\sqrt{\cosh^2\zeta\cosh^2\eta-1}}
\end{eqnarray*}
implying:

\begin{equation} 
 |\det J| = \frac{\cosh \zeta \sinh \zeta}{\sqrt{\cosh^2\zeta\cosh^2\eta-1}}
 \nonumber
\end{equation}
and, consequently: 

\begin{equation}
  \d {\sf L} = \frac{\sinh 2 \zeta}{2} \, \d \zeta \, \d \eta \, \d \phi
\end{equation}
  
\setcounter{equation}{0}
\section{Calculation of the secondary contribution to transverse momentum
spectrum}    

The integration of Eq.~(\ref{spectransf2}) is performed by solving the $\delta$ 
function with respect to the $\varphi$ azimuthal angle according to the well known
formula:

\begin{equation}
 \delta(f(\varphi)) = \sum_i \frac{1}{|f'(\varphi_{0i})|} 
 \delta(\varphi-\varphi_{0i}) 
\end{equation}
where $\varphi_{0i}$ are the zeroes of the $f$ function. In the present case 
(see Eq.~(\ref{epsstar})):

\begin{equation}
  f(\varphi)= x - \gmt \epsilon - \utm p_T \cos \varphi
\end{equation}
thus:

\begin{eqnarray}\label{cosine}
 && \cos \varphi_0= \frac{\gmt \epsilon - x}{\utm p_T}  \\
 && |f'(\varphi_0)| = \utm p_T \sqrt{1-\cos^2\varphi_0} \nonumber   
\end{eqnarray}
There are two angles fulfilling Eq.~(\ref{cosine}) in the $[0,2\pi)$ interval,
provided that the absolute value of its right hand side is $< 1$. For 
these angles, the absolute value of the derivatives in the lower equation 
are equal. Hence, the integration in Eq.~(\ref{spectransf2}) in the azimuthal 
angle yields 2 if $|\gmt \epsilon - x| < \utm p_T$ and 0 otherwise. 
Therefore, Eq.~(\ref{spectransf2}) turns into:

\begin{eqnarray}\label{spectransf3}
\!\!\! && \Big\langle\frac{\d n}{\d p_T}\Big\rangle^{k\rightarrow j} \!\!\!\! = 2 
 \int_{m_j}^{+\infty} \!\!\!\! \d x  \int_{-\infty}^{+\infty} \!\!\!\! \d p_z\; 
 \frac{p_T}{4 \pi \epsilon \sqrt{x^2-m_j^2}} \,\Big\langle\frac{\d n}{\d \epsilon^*}
 \Big\rangle^{k\rightarrow j}\!\!\!\!\!\!(x) \, \nonumber \\
\!\!\! && \times \frac{\theta(\utm p_T - |\gmt \epsilon - x|)}
 {\sqrt{\utms p_T^2 - (\gmt \epsilon - x)^2}}
\end{eqnarray}
The next integration can be done with the change of variable $p_z = m_T \sinh y$, 
i.e. by introducing rapidity, so that $\epsilon = m_T \cosh y$ and 
Eq.~(\ref{spectransf3}) reads:

\begin{eqnarray}\label{spectransf4}
 && \Big\langle\frac{\d n}{\d p_T}\Big\rangle^{k\rightarrow j} \!\!\!\! = 4 
 \int_{m_j}^{+\infty} \!\!\!\! \d x \int_{0}^{+\infty} \!\!\!\! \d y \; 
 \frac{p_T}{4 \pi \sqrt{x^2-m_j^2}} \,\Big\langle\frac{\d n}{\d \epsilon^*} 
 \Big\rangle^{k\rightarrow j}\!\!\!\!\!\!(x) \, \nonumber \\
 && \times \frac{\theta(\utm p_T - |\gmt m_T \cosh y - x|)}
 {\sqrt{\utms p_T^2 - (\gmt m_T \cosh y - x)^2}}
\end{eqnarray}
where advantage has been taken of the fact that the integrand is even 
in rapidity. The last needed change of variable is $z= \cosh y$, leading to:

\begin{eqnarray}\label{spectransf5}
 && \Big\langle\frac{\d n}{\d p_T}\Big\rangle^{k\rightarrow j} \!\!\!\! = 
 4 \int_{m_j}^{+\infty} \!\!\!\! \d x \int_{1}^{+\infty} \!\!\!\! \d z \; 
 \frac{p_T}{4 \pi \sqrt{x^2-m_j^2}} \,\Big\langle\frac{\d n}{\d \epsilon^*} 
 \Big\rangle^{k\rightarrow j}\!\!\!\!\!\!(x) \, \nonumber \\
 && \times \frac{\theta(z-z_-)\theta(z_+-z)}
 {\gmt m_T \sqrt{(z-1)(z+1)(z-z_-)(z_+-z)}}
\end{eqnarray}    
where $z_+$ and $z_-$ are defined in Eq.~(\ref{zetas}) and, obviously, $\gmt=
\sqrt{1+\utms}$. The integral in the $z$ variable is an elliptic integral 
whose solutions depend on the sign of $z_- - 1$; they can be found in 
ref.~\cite{grads}:

\begin{eqnarray}
&& \int_{1}^{+\infty} \!\!\!\! \d z \; \frac{\theta(z-z_-)\theta(z_+-z)}
{\sqrt{(z-1)(z+1)(z-z_-)(z_+-z)}} = \nonumber \\
&& = \frac{2}{\sqrt{(z_+ - z_{\rm min})(z_{\rm max}+1)}} \; 
{\rm F}\left(\frac{\pi}{2},r \right)
\end{eqnarray}
where $z_{\rm min}$, $z_{\rm max}$ and $r$ are quoted in Eq.~(\ref{zetas})
and F is the complete elliptic integral of the first kind.
After the substitution of $x$ with momentum ${\rm p}^* = \sqrt{x^2-m_j^2}$ as 
integration variable in Eq.~(\ref{spectransf5}), the final result is exactly 
as in Eq.~(\ref{second2}).    

\setcounter{equation}{0}
\section{Calculation of the partition function with fixed number of strange
quarks}    

In this section we describe the technique used to handle the numerical
integration of canonical partition functions with four fixed quantum numbers,
namely electric charge $Q$, baryon number $B$, strangeness $S$ and number of
${\rm s}+\bar{\rm s}$ quarks $N_S$. This can be written as in Eq.~(\ref{partfunc}) 
with $n=4$ and $\lambda_j=1$:

\begin{equation}\label{zcan2}
 Z(\Qz) = \frac{1}{(2 \pi)^4} \int_{-\piv}^{\piv} \!\!\! \d^4 \phi \;
 \exp \,[\i \Qz \cdot \phiv + F_{\rm c}(\phiv)] 
\end{equation}  
with $\Qz = (Q,B,S,N_S)$ and:

\begin{equation}
 F_{\rm c}(\phiv) = \frac{\oV}{(2\pi)^3} \! \sum_j (2J_j + 1) \! \int \! \d^3{\rm p} \; 
\log \, [1\pm \e^{-\varepsilon_j/T - \i \qj\cdot \phivs}]^{\pm 1} 
\end{equation}
Straight numerical four-dimensional integration is too time consuming to
allow reasonably quick multiplicity fits: an analytical reduction of the above
integral is indeed necessary. For this purpose, we first integrate out the 
$\phi_B$ variable associated with baryon number. This can be done analitically 
because of the presence of only single-valued baryons in the function $F_{\rm c}$ in 
Eq.~(\ref{zcan2}), neglecting the corrections due to Fermi statistics which 
are very small. Thus:

\begin{equation}\label{fcan}
 F_{\rm c}(\phiv) = F_{\rm c}(\phi_Q,\phi_S,\phi_{N_S})_{\rm mesons} + 
 W_+ \e^{-\i\phi_B} + W_-\e^{\i\phi_B}
\end{equation}
where:

\begin{eqnarray}\label{wcap}
\!\!\!\!\!\!\!\!\!\!\!\!&& W_{\pm} = \sum_{\rm bar. \atop antibar.} z_j \exp \,[-\i(Q_j 
  \phi_Q + S_j \phi_S + N_{Sj} \phi_{N_S})]  \\ 
\!\!\!\!\!\!\!\!\!\!\!\!&& z_j \equiv \frac{\oV(2J_j + 1)}{(2\pi)^3} \int \d^3{\rm p} \; 
  \exp \, [-\varepsilon_j/T]    
\end{eqnarray}
It must be stressed that $W_+ \ne W^*_-$ because the number of strange
quarks is the same for both particles and antiparticles. The integration 
in $\phi_B$ in Eq.~(\ref{zcan2}) can be performed by using the decomposition
(\ref{fcan}) and a series is obtained: 

\begin{equation}\label{barint}
\frac{1}{2 \pi} \!\! \int_{-\pi}^{\pi} \!\!\! \d \phi_B \; \e^{\i B \phi_B}
 \exp \,[W_+ \e^{-\i\phi_B} + W_-\e^{\i\phi_B}] \!
 = \!\! \sum_{k=0}^\infty \frac{W_+^{k+B} W_-^k}{k! (k+B)!}
\end{equation}
Then, the integration in $\phi_{N_S}$ in Eq.~(\ref{zcan2}) is performed. It is 
advantegeous to set $w = \exp[-\i \phi_{N_S}]$ and map the interval $[-\pi,\pi)$ 
onto the unitary circle in the complex plane:

\begin{equation}\label{barcint}
 \frac{1}{2\pi \i} \oint \frac{\d w}{w}\; w^{-{N_S}} \Sigma(w) 
 \exp[{\tilde F}_{\rm M}(w)] 
\end{equation}
where:

\begin{equation}
 {\tilde F}_{\rm M}(w) = F_{\rm c}(\phi_Q,\phi_S,\phi_{N_S})_{\rm mesons} 
\end{equation}
and $\Sigma$ is the series in Eq.~(\ref{barint}). Since mesons can contain at 
most two strange quarks, the function ${\tilde F}_{\rm M}(w)$ can be written
as the sum of three terms:

\begin{equation}
 {\tilde F}_{\rm M}(w) = \alpha + \beta w +\gamma w^2
\end{equation} 
where $\alpha$, $\beta$ and $\gamma$ are the sums $\Sigma_j z_j \exp \, [-\i Q_j 
\phi_Q -\i S_j \phi_S]$ over mesons with 0, 1 and 2 strange quarks respectively.
As baryons may contain at most three strange quarks, the functions $W_+(w)$ and 
$W_-(w)$ in~(\ref{wcap}) are third-degree polynomials in $w$ with conjugate 
coefficients $A_0,\ldots,A_3$ depending on $\phi_Q$ and $\phi_S$, i.e.:

\begin{equation}\label{wcap2}
 W_+(w) = \sum_{n=1}^3 A_n w^n  \qquad  W_-(w) = \sum_{n=1}^3 A^*_n w^n 
\end{equation}
and, consequently, the series $\Sigma(w)$ in Eq.~(\ref{barint}) is an analytic 
function of $w$. Hence, the integrand in Eq.~(\ref{barcint}) has one pole of 
$N_S^{\rm th}$ order in $w=0$ and the integral (\ref{barcint}) turns out to be:

\begin{equation}
 \frac{\exp[\alpha]}{N_S!}\frac{\d^{N_S}}{\d z^{N_S}}\exp \, [\beta w+ 
 \gamma w^2] \Sigma(w) \vl_{w=0}
\end{equation}
The derivative can be calculated along any direction and particularly along
the real axis. Finally, the partition function in Eq.~(\ref{zcan2}) reads: 

\begin{eqnarray}\label{zcan3}
\!\!\!\!&& Z(\Qz) = \frac{1}{(2 \pi)^2} \int_{-\pi}^{\pi} \!\!\! \d \phi_Q
 \int_{-\pi}^{\pi} \!\!\! \d \phi_S \; \e^{\i Q \phi_Q + i S \phi_S 
 +\alpha(\phi_Q,\phi_S)} \nonumber \\
\!\!\!\!&& \times \frac{\D^{N_S} \Sigma(x) \exp \, [\beta(\phi_Q,\phi_S)x + \gamma
 (\phi_Q,\phi_S)x^2]|_{x=0}}{N_S!}
\end{eqnarray}  
which is to be integrated numerically within its two dimensional domain.\\ 
The next problem is how to calculate the derivative in the integrand in 
Eq.~(\ref{zcan3}). We first notice that the series $\Sigma$ can be calculated 
analytically for a real argument because $W_+(x) = W^*_-(x)$ 
(see Eq.~(\ref{wcap2})):

\begin{eqnarray}
 \Sigma(x) & = & \sum_{k=0}^\infty \frac{W_+(x)^{k+B} W^*_+(x)^k}{k! (k+B)!} \nonumber \\
  & = & {\rm I}_B(2 |W_+(x)|) \exp \, [\i B \arg W_+(x)] \nonumber \\
  & = & \frac{{\rm I}_B(2 |W_+(x)|)}{(2 |W_+(x)|)^B} \, (2 W_+(x))^B
\end{eqnarray}
where ${\rm I}_B$ is the modified Bessel function of order $B$. The derivative
of order ${N_S}$ is then expanded:

\begin{eqnarray}
\!\!\!\!\!\!\!\! && \D^{N_S} \Sigma(x) \exp \, [\beta x + \gamma x^2] = \nonumber \\
\!\!\!\!\!\!\!\! && = \sum_{k=0}^{N_S} {N_S \choose k} \D^k \exp \, [\beta x + \gamma x^2] 
 \D^{N_S-k} \Sigma(x)
\end{eqnarray}
It can be shown with a little algebra that:

\begin{equation}
 \D^k \exp[\beta x + \gamma x^2]\vl_{x=0} = k! \sum_{j \ge k/2}^k \frac{\beta^{2j-k} 
 \gamma^{k-j}}{(2j-k)!(k-j)!} 
\end{equation}
The derivative of $\Sigma(x)$ can be further expanded:

\begin{equation}\label{further}
 \D^{k} \Sigma(x) = \sum_{l=0}^k {k \choose l} \D^l \frac{{\rm I}_B(2 |W_+(x)|)}
 {(2 |W_+(x)|)^B} \, \D^{k-l} (2 W_+(x))^B 
\end{equation}  
The derivative of the first factor in the above sum can be calculated 
by taking advantage of recurrence relations for Bessel function derivatives:

\begin{equation}
 \D \frac{{\rm I}_B(2 |W_+(x)|)}{(2 |W_+(x)|)^B} = 
 \frac{{\rm I}_{B+1}(2 |W_+(x)|)}{(2 |W_+(x)|)^{B+1}} \,\, 2 \, \D |W_+(x)|^2
\end{equation}   
while for the second factor:

\begin{equation}
 \D (2W_+(x))^B = B (2W_+(x))^{B-1} \, \D W_+(x)
\end{equation}
The two equations above are the starting point for the numerical calculation
of the derivative in Eq.~(\ref{further}); a recursive algorithm has been 
implemented that computes it for $x=0$, taking into account that (see 
Eq.~(\ref{wcap2})):

\begin{eqnarray}
 && \frac{{\rm I}_B(2 |W_+(0)|)}{(2 |W_+(0)|)^B} = \frac{{\rm I}_B(2 |A_0|)}
 {(2 |A_0|)^B} \nonumber \\
 && \D^l |W_+(x)|^2\vl_{x=0} = l! \sum_{n=0}^l A_{n} A^*_{l-n} \nonumber \\
 && \D^l W_+(x)\vl_{x=0} = l! \, A_l \nonumber 
\end{eqnarray}  
    

\end{document}